\documentclass[aps,prd,preprint,groupedaddress,nofootinbib,showpacs,eqsecnum]{revtex4-1}
\usepackage{graphicx,color,amsmath,hyperref}
\usepackage{mathtools}
\usepackage{array}
\newif\ifdraft
\drafttrue
\newif\ifpreprint
\preprinttrue

\newcommand{\fancyM}{{ \cal K}_{\rm G} \, }
\newcommand{\ts}{{ }}
\newcommand{\fancyA}{{\cal K}_{\rm YM \,}}

\def\fig#1{Fig.~{\ref{#1}}}
\def\Fig#1{Fig.~{\ref{#1}}}
\def\figs#1#2{Figs.~{\ref{#1}} and~{\ref{#2}}}
\def\Figs#1#2{Figs.~{\ref{#1}} and~{\ref{#2}}}
\def\sect#1{Section~{\ref{#1}}}

\def\tab#1{Table~{\ref{#1}}}

\def\spa#1.#2{\left\langle#1\,#2\right\rangle}
\def\spb#1.#2{\left[#1\,#2\right]}
\def\tree{{\rm tree}}

\def\fiveloop{{\rm 5\hbox{-}loop}}

\def\eps{\epsilon}

\def\nn{\nonumber}

\def\E{{\cal E}}

\def\N#1{{N$^{#1}$MC}}
\def\be{\begin{eqnarray}}
\def\ee{\end{eqnarray}}
\def\G{{\rm GR}}

\def\eqn#1{Eq.~(\ref{#1})}

\def\eqns#1#2{Eqs.~(\ref{#1}) and~(\ref{#2})}

\def\NeqOne{{{\cal N}=1}}

\def\NeqFour{{{\cal N}=4}}
\def\NeqFive{{{\cal N} = 5}}

\def\NeqEight{{{\cal N}=8}}

\def\NkMC{{{\rm N}^k{\rm MC}}}

\def\BCJ{{\rm BCJ}}

\def\tn{{\tilde n}}

\def\tJ{{\tilde J}}

\def\E{{\cal E}}

\def\x#1{{\bullet}}

\def\f{\tilde f}

\def\bea{\begin{eqnarray}}
\def\eea{\end{eqnarray}}
\def\ba{\begin{eqnarray}}
\def\ea{\end{eqnarray}}

\def\tree{{\rm tree}}

\newbox\charbox
\newbox\slabox
\def\s#1{{      % Feynman slash
\setbox\charbox=\hbox{$#1$}
\setbox\slabox=\hbox{$/$}
\dimen\charbox=\ht\slabox
\advance\dimen\charbox by -\dp\slabox
\advance\dimen\charbox by -\ht\charbox
\advance\dimen\charbox by \dp\charbox
\divide\dimen\charbox by 2
\raise-\dimen\charbox\hbox to \wd\charbox{\hss/\hss}
\llap{$#1$} }}

\newcommand{\inlineblob}[1]{\, \vcenter{\hbox{\includegraphics[height=4em]{figs/#1}}}}

\newcommand{\inlinefig}[1]{\,\vcenter{\hbox{\includegraphics[height=3.2em]{figs/#1}}}}

\newcommand{\inlineg}[2]{\inlinefig{Vacuum#1loopsV#2.eps}}

\newcommand{\ie}{i.e.\ }

\begin{document}

\ifpreprint
\centerline{UCLA/18/TEP/102 \hfill NSF-ITP-18-031  \hfill Saclay IPhT-T18/024}
\centerline{NORDITA 2018-032 \hfill UUITP-13/18}
\fi

\title{Ultraviolet Properties of \texorpdfstring{$\NeqEight$}{N=8} Supergravity at Five Loops}

\author{Zvi~Bern${}^{a}$, John Joseph Carrasco${}^{b}$, Wei-Ming Chen${}^a$, Alex Edison${}^a$,\\
Henrik Johansson${}^{c,d}$, Julio Parra-Martinez${}^{a}$, Radu Roiban${}^{e}$ and Mao Zeng${}^a$}

\affiliation{
$\null$\\
${}^a$Mani L. Bhaumik Institute for Theoretical Physics\\
Department of Physics and Astronomy \\
University of California at Los Angeles\\
Los Angeles, CA 90095, USA \\
\vskip -4 mm
${}^b$Institute of Theoretical Physics (IPhT), \\
CEA-Saclay and University of Paris-Saclay\\ 
F-91191 Gif-sur-Yvette cedex, France\\
\vskip -4 mm
${}^c$Department of Physics and Astronomy, Uppsala University, 75108 Uppsala, Sweden \\
\vskip -4 mm
${}^d$Nordita, Stockholm University and \\
 KTH Royal Institute of Technology, \\
Roslagstullsbacken 23, 10691 Stockholm, Sweden \\
\vskip -4 mm
${}^e$Institute for Gravitation and the Cosmos, Pennsylvania State University, 
    University Park, PA 16802, USA \\
}

%\date{\today}

\begin{abstract}
We use the recently developed generalized double-copy construction to
obtain an improved representation of the five-loop four-point
integrand of $\NeqEight$ supergravity whose leading ultraviolet
behavior we analyze using state-of-the-art loop-integral expansion and
reduction methods.  We find that the five-loop critical dimension
where ultraviolet divergences first occur is $D_c=24/5$,
corresponding to a $D^8 R^4$ counterterm.  This ultraviolet behavior
stands in contrast to the cases of four-dimensional $\NeqFour$
supergravity at three loops and $\NeqFive$ supergravity at four loops
whose improved ultraviolet behavior demonstrates enhanced
cancellations beyond implications from standard-symmetry
considerations.  We express this $D_c=24/5$ divergence in terms of two
relatively simple positive-definite integrals reminiscent of vacuum
integrals, excluding any additional ultraviolet cancellations at this
loop-order.  We note nontrivial relations between the integrals
describing this leading ultraviolet behavior and integrals describing
lower-loop behavior.  This observation suggests not only a path
towards greatly simplifying future calculations at higher loops, but
may even allow us to directly investigate ultraviolet behavior in
terms of simplified integrals, avoiding the construction of complete
integrands.
\end{abstract}

\pacs{04.65.+e, 11.15.Bt, 11.25.Db, 12.60.Jv \hspace{1cm}}

\maketitle

\vskip -2.5 cm 
\tableofcontents

\section{Introduction}

Since the discovery of supergravity theories~\cite{Supergravity}, a
complete understanding of their ultraviolet properties has remained
elusive. Despite tremendous progress over the years, many properties of
gravitational perturbation theory remain unknown.
  Power counting arguments, driven by the dimensionality of
Newton's constant, suggest that all point-like theories of gravity
should develop an ultraviolet divergence at a sufficiently high loop
order.  However, if a point-like theory were ultraviolet finite, it
would imply the existence of an undiscovered symmetry or structure
that should likely have a fundamental impact on our understanding of
quantum gravity.
Explicit calculations in recent years have revealed the existence of
hidden properties, not readily apparent in Lagrangian
formulations. One might wonder whether these tame the ultraviolet
behavior of point-like gravity theories.  For example, all-loop-order
unitarity cuts exhibit remarkable infrared and ultraviolet
cancellations~\cite{SurprisingCancellations} whose consequences remain
to be fully explored.
Indeed, we know of examples in $\NeqFour$~\cite{NeqFourSugra} and $\NeqFive$~\cite{NeqFiveSugra} supergravity
theories that display ``enhanced
cancellations''~\cite{N4GravThreeLoops, N5GravFourLoops, HalfMaxD5,
  IntegralRelations,NoExplanation}, where quantum corrections exclude
counterterms thought to be consistent with all known symmetries.  In addition, there
are indications that anomalies in known symmetries of supergravity
theories play a role in the appearance of ultraviolet
divergences~\cite{Anomaly,N4GravFourLoop}.  Restoration of these
symmetries in S-matrix elements by finite local counterterms may lead
to the cancellation of known divergences.  
In this paper, we take a step forward by presenting a detailed
analysis of the ultraviolet behavior of the five-loop four-point
scattering amplitude in the maximally supersymmetric theory,
$\NeqEight$ supergravity\footnote{Strictly speaking the maximally
  supersymmetric theory is only recognized as $\NeqEight$ supergravity
  in four dimensions.  While we concern ourselves with mainly higher
  dimensions, in this paper we take the liberty to apply the
  four-dimensional nomenclature.}~\cite{NeqEightSugra}, and observe
properties that should help us determine its four-dimensional
ultraviolet behavior at even higher loops. 

Its many symmetries suggest that, among the point-like theories of
gravity, the maximally supersymmetric theory has the softest
ultraviolet behavior.  These symmetry properties also make it
technically easier to explore and understand its structure.  Over the
years there have been many studies and predictions for the ultraviolet
behavior of $\NeqEight$ supergravity~\cite{GSB,N8Predictions}.  The
current consensus, based on standard symmetry considerations, is that
$\NeqEight$ supergravity in four dimensions is ultraviolet finite up
to at least seven loops~\cite{SevenLoopGravity, BjornssonAndGreen, VanishingVolume}.
Through four loops, direct computation using modern scattering
amplitude methods prove that the critical dimension of $\NeqEight$
supergravity where divergences first occur is~\cite{BDDPR, ThreeFourloopN8, ck4l}
\begin{equation}
D_c = \frac{6}{L} + 4 \,, \hskip 1 cm (2\le L \le 4)
\label{YMCritical}
\end{equation}
where $L$ is the number of loops. This matches the formula~\cite{BDDPR,CompleteFourLoopSYM} for
$\NeqFour$ super-Yang--Mills theory~\cite{N4YM}
which is known to be an ultraviolet finite theory in
$D=4$~\cite{FinitenessN4YM}.  At one loop the critical dimension, for
both $\NeqFour$ super-Yang--Mills theory and $\NeqEight$
supergravity~\cite{GSB}, is $D_c = 8$.   We define the
theories in dimensions $D>4$ via dimensional reduction of $\NeqOne$ supergravity
in $D=11$ and $\NeqOne$ super-Yang--Mills theory in $D=10$~\cite{GSB}.

In this paper we address the longstanding question of whether
\eqn{YMCritical} holds for $\NeqEight$ supergravity at five loops.
Symmetry arguments~\cite{BjornssonAndGreen} suggest $D^8 R^4$ as a
valid counterterm and that the critical dimension for the five-loop
divergence should be $D_c = 24/5$ instead of that suggested by
\eqn{YMCritical}, $D_c = 26/5$. (See also
Refs.~\cite{SevenLoopGravity,VanishingVolume}.) 
Such arguments, however, cannot ascertain whether quantum corrections actually generate
an allowed divergence.  Indeed, explicit three-loop calculations in
$\NeqFour$ supergravity and four-loop calculations in $\NeqFive$
supergravity reveal that while counterterms are allowed by all known
symmetry considerations, none actually exist~\cite{N4GravThreeLoops,
  N5GravFourLoops}.  These enhanced cancellations are nontrivial and
only manifest upon applying Lorentz invariance and a reparametrization
invariance to the loop integrals~\cite{IntegralRelations}.
This implies that the only definitive way to settle the five-loop question is
to directly calculate the coefficient of the potential $D^8 R^4$
counterterm in $D= 24/5$, as we do here.
 This counterterm is of interest because it is the one that would
contribute at seven loops if $\NeqEight$ supergravity were to diverge
in $D=4$. 

Our direct evaluation of the critical dimension of the $\NeqEight$
supergravity theory at five loops proves unequivocally that it first
diverges in $D_c = 24/5$ and no enhanced cancellations are observed.
The fate of $\NeqEight$ supergravity in four-dimensions remains to be
determined. Even with the powerful advances exploited in this current
calculation, direct analysis at seven loops would seem out of
reach. Fortunately the results of our current analysis, when combined
with earlier work at lower loops~\cite{ThreeFourloopN8, ck4l,
  N4GravFourLoop, N4GravThreeLoops,N4SugraMatter,N5GravFourLoops},
reveal highly nontrivial constraints on the subloops of integrals
describing the leading ultraviolet behavior through five loops.  These
patterns suggest not only new efficient techniques to directly determine
the ultraviolet behavior at ever higher loops, but potentially
undiscovered principles governing the ultraviolet consistency.  In
this work we will describe these observed constraints, leaving their
detailed study for the future.

The results of this paper are the culmination of many advances in
understanding and computing gauge and gravity scattering amplitudes at
high-loop orders.  The unitarity
method~\cite{GeneralizedUnitarity,MaximalCutMethod} has been central
to this progress because of the way that it allows on-shell
simplifications to be exploited in the construction of new higher-loop amplitudes. 
We use its incarnation in the maximal-cut
organization~\cite{MaximalCutMethod} to systematically build complete
integrands~\cite{GeneralizedDoubleCopy,FiveLoopN8Integrand}.

The unitarity method combines naturally with double-copy ideas, including
the field-theoretic version of the string-theory Kawai, Lewellen and
Tye (KLT) relations between gauge and gravity tree
amplitudes~\cite{KLT} and the related Bern, Carrasco and Johansson
(BCJ) color-kinematics duality and double-copy
construction~\cite{BCJ,BCJLoop}.  The double-copy relationship reduces
the problem of constructing gravity integrands to that of calculating
much simpler gauge-theory ones.  For our calculation, a
generalization~\cite{GeneralizedDoubleCopy} of the double-copy
procedure has proven invaluable~\cite{FiveLoopN8Integrand}.

The analysis in Ref.~\cite{FiveLoopN8Integrand} finds the first representation of an 
integrand for the five-loop four-point amplitude of $\NeqEight$ supergravity.  The high
power counting of that representation obstructs
the necessary integral reductions needed  to extract its ultraviolet behavior. 
Here we use similar generalized double-copy methods~\cite{GeneralizedDoubleCopy} to construct an 
improved integrand that enormously
simplifies the integration.  The key is starting with an 
improved gauge-theory integrand, which we build by
constraining a manifest-power-counting ansatz  via the
method of maximal cuts. The needed
unitarity cuts are easily obtained from the
gauge-theory integrand of Ref.~\cite{FiveLoopN4}.  

The earlier representation of the supergravity integrand, given in
Ref.~\cite{FiveLoopN8Integrand}, is superficially (though not
actually) quartically divergent in the dimension of interest.  The new
representation shifts these apparent quartic divergences to
contributions that only mildly complicate the extraction of the
underlying logarithmic divergences.  Our construction proceeds as
before except for small differences related to avoiding
certain spurious singularities.  We include the complete gauge and
supergravity integrands in plain-text ancillary
files~\cite{AttachedFile}.

Recent advances in loop integration methods proved essential for
solving the challenges posed by the calculation of ultraviolet
divergences at five loops.  Related issues appeared in the five-loop
QCD beta function calculation, which was completed
recently~\cite{FiveloopQCDBeta}.  For supergravity, higher-rank-tensors
related to the nature of the graviton greatly increase the
number of terms while the absence of subdivergences dramatically
simplifies the calculation.
At high-loop orders the primary method for reducing loop integrals to
a basis relies on integration-by-parts (IBP)
identities~\cite{IBPRefs,SmirnovBook}.  The complexity of such IBP
systems tends to increase prohibitively with the loop order and the
number of different integral types.  Ideas from algebraic geometry
provide a path to mitigating this problem by organizing them in a way
compatible with unitarity methods~\cite{Gluza2010ws, CutIntegrals,
  IBPAdvances, Zhang2016kfo}.  We also simplify the problem by
organizing the IBP identities in terms of an SL($5$) symmetry of the
five-loop integrals~\cite{IntegralRelations}.

The final expression for the leading ultraviolet behavior is
incredibly compact, and exposes, in conjunction with previous
results~\cite{ThreeFourloopN8, ck4l, N4GravFourLoop, N4GravThreeLoops,
  N4SugraMatter, N5GravFourLoops}, simple and striking
patterns.  Indeed, analysis of this leading ultraviolet behavior
indicates the existence of potentially more powerful methods for making
progress at higher loops.

This paper is organized as follows.  In \sect{ReviewSection}, we
review the generalized double-copy construction, as well as the
underlying ideas including BCJ duality and the method of maximal
cuts. We also summarize properties of the previously constructed
five-loop four-point integrand of Ref.~\cite{FiveLoopN8Integrand}.  In
\sect{ImprovedIntegrandSection}, we construct new $\NeqFour$
super-Yang--Mills and $\NeqEight$ supergravity integrands with
improved power-counting properties.  Then, in
\sect{VacuumExpansionSection} describe our procedure for expanding the
integrands for large loop momenta, resulting in integrals with no external momenta,
which we refer to as vacuum integrals.  In
\sect{TopLevelIntegrationSection}, as a warm up to the complete
integral reduction described in \sect{AllIntegrationSection}, we
simplify the integration-by-parts system of integrals by assuming that the only
contributing integrals after expanding in large loop momenta are those 
with maximal cuts.  The results for the
five-loop ultraviolet properties are given in these sections.  In
\sect{PatternsSection}, by collecting known results for the leading
ultraviolet behavior in terms of vacuum
integrals we observe and comment on the intriguing and nontrivial
consistency for such integrals between higher and lower loops.
We present our conclusions in \sect{ConclusionSection}.

%%%%%%%%%%%%%%%%%%%%%%%%%%%%%
\section{Review}
\label{ReviewSection}

The only known practical means for constructing higher-loop gravity
integrands is the double-copy procedure that recycles gauge-theory
results into gravity ones.
Whenever gauge-theory integrands are available in forms that manifest the BCJ
duality between color and kinematics~\cite{BCJ,BCJLoop}, the
corresponding (super)gravity integrands are obtained by
replacing color factors with the kinematic numerators of the same or
of another gauge theory.
Experience shows that it is sometimes difficult to find such
representations of gauge-theory integrands.  In some cases this
can be overcome by increasing the power count of individual
terms~\cite{OConnelHighPowerocunt}, or by introducing
nonlocalities in integral coefficients~\cite{FivePointN4}. Another
possibility is to find an integrand where BCJ duality holds 
on every cut, but does not hold with cut conditions removed~\cite{BCJonCuts}.
Unfortunately, these ideas have not, as yet, led to a BCJ
representation of the five-loop four-point integrand of $\NeqFour$
super-Yang--Mills theory.

To avoid this difficulty, a generalized version of the BCJ double-copy
construction has been developed. Although
relying on the existence of BCJ duality at tree level, the generalized
double-copy construction does not use any explicit representation of
tree- or loop-level amplitudes that satisfies BCJ duality. It instead
gives an algorithmic procedure which converts generic gauge-theory
integrands into gravity ones~\cite{GeneralizedDoubleCopy}.
This is used in Ref.~\cite{FiveLoopN8Integrand} to construct 
an integrand for the five-loop four-point amplitude of $\NeqEight$ 
supergravity. 

In this section we give an overview of the ingredients and methods
used in the construction of the five-loop integrand.  We begin with a
brief review of BCJ duality and the maximal-cut method which underlies
and organizes the construction, and then proceed to reviewing the
generalized double copy and associated formulae. We then summarize
features of the previously constructed
integrand~\cite{FiveLoopN8Integrand} for the five-loop four-point
amplitude of $\NeqEight$ supergravity.  In
\sect{ImprovedIntegrandSection} we use the generalized double copy to
find a greatly improved integrand for extracting ultraviolet
properties, which we do in subsequent sections.

\subsection{BCJ duality and the double copy}

The BCJ duality~\cite{BCJ,BCJLoop} between color and kinematics is a
property of on-shell scattering amplitudes which has so far been
difficult to discern in a Lagrangian formulation of Yang-Mills field
theories~\cite{Square,WeinzierlBCJ}.
Nevertheless various tree-level proofs exist~\cite{BjerrumMomKernel}.

The first step to construct a duality-satisfying representation of
amplitudes is to organize them in terms of graphs with only cubic (trivalent)
vertices.  This process works for any tree-level amplitude in any
$D$-dimensional gauge theory coupled to matter fields.  For the
adjoint representation case, an $m$-point tree-level amplitude may be
written as
\begin{equation}
\mathcal{A}^{\tree }_{m} = g^{m-2} \sum_{j}
\frac{c_{j} n_{j}}{\prod_{\alpha_{j}}p^{2}_{\alpha_{j}}} \,,
\label{CubicRepresentation}
\end{equation}
where the sum is over the $(2m-5)!!$ distinct tree-level graphs with
only cubic vertices.  Such graphs are the only ones needed
because the contribution of any diagram with quartic or higher-point
vertices can be assigned to a graph with only cubic vertices by
multiplying and dividing by appropriate propagators.
The nontrivial kinematic information is contained in the kinematic
numerators $n_{j}$; they generically depend on momenta, polarization,
and spinors.
The color factors $c_j$ are obtained by dressing every vertex in graph
$j$ with the group theory structure constant,
$\tilde{f}^{abc}=i\sqrt{2}f^{abc} = \mathrm{Tr}([T^{a},T^{b}] T^{c})$,
where the Hermitian generators of the gauge group are normalized via
$\mathrm{Tr}(T^{a}T^{b})=\delta^{a b}$. The denominator is given by
the product of the Feynman propagators of each graph $j$.

%%%%%%%%%%%%%% FIGURE %%%%%%%%%%%%% 
\begin{figure}
\includegraphics[scale=.41]{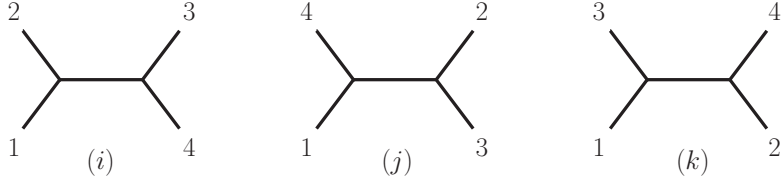}
\vskip -.3 cm 
\caption{The three four-point diagrams participating in 
either color or numerator Jacobi identities.
   }
\label{JacobiFigure}
\end{figure}
%%%%%%%%%%%%%%%%%%%%%%%%%%%    

The kinematic numerators of an amplitude in a BCJ representation obey
the same algebraic relations as the color factors~\cite{BCJ,BCJLoop,
  ck4l, Review}.  The key property is the requirement that all Jacobi
identities obeyed by color factors are also obeyed by the kinematic numerators,
\begin{equation}
c_{i}+c_{j}+c_{k} = 0\ \Rightarrow\ n_{i}+n_{j}+n_{k} = 0 \, ,
\label{BCJDuality}
\end{equation}
where $i$, $j$, and $k$ refer to three graphs which are identical
except for one internal edge.  \Fig{JacobiFigure} shows three basic
diagrams participating in the Jacobi identity for color or numerator
factors. They can be embedded in a higher-point diagram.
Furthermore, the kinematic numerators should obey the same
antisymmetry under graph vertex flips as the color factors.
A duality-satisfying representation of an amplitude can be obtained
from a generic one through generalized gauge transformations---shifts
of the kinematic numerators,
\begin{equation}
 n_i \rightarrow n_i + \Delta_i \,,
\label{GGT}
\end{equation} 
which are constrained not to change the amplitude.  When the duality
is manifest, the kinematic Jacobi relations~\eqref{BCJDuality} express
all kinematic numerators in terms of a small set of ``master'' numerators. 
While there is a fairly large freedom in choosing
them, only the numerators of certain graphs can form such a basis.

Once gauge-theory tree amplitudes have been arranged into a form where
the duality is manifest~\cite{BCJ,BCJLoop}, we obtain corresponding
gravity amplitudes simply by replacing the color factors of one
gauge-theory amplitude with the kinematic numerators of another
gauge-theory amplitude,
\begin{equation}
c_{i}\ \rightarrow\ \tilde{n}_{i}\,,
\label{ColorSub}
\end{equation}
as well as readjusting the coupling constants. 
This replacement gives the double-copy form of a gravity tree amplitude,
\begin{equation}
\mathcal{M}^{\tree}_{m} = i \left(\frac{\kappa}{2}\right)^{m-2}
\sum_{j}\frac{\tilde{n}_{j}n_{j}}{\prod_{\alpha_{j}}p^{2}_{\alpha_{j}}} \,,
\label{DoubleCopy}
\end{equation}
where $\kappa$ is the gravitational coupling and $\tilde{n}_j$ and
$n_j$ are the kinematic numerator factors of the two gauge theories.
The gravity amplitudes obtained in this way depend on the specific input
gauge theories. 
As discussed in Refs.~\cite{BCJLoop,Square}, \eqn{DoubleCopy} holds
provided that at least one of the two amplitudes satisfies the duality
(\ref{BCJDuality}) manifestly.  The other may be in an arbitrary
representation.

An earlier related version of the double-copy relation valid at tree
level is the KLT relations between gauge and gravity
amplitudes~\cite{KLT}.  Their general form in terms of a basis of
gauge-theory amplitudes is,
\begin{align}
\mathcal{M}^{\tree }_{m}= & \null
 i \left(\frac{\kappa}{2}\right)^{m-2}
\smashoperator{\sum_{\tau,\rho \in {\cal S}_{m-3}}}K(\tau|\rho)
\tilde A^{\tree}_{m}\left(1,\rho_2, \ldots, \rho_{m-2}, m, (m-1)\right)\nonumber
 \\
& \hskip 3.4 cm \times
A^{\tree}_{m}\left(1,\tau_2, \ldots, \tau_{m-2}, (m-1), m\right) \,.
\label{KLTRelations}
\end{align}
Here the $A^{\tree}_{m}$ are color-ordered tree amplitudes with the
indicated ordering of legs and the sum runs over $(m-3)!$ permutations
of external legs.  The KLT kernel $K$ is a matrix with indices
corresponding to the elements of the two orderings of the relevant
partial amplitudes.  It is also sometimes referred to as the momentum kernel.  Compact
representations of the KLT kernel are found in
Refs.~\cite{OneloopN8,BjerrumMomKernel,abelianZ}.

%%%%%%%%%%%%%%%%%%%%%%%%%%%%%%%%%%

At loop-level, the duality between color and kinematics
(\ref{BCJDuality}) remains a conjecture~\cite{BCJLoop}, although
evidence continues to
accumulate~\cite{FivePointN4,OtherExamples,ck4l,BCJDifficulty}.
As at tree level, loop-level amplitudes in a gauge theory coupled to
matter fields in the adjoint representation can be expressed as a sum
over diagrams with only cubic (trivalent) vertices:
\begin{equation}
\mathcal{A}^{L\hbox{-}\mathrm{loop}}_{m}=i^{L} g^{m-2+2 L}
\sum_{\mathcal{S}_{m}}\sum_{j}\int\prod_{l=1}^{L}\frac{d^{D}p_{l}}{(2\pi)^{D}}
\frac{1}{S_{j}}\frac{c_{j} n_{j}}{\prod_{\alpha_{j}}p^{2}_{\alpha_{j}}}\,.
\label{CubicRepLoop}
\end{equation}
The first sum runs over the set $\mathcal{S}_{m}$  of $m!$ permutations of the external legs.  The second sum runs over the distinct
 $L$-loop $m$-point graphs with only cubic
vertices; as at tree level, by multiplying and dividing by propagators
it is trivial to absorb numerators of contact diagrams that contain
higher-than-three-point vertices into numerators of diagrams with
only cubic vertices.
The symmetry factor $S_{j}$ counts the number of automorphisms of
the labeled graph $j$ from both the permutation sum and from any internal
automorphism symmetries.  This symmetry factor is not
included in the kinematic numerator.

The generalization of BCJ duality to loop-level amplitudes amounts to
demanding that all diagram numerators obey the same algebraic
relations as the color factors~\cite{BCJLoop}.  The Jacobi identities
are implemented by embedding the three diagrams in \fig{JacobiFigure}
into loop diagrams in all possible ways and demanding that
identities of the type in \eqn{BCJDuality} hold for the loop-level numerators
as well.  In principle, given any representation of an amplitude, one
may attempt to construct a duality-satisfying one by modifying the
kinematic numerators through generalized gauge
transformations~\eqref{GGT}; however, a more systematic approach is to
start with an ansatz exhibiting certain desired properties and
impose the kinematic Jacobi relations.
As at tree level, when the duality is manifest all
kinematic numerators are expressed in terms of those of a small number of ``master
diagrams''~\cite{FivePointN4,ck4l}.

Just like with tree numerators, once gauge-theory numerator factors
which satisfy the duality are available, replacing the color factors by
the corresponding numerator factors~(\ref{ColorSub}) yields the
double-copy form of gravity loop integrands,
\begin{align}
\mathcal{M}^{L\hbox{-}\mathrm{loop}}_{m} =
i^{L+1} \left(\frac{\kappa}{2}\right)^{m-2+2 L}
\sum_{\mathcal{S}_{m}}\sum_{j}\int\prod_{l=1}^{L}\frac{d^{D}p_{l}}{(2\pi)^{D}}
\frac{1}{S_{j}}\frac{\tilde{n}_{j}n_{j}}{\prod_{\alpha_{j}}p^{2}_{\alpha_{j}}}
\,,
\label{DoubleCopyLoop}
\end{align}
where $\tilde{n}_j$ and $n_j$ are gauge-theory numerator factors.  The
theories to which the gravity amplitudes belong are dictated by the
choice of input gauge theories.

Thus, the double-copy construction reduces the problem of constructing
loop integrands in gravitational theories to the problem of finding BCJ
representations of gauge-theory amplitudes.\footnote{Through four
  loops, there exist BCJ representations of $\NeqFour$
  super-Yang--Mills amplitudes that exhibit the same graph-by-graph
  power counting as the complete amplitude, {\it i.e.} all ultraviolet
  cancellations are manifest. It is an interesting open problem whether
  this feature will continue at higher loops.}
Apart from offering a simple means for obtaining loop-level scattering
amplitudes in a multitude of (super)gravity theories, the double-copy
construction has also been applied to the construction of black-hole
and other classical solutions~\cite{ClassicalSolutions} including
those potentially relevant to gravitational-wave
observations~\cite{RadiationSolutions}, corrections to gravitational
potentials~\cite{Donoghue}, and the relation between symmetries of
supergravity and gauge theory~\cite{SugraSyms, DoubleCopyTheories,
  DoubleCopyTheoriesFund}.  The duality underlying the double copy has
also been identified in a wider class of quantum field and string
theories~\cite{BLGBCJ,NLSMBCJ,abelianZ,NLSMaction,
  DiskandHeteroticStringBCJ, ConformalGravity}, including those with fundamental
representation matter~\cite{FundMatter}.  For recent reviews, see
Ref.~\cite{Review}.

When it turns out to be difficult to find a duality-satisfying
representation of a gauge-theory amplitude, as in the case for the
five-loop four-point amplitude of $\NeqEight$ supergravity, an
alternative method is available.  We use the generalized double-copy
procedure~\cite{GeneralizedDoubleCopy} that relies only on the
existence of duality-consistent properties at tree-level.
This type of approach may also potentially aid applications of
BCJ duality to problems in classical gravity.

%%%%%%%%%%%%%%%%%%%%%%%%%%%
\subsection{Method of maximal cuts}

The generalized double-copy construction of
Refs.~\cite{GeneralizedDoubleCopy,FiveLoopN8Integrand} relies on the
interplay between the method of maximal cuts~\cite{MaximalCutMethod}
and tree-level BCJ duality.  The maximal-cut method is a refinement of
the generalized-unitarity method~\cite{GeneralizedUnitarity}, designed
to construct the integrand from the simplest set of generalized
unitarity cuts.  In the generalized double-copy approach we apply the
maximal-cut method in a constructive way, assigning missing
contributions to new higher-vertex contact diagrams as necessary.

%%%%%%%%%%%%%% FIGURE  %%%%%%%%%%%%%%                                          
\begin{figure}[t]
  \includegraphics[clip,scale=0.54]{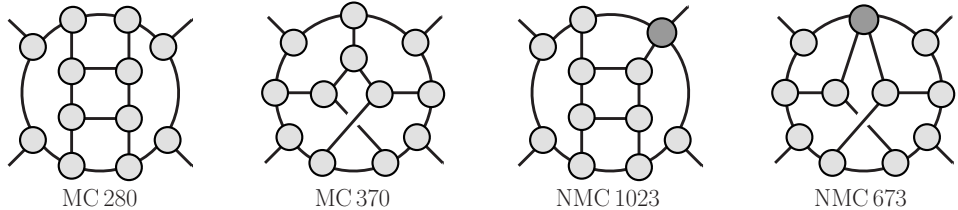}
\vskip -.2 cm
\caption[a]{Sample maximal and next-to-maximal cuts.  The exposed
  lines connecting the blobs are taken to be on shell delta-functions.
   }
\label{MCandNMCFigure}
\end{figure}
%%%%%%%%%%%%%%%%%%%%%%%%%%%%%%%%%%%%%

%%%%%%%%%%%%%% FIGURE  %%%%%%%%%%%%%%
\begin{figure}[t]
  \includegraphics[clip,scale=0.54]{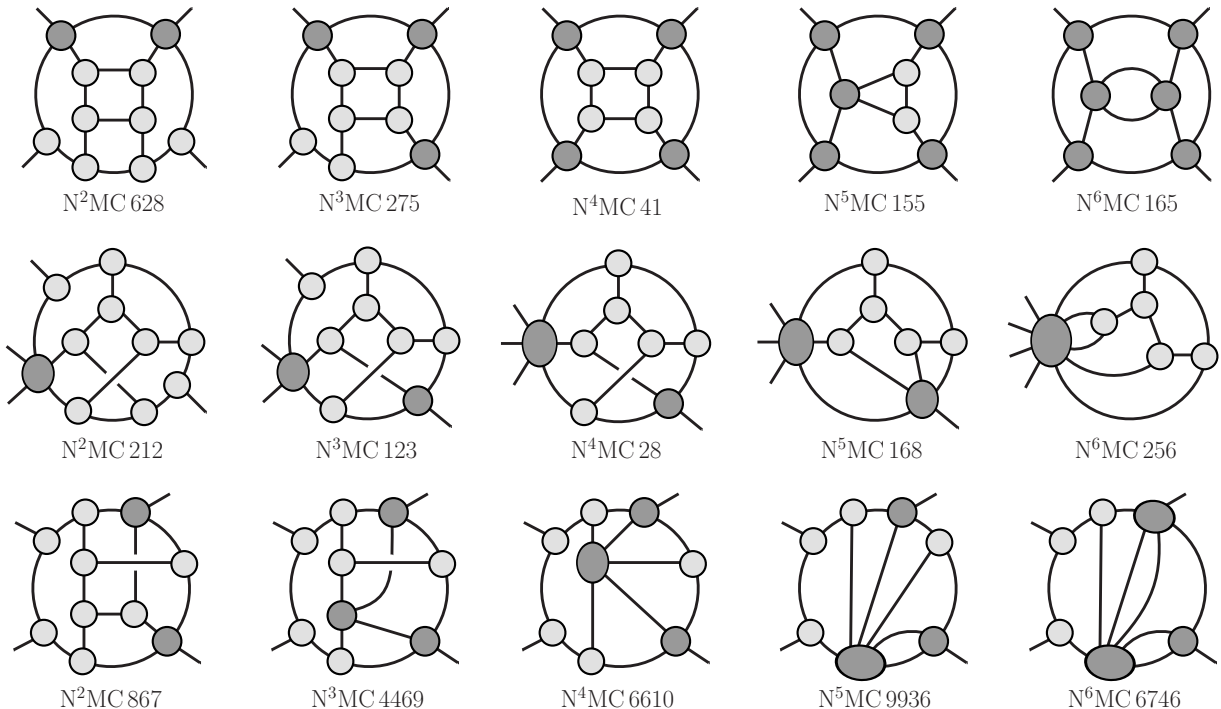}
\vskip -.2 cm
\caption[a]{Sample \N{k}s used in the construction of five-loop four-point amplitudes.
   The exposed lines connecting the blobs are taken to be on-shell delta-functions. 
}
\label{NkMaxSampleCutsFigure}
\end{figure}
%%%%%%%%%%%%%%%%%%%%%%%%%%%%%%%%%%%%%

In both gauge and gravity theories, the method of maximal
cuts~\cite{MaximalCutMethod} constructs multiloop integrands from
generalized-unitarity cuts that decompose loop integrands into
products of tree amplitudes,
\begin{align}
{\cal C}^\NkMC & = \sum_{\rm states} {\cal A}_{m(1)}^{\rm tree} \cdots {\cal A}_{m(p)}^{\rm tree}\,,
&   k \equiv \sum^p_{i=1}m(i)-3p \,,
\label{GeneralizedCut}
\end{align}
where the ${\cal A}_{m(i)}^{\rm tree}$ are tree-level
$m(i)$-multiplicity amplitudes corresponding to the blobs illustrated
for various five-loop examples in \figs{MCandNMCFigure}{NkMaxSampleCutsFigure}.  
We organize these cuts according to levels that correspond to the number $k$ of internal 
propagators that remain off shell.  

When constructing gauge-theory amplitudes, we use tree amplitudes
directly as in \eqn{GeneralizedCut}.  For $\NeqFour$ super-Yang--Mills
it is very helpful to use a four-dimensional on-shell
superspace~\cite{Nair} to organize the state sums~\cite{SuperSums}.
Some care is needed to ensure that the obtained expressions are valid
in $D$ dimensions, either by exploiting cuts whose supersums are valid
in $D \le 10$ dimensions~\cite{CompleteFourLoopSYM,FiveLoopN4} or
using six-dimensional helicity~\cite{SixDimSuperSpace}.  Once we have
one version of a gauge-theory integrand, we can avoid re-evaluating
the state sums to find new representations, simply by using the cuts
of the previously constructed integrand instead of
\eqn{GeneralizedCut} to construct target expressions.  In the same
spirit, for $\NeqEight$ supergravity we can always bypass
\eqn{GeneralizedCut} by making use of the KLT tree relations
(\ref{KLTRelations}).  The state sums also factorize allowing us to
express the $\NeqEight$ supergravity cuts directly in terms of
color-order $\NeqFour$ super-Yang--Mills cuts.  (See Section 2 of
Ref.~\cite{FiveLoopN8Integrand} for further details).

\Figs{MCandNMCFigure}{NkMaxSampleCutsFigure} give examples of cuts
used in the construction of the integrands of five-loop four-point
amplitudes.  At the maximal-cut (MC) level, e.g.~the first two
diagrams of \fig{MCandNMCFigure}, the maximum number of internal lines
are placed on shell and all tree amplitudes appearing in
\eqn{GeneralizedCut} are three-point amplitudes.
At the next-to-maximal-cut (NMC) level, e.g.~the third and fourth
diagrams of \fig{MCandNMCFigure}, all except one internal line are
placed on shell shell; all tree amplitudes are three-point amplitudes
except one which is a four-point amplitude. Similarly, for an \N2,
two internal lines are kept off shell and so forth, as illustrated in
\fig{NkMaxSampleCutsFigure}.

In the method of maximal cuts, integrands for loop amplitudes are
obtained by first finding an integrand whose
maximal cuts reproduce the direct calculation of maximal cuts
in terms of sums of products of three-point tree-level amplitudes.
This candidate integrand is then corrected by adding to it contact terms such
that all NMCs are correctly reproduced and systematically proceeding
through the next$^k$-maximal cuts (\N{k}s), until no further
corrections are necessary.  The level where this happens is determined
by the power counting of the theory and by choices made at earlier
levels.
For example, for five-loop amplitudes in $\NeqFour$ super-Yang--Mills
theory, cuts through the \N{3} level are needed, though as we
describe in the next section, it is useful to skip certain ill-defined
cuts at the \N{2} and \N{3} level and then recover the missing
information by including instead certain \N{4} level cuts.  For the
four-point $\NeqEight$ supergravity amplitude at the same loop order,
cuts through the \N{6} level are necessary.
In general, it is important to evaluate more cuts than the spanning
set (necessary for constructing the amplitude) to gain nontrivial
crosschecks of the results.  For example, in
Ref.~\cite{FiveLoopN8Integrand} all \N{7} cuts and many \N{8} cuts
were checked, confirming the construction.

%%%%%%%%%%%%%% FIGURE  %%%%%%%%%%%%%%
\begin{figure}[t]
  \includegraphics[clip,scale=0.54]{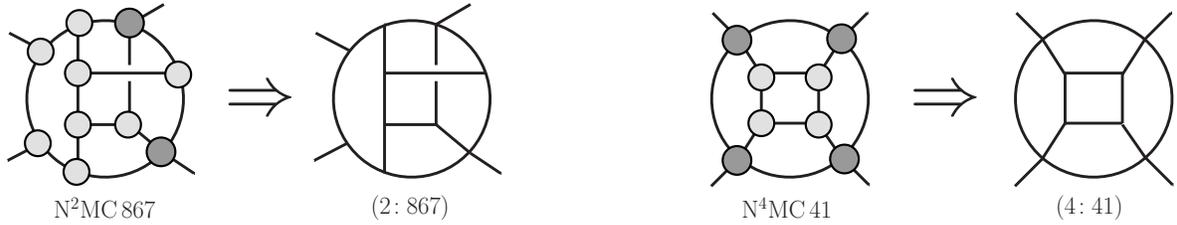}
\vskip -.2 cm
\caption[a]{New contribution found via the method of maximal cuts can
  be assigned to contact terms. The labels (X:\,Y) correspond to the
  labeling of Ref.~\cite{FiveLoopN8Integrand} and refer to the level
  and contact diagram number.  }
\label{CutToContactsFigure}
\end{figure}
%%%%%%%%%%%%%%%%%%%%%%%%%%%%%%%%%%%%%

To make contact with color/kinematics-satisfying representations of
gauge-theory amplitudes it is convenient to absorb all contact terms
into diagrams with only cubic vertices~\cite{ThreeFourloopN8, ck4l,
  N4GravFourLoop, N4GravThreeLoops, N4SugraMatter,
  N5GravFourLoops,BCJDifficulty}.  For problems of the complexity of
the five-loop supergravity integrand, however, it can be more
efficient to assign each new contribution of an \N{k} to a contact
diagram instead of to parent diagrams, consisting of ones
with only cubic vertices. These new contributions are, by
construction, contact terms---they contain only the propagators of the
graph with higher-point vertices---because any contribution that can
resolve these vertices into propagator terms is already accounted for
at earlier levels.
In this organization each new contact diagram can be determined
independently of other contact diagrams at the same level and depends
only on choices made at previous levels. 
More explicitly, as illustrated in \fig{CutToContactsFigure}, a new
contribution arising from an \N{k} is assigned to a contact diagram
obtained from that cut by replacing the blobs representing tree-level
amplitudes by vertices with the same multiplicity.
The contact terms should be taken off shell by removing the cut
conditions in a manner that reflects the diagram symmetry.  Off-shell
continuation necessarily introduces an ambiguity since it is always
possible to include terms proportional to the inverse propagators that
vanish by the cut condition; such ambiguities can be absorbed into
contact terms at the next cut level.

%%%%%%%%%%%%%%%%%%%%%%%%%%%
\subsection{Generalized double-copy construction}
Whenever gauge-theory amplitudes are available in a form that obeys
the duality between color and kinematics, the BCJ double-copy
construction provides a straightforward method of obtaining the
corresponding (super)gravity amplitudes. If a duality-satisfying
representation is expected to exist but is nonetheless unavailable,
the generalized double-copy construction supplies the additional
information necessary for finding the corresponding (super)gravity
amplitude.  Below we briefly summarize this procedure.  A more
thorough discussion can be found in Ref.~\cite{FiveLoopN8Integrand}.

The starting point of the construction is a ``naive double copy''of
two (possibly distinct) gauge-theory amplitudes written in terms of
cubic diagrams obtained by applying the double-copy substitution
\eqref{ColorSub} to these amplitudes despite none of them manifesting the
BCJ duality between color and kinematics.
While the resulting expression is not a (super)gravity amplitude, it
nonetheless reproduces the maximal and next-to-maximal cuts of the
desired (super)gravity amplitude as the three- and four-point
tree-level amplitudes entering these cuts obey the duality between
color and kinematics.  Contact term corrections are necessary to
satisfy the \N{k}  with $k\ge 2$; the method of maximal cuts can
be used to determine them.
For \N2 and \N3 at five loops, whose associated contact terms are the most 
complicated~\cite{GeneralizedUnitarity,FiveLoopN4}, it is advantageous
to obtain these corrections using formulas that express the cuts in
terms of violations of the BCJ relations~\eqref{BCJDuality}.

%%%%%%%%%%%%%% FIGURE %%%%%%%
\begin{figure}
\includegraphics[scale=.54]{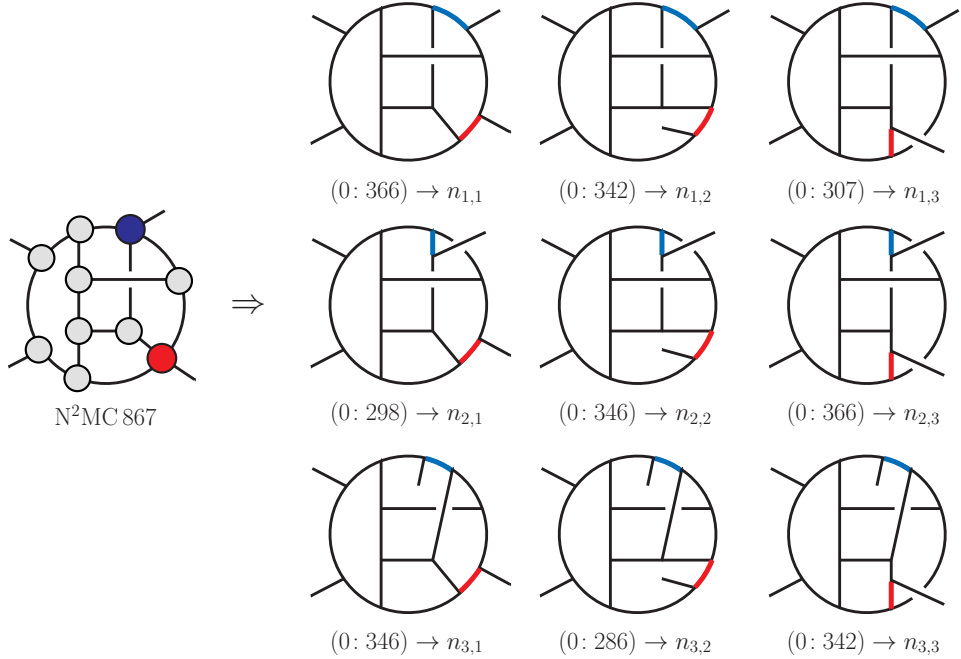}
\caption{An example illustrating the notation in \eqn{DoubleCopyCut}.
  Expanding each of the two four-point blob gives a total of nine
  diagrams.  The label \N2{\,867} refer to 867th diagram of the 2nd level cuts, and
  the $n_{i,j}$ correspond to labels used in the cut. The shaded thick
  (blue and red) lines are the propagators around which BCJ
  discrepancy functions are defined.  }
\label{FourxFourBlobFigure}
\end{figure}
%%%%%%%%%%%%%%%%%%%%%%%%%%%

The existence of BCJ representations at tree level implies that
representations should exist for all cuts of gauge-theory amplitudes
that decompose the loop integrand into products of tree amplitudes to
any loop order.
This further suggests that the corresponding cuts of the gravity amplitude can
be expressed in double-copy form,
\begin{equation}
{\cal C}_\G = \sum_{i_1,\dots,i_q}
\frac{n_{i_1, i_2, ...i_q}^\text{BCJ} \tn_{i_1, i_2, ...i_q}^\text{BCJ}} 
{D_{i_1}^{(1)}\dots D_{i_q}^{(q)} }\,,
\label{DoubleCopyCut}
\end{equation}
where the $n^\BCJ$ and $\tn^\BCJ$ are the BCJ numerators associated
with each of the two copies. In this expression the cut conditions are
understood as being imposed on the numerators.  Each sum runs over the
diagrams of each blob and $D_{i_m}^{(m)}$ are the product of the uncut
propagators associated to each diagram of blob $m$.
This notation is illustrated in \fig{FourxFourBlobFigure} for an \N2.
In this figure, each of the two four-point blobs is expanded into
three diagrams, giving a total of nine diagrams.  For example, the
indices $i_1 = 1$ and $i_2=1$ refers to the five-loop diagram produced
by taking the first diagram from each blob and connecting it to the
remaining parts of the five-loop diagram.  The denominators in
\eqn{DoubleCopyCut} correspond to the thick (colored) lines in the
diagrams.

The BCJ numerators in \eqn{DoubleCopyCut} are
related~\cite{BCJLoop,Square} to those of an arbitrary representation
by a generalized gauge transformation~\eqref{GGT}; the shift
parameters follow the same labeling scheme as the numerators
themselves,
\begin{eqnarray}
n_{i_1, i_2, ...i_q} &=& n_{i_1, i_2, ...i_q}^\text{BCJ}
   + \Delta_{i_1, i_2, ...i_q} \, .
\label{GeneralizedGaugeTrans}
\end{eqnarray}
The shifts $ \Delta_{i_1, i_2, ...i_q}$ are constrained to leave the
corresponding cuts of the gauge-theory amplitude unchanged.  Using
such transformations we can reorganize a gravity cut in terms
of cuts of a naive double copy and an additional contribution,
\begin{equation}
 {\cal C}_{\G} = \sum_{i_1,\dots,i_q} \frac{n_{i_1, i_2, ...i_q} \tn_{i_1, i_2, ...i_q} }
  {D_{i_1}^{(1)}\dots D_{i_q}^{(q)}} +{\cal E}_\G\left(\Delta\right)\, ,
\label{CutNDC}
\end{equation}
where the cut conditions are imposed on the numerators.
Rather than expressing the correction ${\cal E}_\G$ in terms of the
generalized-gauge-shift parameters, it is useful to re-express the
correction terms as bilinears in the violations of the kinematic
Jacobi relations~\eqref{BCJDuality} by the generic gauge-theory
amplitude numerators. These violations are known as BCJ
discrepancy functions.

As an example, the cut in \fig{FourxFourBlobFigure} is
composed of two four-point tree amplitudes and the rest are
three-point amplitudes. For any cut of this structure, two
four-point trees connected to any number of three-point trees, 
the correction has a simple expression,
\begin{equation}
{\E}^{4\times 4}_\G =
- \frac{1}{d^{(1,1)}_{1} d^{(2,1)}_{1}}
     \Bigr(J_{\x1, 1}\tJ_{1, \x2}
        + J_{1,\x2} \tJ_{\x1,1} \Bigr) \,,
\label{Extra4x4Simple}
\end{equation}
where $d^{(b,p)}_{i}$ is the $p$th propagator of the $i$th diagram
inside the $b$th blob and
\begin{align}
J_{\x1, i_2} \equiv \sum_{i_1=1}^3 n_{i_1 i_2} \,,
\hskip 1. cm
J_{i_1, \x2} \equiv \sum_{i_2=1}^3 n_{i_1 i_2}\,,
\hskip 1. cm
\tJ_{\x1, i_2} \equiv \sum_{i_1=1}^3 \tn_{i_1 i_2} \,,
\hskip 1. cm
\tJ_{i_1, \x2} \equiv \sum_{i_2=1}^3 \tn_{i_1 i_2}\,.
\end{align}
are BCJ discrepancy functions.
Notably, these discrepancy functions vanish whenever the numerators
involved satisfy the BCJ relations, even if the representation as a whole
does not satisfy them.
Such expressions are not unique and can be rearranged using various
relations between $J$s~\cite{JConstraints, GeneralizedDoubleCopy,
  FiveLoopN8Integrand}.  For example, an alternative version, equivalent to
\eqn{Extra4x4Simple}, is
\begin{equation}
{\E}^{4\times 4}_{\G} =
- \frac{1}{9} \sum_{i_1,i_2 =1}^3 \frac{1}{d^{(1,1)}_{i_1} d^{(2,1)}_{i_2}}
  \Bigl(J_{\x1, i_2} \tJ_{i_1, \x2}
     + J_{i_1, \x2} \tJ_{\x1, i_2} \Bigr) \,.
\label{Extra4x4Symmetric}
\end{equation}

Similarly, a cut with a single five-point tree amplitude and the
rest three-point tree amplitudes is given by
\begin{align}
{\cal C}^{5}_\G =\sum_{i=1}^{15} \frac{n_{i}  \tn_{i}}{d^{(1)}_{i} d^{(2)}_{i} }
 + {\E}^{5}_\G
 \qquad
 \text{with}
 \qquad
 {\E}^{5}_\G & = - \frac{1}{6}\sum_{i=1}^{15}
 \frac{J_{\{i,1\}} \tJ_{\{i,2\}} + J_{\{i,2\}} \tJ_{\{i,1\}} }
  { d^{(1,1)}_{i} d^{(1,2)}_{i}} \,,
\label{Extra5}
\end{align}
where $J_{\{i,1\}}$ and $J_{\{i,2\}}$ are BCJ discrepancy functions
associated with the first and second propagator of the $i$th diagram.
(See Ref.~\cite{FiveLoopN8Integrand} for further details.)

As the cut level $k$ increases the formulas relating the amplitudes'
cuts with the cuts of the naive double copy become more intricate, but
the basic building blocks remain the BCJ discrepancy functions.
The formulas often enormously simplify the computation of the contact
term corrections and are especially helpful at five loops at the \N2
and \N3 level, where calculating the contact terms via the maximal-cut
method can be rather involved.
Beyond this level the contact terms become much simpler due to a
restricted dependence on loop momenta and are better dealt with using the
method of maximal cuts and KLT relations~\cite{KLT}, as described in
Ref.~\cite{FiveLoopN8Integrand}.

%%%%%%%%%%%%%%%%%%%%%%%%%%%
\subsection{Previously Constructed Five-Loop Four-Point Integrands}

Five-loop four-point integrands have previously been constructed for
$\NeqFour$ super-Yang--Mills \cite{FiveLoopN4} and $\NeqEight$
supergravity \cite{FiveLoopN8Integrand}.  Here we review some of their
properties which serve as motivation for the
construction in \sect{ImprovedIntegrandSection} of new $\NeqFour$ super-Yang--Mills and $\NeqEight$
supergravity integrands with better manifest ultraviolet properties.

%%%%%%%%%%%%%% FIGURE  %%%%%%%%%%%%%%%%
\begin{figure}[t]
\includegraphics[clip,scale=0.4]{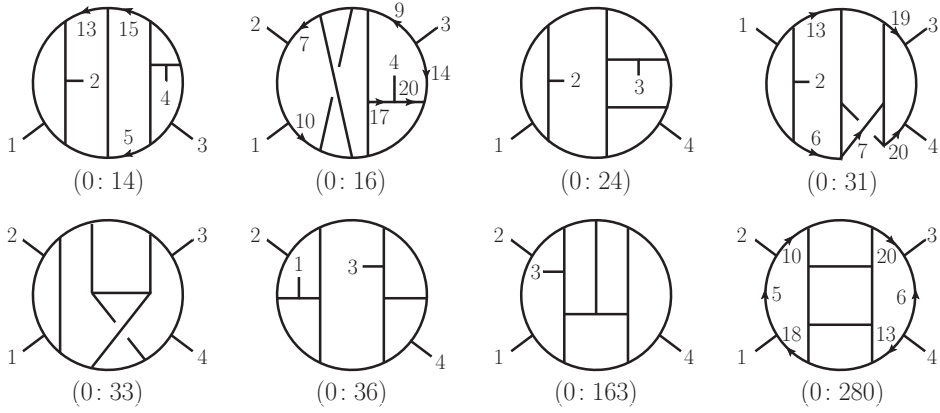}
\caption[a]{Sample graphs for the five-loop four-point $\NeqFour$
  super-Yang--Mills amplitude. The graph labels correspond to the ones
  in Ref.~\cite{FiveLoopN8Integrand} and here. }
\label{5loopImprovedUVdiagsFigure}
\end{figure}
%%%%%%%%%%%%%%%%%%%%%%%%%%%%%%%%%%%%%%%%%%

The five-loop four-point integrand of $\NeqEight$ supergravity
constructed in Ref.~\cite{FiveLoopN8Integrand} is obtained through
the generalized double-copy procedure, starting from a slightly
modified form of the corresponding $\NeqFour$ super-Yang--Mills
integrand of Ref.~\cite{FiveLoopN4}. This modified super-Yang--Mills
representation is given explicitly in an ancillary file of
Ref.~\cite{FiveLoopN8Integrand}.

All representations of the five-loop four-point $\NeqFour$
super-Yang--Mills amplitude that we use contain solely diagrams with
only cubic (trivalent) vertices, so can be written using
\eqn{CubicRepLoop} as
\begin{equation}
\hskip -.01 cm 
{\cal A}_4^{(5)} = i g^{12} st A_4^\tree \sum_{{\cal S}_4} \sum_{i =1}^{N_D}
\! \int \!  \prod_{j=5}^9\frac{d^D \ell_j} {(2\pi)^D} \frac{1}{S_i}
 \frac{c_i\, N_i}{\prod_{m_i=5}^{20} \ell_{m_i}^2} \,,
\label{AmplitudeGraphs}
\end{equation}
where we have explicitly extracted an overall crossing symmetric
prefactor of $st A_4^\tree$ from the kinematic numerators when
compared to~\eqn{CubicRepLoop}.  The gauge coupling is $g$, the
color-ordered $D$-dimensional tree amplitude is
$A_4^\tree \equiv A_4^\tree(1,2,3,4)$, and $s= (k_1 + k_2)^2$ and
$t =(k_2 + k_3)^2$ are the standard Mandelstam invariants.
We denote external momenta by $k_i$ with $i=1,\ldots,4$ and the five
independent loop momenta by $\ell_j$ with $j=5,\ldots,9$.  The
remaining momenta $\ell_j$ with $10\le j\le 20$ of internal lines are linear combinations
of the five independent loop momenta and external momenta.  As always,
the color factors $c_i$ of all graphs are obtained by dressing every
three-vertex in the graph with a factor of $\f^{abc}$.

The number $N_D$ of diagrams that we include depends on the particular
representation we choose.  The form given in Ref.~\cite{FiveLoopN4}
has 416 diagrams, while the one used in
Ref.~\cite{FiveLoopN8Integrand} has 410 diagrams.  Some sample graphs
from this list of 410 diagrams are shown in
\fig{5loopImprovedUVdiagsFigure}.

It is useful to inspect some of the numerators associated with the
sample diagrams.  Choosing as examples diagrams 14, 16, 31 and 280
from the 410 diagram representation of
Ref.~\cite{FiveLoopN8Integrand}, we have the $\NeqFour$
super-Yang--Mills numerators
\begin{align}
N_{14} & = s \Bigl(s^2 s_{3,5} - \frac{5}{2}  \ell_5^2 \ell_{13}^2 \ell_{15}^2\Bigr)\,, \nn \\
N_{16} & = -s \Bigl( s^3 + s^2 \tau_{3, 15} -
\frac{3}{2} s  \ell_7^2 \ell_{10}^2 + \frac{3}{2} \ell_7^2 \ell_{10}^2 
         ( \tau_{1, 15} + \tau_{2, 15} + \tau_{4, 15} + \ell_9^2 - \ell_{14}^2 -
          \ell_{17}^2 + \ell_{20}^2 ) \Bigr)\,, \nn\\
N_{31} & = s \Bigl(
        s \bigl(- s^2 -  \ell_{13}^2 \ell_{20}^2 +
         s ( \tau_{6, 19} +  \ell_{13}^2 + \frac{1}{2} \ell_{20}^2) +
              \ell_{6}^2 ( \ell_{20}^2 - \ell_{19}^2) \bigr)
            - \frac{1}{2} \ell_6^2 \ell_7^2 \ell_{19}^2 \Bigr) \,, \nn \\
N_{280} & =  s^4 +  s^3 (\tau_{10, 13} + \tau_{18, 20}) 
     + \frac{1}{2} s^2 (\tau_{10, 13}^2 + \tau_{18, 20}^2) + 2 t (\ell_5^2 + \ell_6^2)
          ( \ell_{13}^2 \ell_{18}^2  +  \ell_{10}^2 \ell_{20}^2 ) \,,
\label{Neq4SYMNumerSamples}
\end{align}
where $s$ and $t$ are the usual Mandelstam invariants and
\begin{equation}
s_{i,j} = (\ell_i + \ell_j)^2\, , \hskip 2 cm 
\tau_{i, j} = 2 \ell_i \cdot \ell_j \, .
\end{equation}
The corresponding naive double-copy numerators are obtained by simply
squaring these expressions.

The $\NeqEight$ integrand found in Ref.~\cite{FiveLoopN8Integrand}
suffers from poor graph-by-graph power counting, which obstructs the
extraction of its leading ultraviolet behavior. Many of its diagrams
in the naive double-copy part contain spurious quartic power
divergences in $D=24/5$, which are equivalent to logarithmic
divergences in $D=4$.  As discussed in~\cite{BjornssonAndGreen,
  SevenLoopGravity,VanishingVolume}, such divergences are spurious and
should cancel out.
The difficulties raised by the spurious power counting are two
fold. First, we will see in \sect{VacuumExpansionSection} that their
presence causes a rapid growth in the number of terms in the series
expansion of the integrand necessary to isolate the potential
logarithmic divergence in $D=24/5$. Second, this expansion yields
graphs with propagators raised to a high power, which leads to an IBP
system with billions of integrals.

There are two distinct ways to overcome these difficulties.  The first
is to construct a new super-Yang--Mills integrand which improves the
power counting of the naive double copy. This in turn minimizes the
number of integrals and equations in the full IBP system.
We will give
the construction of this new representation of the $\NeqFour$
super-Yang--Mills integrand as well as of the $\NeqEight$ supergravity
integrand that follows from it in the next section.  This represents a
complete solution.  Still it is useful to have a separate check. Our
second resolution is to make simplifying assumptions on the type of
integrals that can contribute to the final result after applying IBP
integral identities.  This approach will be discussed in
\sect{TopLevelIntegrationSection} and will allow us to integrate the
more complicated integrand of Ref.~\cite{FiveLoopN8Integrand}. The
agreement between the results of these two approaches represents a
highly non-trivial confirmation of both the integrands and the
integration procedure.

%%%%%%%%%%%%%%%%%%%%%%%%%%%%%
\section{Improved integrands}
\label{ImprovedIntegrandSection}

In this section we describe the construction of a new form of the five-loop
four-point integrand for $\NeqFour$ super-Yang--Mills theory and then
use it to construct an improved $\NeqEight$ supergravity integrand.
The $\NeqEight$ integrand we obtain still exhibits power
divergences in $D=24/5$ but, as we shall see, their structure is such
that they do not lead to a dramatic increase in the number of
integrals needed for the extraction of the leading logarithmic
ultraviolet behavior of the amplitude.
In \sect{AllIntegrationSection} we extract the
ultraviolet properties using this improved $\NeqEight$ five-loop
integrand without making any assumptions on the final form of the
large-loop momentum integrals.

\subsection{Construction of improved \texorpdfstring{$\NeqFour$}{N=4} super-Yang--Mills integrand}

%%%%%%%%%%%%%% FIGURE  %%%%%%%%%%%%%%%%
\begin{figure}[t]
\includegraphics[clip,scale=0.45]{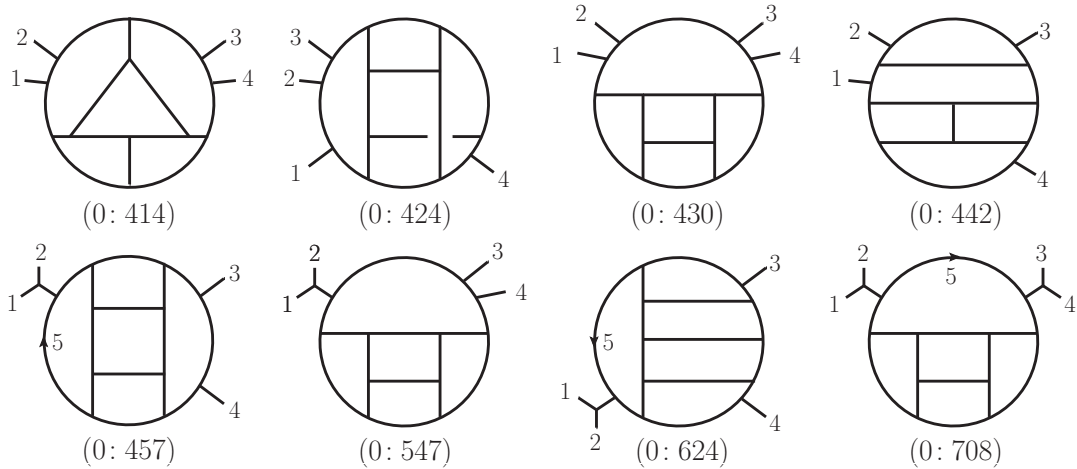}
\caption[a]{Some of the additional graphs for the improved representation
  of the integrand of the five-loop four-point $\NeqFour$ super-Yang--Mills amplitude.  These graphs
  were not needed in earlier constructions~\cite{FiveLoopN4, FiveLoopN8Integrand}.  The 
  labeling scheme is to the contact level and then the diagram number corresponding
  to the labels of the ancillary files~\cite{AttachedFile}.}
\label{5loopNewdiagsFigure}
\end{figure}
%%%%%%%%%%%%%%%%%%%%%%%%%%%%%%%%%%%%%%%%%%

The key power-counting requirement we demand of every term of the
improved Yang--Mills representation is that its naive double copy, as
described in \sect{ReviewSection}, has no worse than a logarithmic
divergence in $D=24/5$.
This translates to a representation with no more than four powers of
loop momenta in the kinematic numerator of any
one-particle-irreducible diagram. These conditions require us to
introduce new diagrams of the type illustrated in
\fig{5loopNewdiagsFigure}.  These graphs are characterized by the
vanishing of their maximal cuts. For these diagrams, this
implies that the poles due to the propagators independent of loop
momenta (to which we will refer to as ``dangling trees'') are spurious.
It also turns out that their numerators have fewer than four powers of loop momenta.  Such
dangling tree diagrams are crucial for obtaining ultraviolet-improved
supergravity expressions via the generalized double-copy procedure.
The general pattern is that, to improve the double-copy expression,
the terms with the highest power counting in the super-Yang--Mills integrand should
come from diagrams with dangling trees. Due to the reduced number of
possible loop-momentum factors in their kinematic numerators, the
squaring of the numerator (naive double copy) of such diagrams keeps
the superficial power counting under control.

To construct such a representation of the five-loop four-point $\NeqFour$  super-Yang--Mills integrand
we apply the maximal-cut method to  an ansatz that has the desired power counting  properties.
Inspired by the structure of the lower-loop amplitudes~\cite{BRY,BDDPR, BCJLoop, ck4l} we 
further simplify the ansatz and improve the power-counting properties of the naive double copy by imposing the following 
constraints:
\begin{itemize}
\item Each numerator is a polynomial of degree eight in momenta, of which no more than four can be loop momenta.
\item Every term in every numerator contains at least one factor of an 
         external kinematic invariant, $s$ or $t$. 
\item No diagram contains a one-loop tadpole, bubble or triangle subdiagram. Also, two-point two- and three-loop subdiagrams, and three-point two-loop subdiagrams, are excluded. 
\item For each one-loop $n$-gon the maximum power of the corresponding loop momentum 
is $n-4$. In particular, this means that numerators do not depend on the loop 
 momenta of any box subdiagrams. 
\item Diagram numerators respect the diagram symmetries.
\item The external state dependence is included via an overall factor of the tree amplitude.
\end{itemize}
Such simplifying conditions can always be imposed as long as the system of equations 
resulting from matching the cuts of the ansatz with those of the amplitude still has solutions.
The conditions above turn out to be incompatible with a
representation where BCJ duality holds globally on the fully off-shell
integrand.   They are
nevertheless compatible with all two-term kinematic Jacobi relations
(meaning where one of the three numerators of the Jacobi relation
\eqref{BCJDuality} vanishes by the above constraints), which we impose {\it a posteriori}:
\begin{itemize}
\item The solution to cut conditions is such that the ansatz obeys all
  two-term kinematic Jacobi relations.
\end{itemize}

Similarly with the earlier representation of the five-loop four-point $\NeqFour$ super-Yang--Mills amplitude,  we organize the 
integrand in terms of diagrams with only cubic vertices; the numerators have the structure shown in~\eqn{AmplitudeGraphs}.  
In the present case we have 752 diagrams.  The first 410 diagrams are
the same as for the previous integrand~\cite{FiveLoopN8Integrand},
some of which are displayed in \fig{5loopImprovedUVdiagsFigure}.
There are an additional 342 diagrams, a few of which are displayed in
\fig{5loopNewdiagsFigure}.  In addition to the dangling tree graphs discussed
above, this includes other diagrams such the ones on the first line of \fig{5loopNewdiagsFigure}.

For each diagram we write down an ansatz for the $N_i$ which is a
polynomial of fourth degree in the independent kinematic invariants,
subject to the constraints above. Each independent term is assigned an
arbitrary parameter.  This ansatz is valid for all external
states, as encoded in the overall tree-level amplitude factor in
\eqn{AmplitudeGraphs}.  This simple dependence on external states is
expected only for the four-point amplitudes.\footnote{For higher-point
  amplitudes the necessary ansatz is more
  involved~\cite{FivePointN4} and it will not exhibit a clean
  separation between external state data and loop kinematics.} The
most general ansatz that obeys the first four constraints above
has $535,146$ terms; requiring that each numerator respects the
graph's symmetries and also imposing the maximal cuts of the amplitude
reduces this to a more managable size.

The parameters of the ansatz are determined via the method of maximal
cuts. Rather than constructing unitarity cuts directly from their
definition as products of tree-level amplitudes, it is far more
convenient to use the previously constructed
versions~\cite{FiveLoopN4,FiveLoopN8Integrand} of the amplitude
integrand as input. This approach circumvents the need for
supersymmetric state sums~\cite{SuperSums} (which become nontrivial at
high-loop orders and in arbitrary dimensions) and recycles the simplifications which have already
been carried out for the construction of that integrand. Moreover, it
makes full use of the $D$-dimensional validity of that integrand, which is
confirmed in Ref.~\cite{FiveLoopN4}.

The maximal cuts impose simple constraints on the free parameters; it
is convenient to replace them in the ansatz.  Next, NMC conditions are
solved; as their
solution is quite involved, it is impractical to
plug it back directly into the ansatz. To proceed, we introduce the notion of
a presolution of a given \N{k} as the solution of all constraints imposed by all 
lower-level cuts which overlap with the
given cut. The advantage of using presolutions is that they account
for a large part of the lower-level cut constraints on the parameters
entering the given cut without the complications ensuing from 
simultaneously solving all the lower-level cut conditions and replacing the
solution in the ansatz.  Thus, instead of simultaneously solving all
the NMC cut constraints and evaluating the ansatz on the solution
before proceeding to the \N{2} cuts, we construct all the \N{2}
presolutions and then solve each of them simultaneously with the \N{2}
cut condition.
We proceed recursively in this way through all relevant cut levels.
The integrand of the amplitude is then found by simultaneously
re-solving all the new constraints on the parameters of the ansatz
derived at each level.
While this is equivalent to adding contact terms, the ansatz approach effectively distributes  them in the diagrams of the ansatz and prevents 
the appearance of any terms with artificially high power count.

%%%%%%%%%%%%% FIGURE  %%%%%%%%%%%%%%%%
\begin{figure}[tb]
  \includegraphics[clip,scale=0.48]{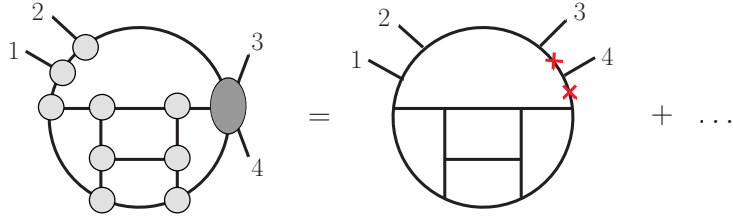}
\vskip -.2 cm
\caption[a]{This cut is not considered as it contains a singular
  diagram;  instead we recover the missing information from higher level cuts.
  The shaded (red) ``$\times$'' mark complete propagators (not replaced by delta functions), the other 
  exposed propagators are all placed on shell (replaced by delta functions).
}
\label{CutsDoubleProp}
\end{figure}
%%%%%%%%%

In carrying out this application of the method of maximal cuts we encounter a technical complication with diagrams 
with four-loop bubble subdiagrams, three of which are illustrated in \fig{5loopNewdiagsFigure}: (0:\,430),  (0:\,547) 
and (0:\,708). 
The main difficulty stems from the fact that both propagators connecting the bubble to the rest of the diagram carry 
the same momentum so the diagram effectively exhibits a doubled propagator.
While such double propagators are spurious  and can in principle be algebraically eliminated
since the representations of Refs.~\cite{FiveLoopN4,FiveLoopN8Integrand} does not have them, they nevertheless 
make difficult the evaluation of the cuts.
It moreover turns out that, with our strict power counting requirements, there is no solution that
explicitly eliminates the double poles from all diagrams, even though they cancel in all cuts. 
Such graphs cause certain cuts to be ill-defined without an additional prescription. Indeed, if only one of the two 
equal-momentum propagators is cut the tree amplitude containing the second one becomes singular unless a 
specific order of limits is taken. This phenomenon is illustrated  in \fig{CutsDoubleProp};  by replacing the propagator on 
one side of the bubble subdiagram with an on-shell delta-function, the propagator on the other side, marked by a shaded (red) ``$\times$'', becomes singular.

%%%%%%%%%%%%% FIGURE  %%%%%%%%%%%%%%
\begin{figure}[tb]
  \includegraphics[clip,scale=0.45]{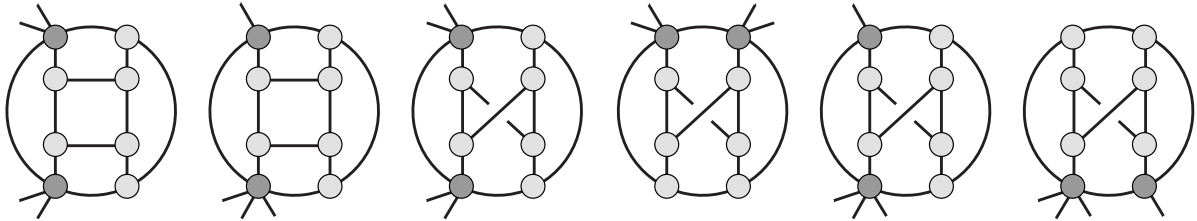}
\vskip -.2 cm
\caption[a]{The list of additional N$^4$MCs that are needed to fix the diagrams with doubled propagators.
}
\label{N4constraintsFigure}
\end{figure}
%%%%%%%%%

One can devise a prescription that realizes the expected cancellation
of such $1/0$ terms among themselves.  It is, however, more convenient
to simply skip the singular cuts altogether and recover the missing
information from higher-level cuts that overlap with the skipped ones
({\it i.e.} cuts in which the doubled propagator is not cut).  In the
absence of doubled propagators, cuts through \N3 level contain all the
information necessary for the construction of the amplitude, as seen
in \cite{FiveLoopN8Integrand}, because the power counting of the
theory implies that numerators can have at most three inverse
propagators and thus there can be at most N${}^3$ contact terms.  In
our case, to recover cut constraints absent due to the unevaluated
singular cuts we must include certain \N4 cuts; the complete list is
shown in \fig{N4constraintsFigure}. All other \N4 as well as some \N5
cuts serve as consistency checks of our construction.

Our new representation for the five-loop four-point integrand is given in an ancillary 
file~\cite{AttachedFile}. 
Generalized gauge invariance implies that there is no unique form of the integrand; indeed, the 
global solution of the cut conditions and of the two-term Jacobi relations leaves 10607  free parameters.
They ``move'' terms between diagrams without affecting any of the unitarity cuts.
These parameters should not affect any observable; in particular, they should drop out of the gravity amplitude (after nontrivial algebra)
resulting from the generalized double-copy construction based on this amplitude.
To simplify the expressions we  set them to zero.  

It is instructive to see how the power counting of the new representation differs from that of the previous one~\cite{FiveLoopN8Integrand}. 
Setting the free parameters to zero, the counterparts of the numerators $N_{14}, N_{16}, N_{31}$ and $N_{280}$ shown for the previous representation 
in \eqn{Neq4SYMNumerSamples} are
\begin{align}
N_{14} & = \frac{1}{2} s^3 \bigl( \tau_{3, 5} - \tau_{4, 5} -s \bigr)\,, \nn \\
N_{16} & = N_{14} \,, \nn \\
N_{31} & = \frac{1}{2} s^3 \bigl(\tau_{1, 5} + \tau_{1, 6} 
      + \tau_{2, 5} + \tau_{2, 6} + 2 \tau_{3, 6} + 2 \tau_{5, 6} -s \bigr) \,, \nn\\
N_{280} & = s^4 + 2 s^3 u - u \tau_{2, 5} \tau_{3, 5} \ell_6^2
      + s \tau_{3, 5}^2 \ell_6^2 + \cdots + 8 u^2 \ell_5^2 \ell_6^2 \,,
\label{Neq4SYMNewNumerSamples}
\end{align}
where in $N_{280} $ we have kept only a few terms, since it is somewhat lengthy.  The complete
list of kinematic numerators is contained in the ancillary file ~\cite{AttachedFile}.  
Compared to the super-Yang--Mills numerators in \eqn{Neq4SYMNumerSamples}, the maximum
number of powers of loop momenta dropped from six to one in the first three numerators and to four powers in $N_{280}$.  
Consequently, the naive double-copy numerators have only up to eight powers of  loop momenta.
The naive double-copy numerators also inherit the property that every term carries at least two powers of $s$ or $t$,
a property that all contact term corrections share by construction.

Similarly, the additional diagrams in \fig{5loopNewdiagsFigure} are also very well-behaved at large loop momenta. An illustrative sample of 
the additional numerators is
\begin{align}
N_{547} & =  \frac{3}{2} s \ell_5^2 (t \tau_{1, 5}  
           - u \tau_{2, 5} - 3 s \tau_{3, 5} - 6 u \tau_{3, 5})\,, \nn\\
N_{624} & =  - \frac{61}{10} s^3 (u - t + \tau_{1, 5} - \tau_{2, 5} )\,, \nn\\
N_{708} & =  6 s^2 (t-u) \ell_5^2 \,, 
\label{SampleExtraNumerators}
\end{align}
where the labels correspond to those in \fig{5loopNewdiagsFigure}.

The naive double copy of all 752 diagrams gives diagrams that are
completely ultraviolet finite in $D=22/5$.  In $D = 24/5$ it exhibits
no power divergences, in contrast to the double copy of
the earlier representation of the super-Yang--Mills amplitude.
As we will see below, the contact term corrections needed to obtain
the $\NeqEight$ supergravity amplitude will lead to contributions that
individually have power divergences but, as we will discuss in
\sect{VacuumExpansionSection}, it is such that it
that does not increase the number of integrals that must be
evaluated. Furthermore, as we note in \sect{AllIntegrationSection}, in
$D=22/5$ the contact term contributions all cancel after IBP
reduction, leaving a completely ultraviolet finite result.

To confirm our construction, we have performed the standard checks of
verifying cuts beyond those needed for the construction, such as all
non-singular cuts at the \N4 and \N5 levels.
We have confirmed that our improved $\NeqFour$ super-Yang--Mills integrand 
generates exactly the same ultraviolet divergence in the critical dimension $D_c = 26/5$ as obtained in 
Ref.~\cite{FiveLoopN8Integrand} using the earlier representation of the amplitude.  
To carry out this check  we followed the same procedure explained in that paper for extracting
the ultraviolet divergence, using the same integral identities.

\subsection{Improved \texorpdfstring{$\NeqEight$}{N=8} supergravity integrand}
\label{ImprovedN8Section}

Armed with the new five-loop four-point integrand of $\NeqFour$
super-Yang--Mills theory we now proceed to the construction of the corresponding improved integrand of $\NeqEight$ supergravity,
following the generalized double-copy
construction~\cite{GeneralizedDoubleCopy} outlined in
\sect{ReviewSection}.  Our construction essentially follows the same steps as
in Ref.~\cite{FiveLoopN8Integrand}, so we will not repeat the details.  
We obtain a set of contact terms, organized according to levels, which correct the naive double copy
to an integrand for the $\NeqEight$ supergravity amplitude.
As a consequence of the improved term-by-term ultraviolet behavior of the gauge-theory 
amplitude,  the individual terms of the resulting supergravity integrand  are also better behaved
at large loop momenta.

%%%%%%%%%%%%% FIGURE  %%%%%%%%%%%%                                     
\begin{figure}[tb]
 \includegraphics[clip,scale=0.39]{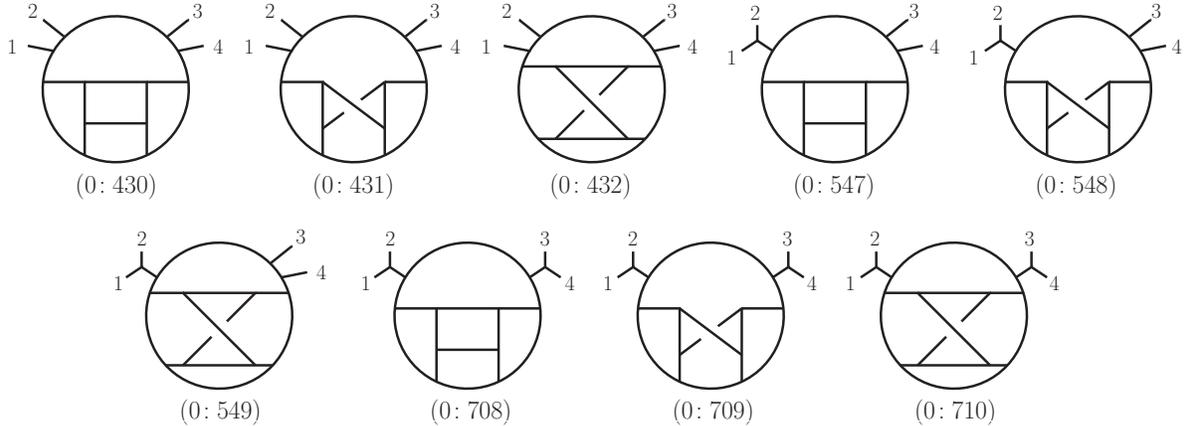}
%\vskip -.2 cm
\caption[a]{The diagrams whose numerators were set to zero, to simplify the supergravity construction
by avoiding doubled propagators.
}
\label{VanishingDiagramsFigure}
\end{figure}
%%%%%%%%%%%%%%%%%%%%%%%%%%%%%

The difference with the construction in
Ref.~\cite{FiveLoopN8Integrand} is related to the existence of the
diagrams with doubled propagators in the super-Yang--Mills amplitude,
such as (0:\,430), (0:\,547) and (0:\,708) of
\fig{5loopNewdiagsFigure}. Unlike the gauge-theory construction, here
we can avoid needing to identify and skip cuts with ill-defined
values.
To this end we notice that, since the maximal cuts of these diagrams
vanish, they contribute only contact terms even in the naive double
copy. We may therefore simply set to zero these diagrams in the naive
double copy and recover their contributions directly as contact terms
at the relevant level. For the same reason we can also set to
zero in the naive double copy other diagrams with vanishing maximal
cuts.
The consistency of this reasoning is checked throughout the
calculation by the absence of ill-defined cuts as well as by the
locality of all contact term numerators. Had the latter not be the
case it would imply the violation of some lower-level cuts.  This in turn
would have meant that some term we set to zero contributed more than merely 
contact terms to the amplitude.   The net effect is that we can build
the complete integrand by using cuts through the \N6 level, just as
in the previous construction~\cite{FiveLoopN8Integrand}, 
and there is no need to go beyond this, except to verify the completeness of the 
result.

As discussed in \sect{ReviewSection}, the cuts of the supergravity
amplitude can be computed in terms of the BCJ discrepancy functions of
the full gauge-theory amplitude rather than from the discrepancy
functions of the amplitude with the doubled-propagator diagrams set to
zero.
It turns out that the cuts touching the doubled-propagator diagrams are sufficiently 
simple to be efficiently evaluated using KLT relations on the cuts.
The completeness of the construction is guaranteed by verifying all (generalized)
unitarity cuts.

%%%%%%%%%%%%%%%%% TABLE %%%%%%%%%%%%%%%%%%%%%%%%%%%%%%%%%
\begin{table}[tb]
\begin{center}
\begin{tabular}{|c|c|c|}
\hline
\ Level\ & \ No. diagrams \ &\ No. nonvanishing diagrams \ \\
\hline
0 &     752   & 649 \\
1 &   2,781   & 0 \\
2 &   9,007   & 1,306 \\
3 &  17,479   & 2,457 \\
4 &  22,931   & 2,470 \\
5 &  20,657   & 1,335 \\
6 &  13,071   & 256 \\
\hline
total & 86,678 & 8,473 \\
\hline
\end{tabular}
\caption{The number of diagrams at each contact-diagram level as well as 
the number of diagrams at each level with nonvanishing numerators.
}
 \label{NumberDiagramsTable}
\end{center}
\end{table}
%%%%%%%%%%%%%%%%% TABLE %%%%%%%%%%%%%%%%%%%%%%%%%%%%%%%%%

The complete amplitude is given by a sum over the 752 diagrams of the
naive double copy and the 85,926 contact term diagrams,
\begin{equation}
{\cal M}_4^{\fiveloop} = i \Bigl(\frac{\kappa}{2}\Bigr)^{12}
 s t u M_4^{\rm tree} \sum_{k=0}^6 \sum_{{\cal S}_4} \sum_{i=1}^{T_k}
 \int \prod_{j = 5}^9 \frac{d^D \ell_j}{(2 \pi)^D} \frac{1}{S_i} \frac{{\cal N}_i^{(k)}}
 { \prod_{m_i=5}^{20-k} \ell_{m_i}^2 } \, ,
\label{GRContactFiveLop}
\end{equation}
where $M_4^\tree$ is the four-point $\NeqEight$ supergravity tree amplitude
and $u = -s-t$.
Here $T_k$ is the total number of diagrams at level $k$; they are
given in \tab{NumberDiagramsTable}.
The diagram count at each level differs somewhat from the earlier
construction~\cite{FiveLoopN8Integrand} because here we include all
the daughter diagrams that arise collapsing propagators of any of the
752 parent diagrams of the naive double copy instead of those obtained
only from the first 410 diagrams.  The parent-level diagrams are
obtained from the improved representation of the $\NeqFour$
super-Yang--Mills four-point amplitude through the double-copy
substitution~\eqref{ColorSub} and setting to zero the numerators of
the diagrams shown in \fig{VanishingDiagramsFigure}.  The contact
terms are generated using the procedures summarized above.  We collect
the results for all diagrams, numerators ${\cal N}_i^{(k)}$ and symmetry
factors, $S_i$, at each level in the plain-text {\sc Mathematica}-readable
ancillary files~\cite{AttachedFile}.

%%%%%%%%%%%%% FIGURE  %%%%%%%%
\begin{figure}[tb]
 \includegraphics[clip,scale=0.54]{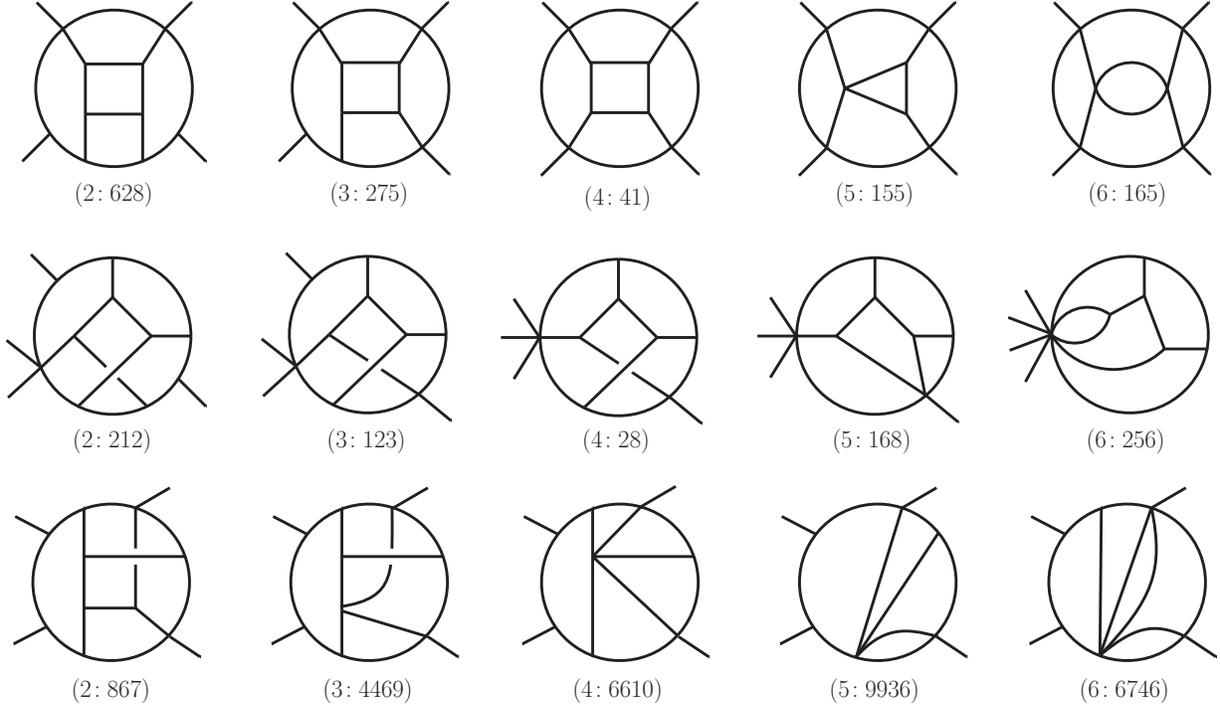}
\caption[a]{Sample contact term diagrams corresponding to the cuts in \fig{NkMaxSampleCutsFigure}. The labels (X:\,Y) 
refer to the level and contact diagram number. The final four diagrams have vanishing numerator; the first eleven are 
nonvanishing.
}
\label{NkMaxContactsFigure}
\end{figure}
%%%%%%%%%%%%%%%%%%%%%%%%%%%%%%

A striking property of the supergravity contact terms, which is
obvious from \tab{NumberDiagramsTable}, is that most of them vanish.
The precise number of vanishing diagrams depends on the particular
starting point used in the naive double copy and on details of the
off-shell continuation of the contact terms at each level. As for the
previously-constructed integrand in Ref.~\cite{FiveLoopN8Integrand},
this is a consequence of the many kinematic Jacobi identities that
hold for the super-Yang--Mills amplitude used in our
construction. This effect is even more clear here, where the
$\NeqFour$ super-Yang--Mills integrand obeys all the two-term kinematic Jacobi
relations. While this integrand does not support a solution for all
three-term Jacobi relations, it may be possible to further reduce the
number of supergravity contact terms by imposing a judiciously-chosen
subset of these relations.

%%%%%%%%%%%%%%%%%%%%%%%%%%%%%
\section{Ultraviolet vacuum integral expansion}
\label{VacuumExpansionSection}

In previous sections we reviewed the integrand of the five-loop four-point amplitude of $\NeqEight$ supergravity 
found in Ref.~\cite{FiveLoopN8Integrand} and constructed a new one, with certain improved power-counting properties. 
In this section we expand these integrands in the ultraviolet, {\it i.e.} for external momenta small compared to the loop momenta,
and point out key features of the new integrand.  This expansion generates integrals reminiscent of vacuum integrals with no external momenta;  we 
call such integrals ``vacuum integrals'' as well.
While we are interested in the logarithmic divergence in $D=24/5$, both integrands also exhibit spurious quadratic and quartic divergences in this dimension. Finiteness of the  five-loop amplitude in $D <24/5$ guarantees that they should cancel out. However, the graph-by-graph presence of spurious singularities both in the naive double-copy part and in the contact terms of the integrand of Ref.~\cite{FiveLoopN8Integrand}  leads to a rapid increase in the number of terms when extracting the logarithmic divergence. 
By construction, the new integrand can have power divergences only through its contact terms. Moreover, their structure is such that 
the number of different integrals which appear in the ultraviolet expansion is substantially decreased compared to the earlier integrand.

\subsection{Vacuum expansion of integrands}

The basic challenge is to extract logarithmic divergences underneath spurious power divergences. 
To do so we follow the standard method of series expanding the integrand in the ultraviolet region~\cite{Vladimirov}, where 
the external momenta are much smaller than loop momenta, which are commensurate.  
This strategy was applied to various supergravity calculations in Refs.~\cite{N4GravThreeLoops, N4SugraMatter, N4GravFourLoop, N5GravFourLoops}.
The different orders in this expansion are expressed as vacuum integrals with different degrees of ultraviolet divergence. In dimensional regularization, only logarithmically divergent vacuum integrals can result in a pole. Logarithmically-divergent terms in lower dimensions are power divergent in higher dimensions. Thus, by integrating all logarithmically-divergent terms in $D<24/5$, 
we are checking that power divergences cancel in $D=24/5$. Indeed, as we explain in \sect{AllIntegrationSection}, 
we explicitly verify that in $D=22/5$ all the divergences cancel.  This also proves that any power divergences in $D=24/5$
are artifacts of our representations.
While we do not have representation of the integrand that exhibits only logarithmic divergences in this dimension, the naive double-copy contributions in our new representation were constructed to have this property. 

Dimensional analysis shows that the local term\footnote{This is the same term
  that may appear at seven loops in $D=4$, though the appearance of
  the former of course does not immediately imply the presence of the
  latter.} in the effective action
that corresponds to a logarithmic divergence in $D=24/5$ at five loops
has the generic structure $D^8 R^4$.
Its momentum space form has 16 momentum factors; of them, eight
correspond to the $(st A^\tree)^2 = stu M_4^\tree$ prefactor of the amplitude.  Thus,
the logarithmically-divergent part of each integral has eight factors
of external momenta.  Because every term in every supergravity numerator ${\cal N}$
has at least two powers of $s$ or $t$, we need to expand the integrand
to at most fourth order in small external momenta.

\begin{figure}[tb]
  \includegraphics[scale=0.6]{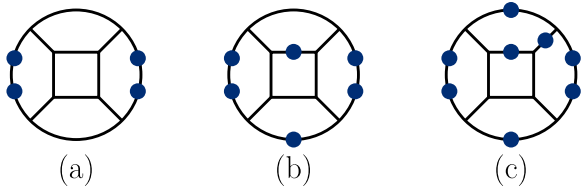}
\vskip -.2 cm 
\caption[a]{\small After series expanding one encounters vacuum diagrams with up to 
8 additional propagators, as well as numerators which are suppressed here.
Each (blue) dot corresponds to a repeated propagator.  Diagram (a), (b) and (c) are examples
with four, six and eight higher-power propagators.
  }
\label{Vacuum5loopsYMVdotsFigure}
\end{figure}

The dependence of the numerator polynomial on external momenta determines the order to which each term must be expanded.  
It is therefore useful to decompose each numerator into expressions ${\cal N}^{(m)}$ with fixed number $m$ of external momenta (and $16-m$ powers of loop momentum)
\begin{equation}
{\cal N} = {\cal N}^{(4)} + {\cal N}^{(5)} + {\cal N}^{(6)} + \cdots +  {\cal N}^{(16)} \, .
\end{equation}
There is freedom in this decomposition, including that induced by the choice of independent loop momenta.
Terms with more than eight powers of external momenta in the numerator are ultraviolet finite in $D=24/5$ and can therefore 
be ignored.  
For terms ${\cal N}^{(8)}$ with exactly eight powers of external momentum in the numerator we need
only the leading terms in the expansion of the propagators as higher-order terms are finite. It suffices therefore to set 
to zero all external momenta in propagators, {\it e.g.} for the ${\cal N}^{(8)}$ terms in the diagram shown in~\Fig{Vacuum5loopsYMVdotsFigure}(a)
\begin{align}
& \frac{{\cal N}^{(8)}}  {
   (\ell_5^2)^3 \, (\ell_6^2)^3 \,\ell_7^2 \,\ell_8^2 \,\ell_9^2
   ( \ell_5 + \ell_7)^2 (\ell_5 - \ell_9)^2 (\ell_5 + \ell_6 + \ell_7)^2  (- \ell_5 - \ell_6 + \ell_9)^2 }\nn \\
& \times \frac{1}{(\ell_5 + \ell_6 + \ell_8 - \ell_9)^2  ( \ell_5 + \ell_6 + \ell_8)^2  (\ell_5 + \ell_6 + \ell_7 + \ell_8)^2  }\, .
\end{align}
The leading divergence of terms with $4\le m \le 7$ is power-like. The extraction of the logarithmic divergence underneath requires that propagators be expanded to $(8-m)$-th order
in the momenta $k_i$:
\be
\frac{{\cal N}^{(m)}}{\prod_{i=1}^{I} d_i}\rightarrow \frac{ {\cal N}^{(m)}}{(8-m)!}
\sum_{i_1,\dots, i_{8-m}=1}^3  k_{i_1}^{\mu_1}\dots k_{i_{8-m}}^{\mu_{8-m}}
\biggl( \frac{\partial}{\partial k^{\mu_1}_{i_1}} \dots  \frac{\partial}{\partial k_{i_{8-m}}^{\mu_{8-m}}} \frac{1}{\prod_{i=1}^{I} d_i}\biggr|_{k_j=0} \,\biggr) \, ,
\label{expansion_general}
\ee
where $I$ is the number of internal lines of the diagram and $d_i$ the corresponding inverse propagators.
The action of derivatives leads to propagators raised to higher
powers---{\it i.e.} to repeated propagators---which we denote by dots,
one for each additional power. Up to four further dots appear when
derivatives act four times and external momenta are set to zero. Examples, with numerators
suppressed, are included in diagrams (b) and (c) of
\fig{Vacuum5loopsYMVdotsFigure}. The increase in the number of classes
of vacuum integrals (as specified by the number of dots) leads in turn
to an increase in the complexity of the IBP system necessary to reduce
them to master integrals.
The expansion also leads to higher-rank tensor vacuum integrals, which
appear as integrals with numerators containing scalar products of loop
and external momenta. We discuss dealing with such integrals below.

It is instructive to contrast, from the standpoint of the vacuum expansion, the old and new four-point five-loop $\NeqEight$ 
supergravity integrands; we will choose  the level-0 diagrams 14, 16, 31, 280 shown in \fig{5loopImprovedUVdiagsFigure} as 
illustrative examples.
The numerators of these diagrams are, respectively, the naive double copies ({\it i.e.} squares) of the numerator factors of the old 
representation of the $\NeqFour$ super-Yang--Mills amplitude, given  in \eqn{Neq4SYMNumerSamples}, and the new representation, 
given in \eqn{Neq4SYMNewNumerSamples}.
In the old representation, $ {\cal N}^{(4)}_{0:\,14}$, ${\cal
  N}^{(4)}_{0:\,16}$, ${\cal N}^{(4)}_{0:\,31}$, ${\cal
  N}^{(4)}_{0:\,280}$ are all nonvanishing and, for these terms, the
logarithmic divergence is given by \eqn{expansion_general} with $m=4$.
The resulting vacuum diagrams exhibit up to eight dots. \footnote{The
  leading term in the small momentum expansion is quartically
  divergent and corresponds to a logarithmic divergence in $D=4$ which
  should cancel on general grounds when all contributions are
  collected.}
In the improved representation constructed in
\sect{ImprovedIntegrandSection}, the first nonvanishing terms in
the decomposition of supergravity numerators are $ {\cal
  N}^{(8)}_{0:\,14}$, ${\cal N}^{(8)}_{0:\,16}$, ${\cal
  N}^{(8)}_{0:\,31}$, ${\cal N}^{(8)}_{0:\,280}$.  Thus, no expansion of
propagators is needed and the leading term obtained by setting to zero
external momenta in the propagators gives the logarithmic divergence
in $D=24/5$. The corresponding vacuum integrals have four dots.

%%%%%%%%% FIGURE %%%%%%%%%%%%%%%
%%%%%%%%%%%%%%%%%%%%%%%%%%%%%%%%

Because of the complexity of the expressions, essentially all combinations of repeated propagators---up to the maximally-allowed number of dots---and numerators can appear either in the expansion itself or as part of the IBP system. Thus, a clear requirement 
to simplify the integration is to reduce the maximal number of dots.
As discussed above, we would naively expect up to eight dots from the expansion of the naive double copy (level-0) diagrams in 
the representation  of Ref.~\cite{FiveLoopN8Integrand}. It turns out however that,  upon reduction of tensor integrals, all seven- 
and eight-dot vacuum  integrals drop out diagram by diagram. This is a consequence of the structure of the representation of the gauge-theory amplitude.
As will be seen in \sect{AllIntegrationSection}, the IBP system does not close unless it includes integrals with an extra dot compared 
to the desired ones.  Thus, for the old representation we need vacuum integrals with up to seven dots.  There 
are  1,292,541,186  different such vacuum integrals of which 16,871,430 are distinct integrals. It is nontrivial to construct and solve 
the relevant complete IBP system.

For the improved representation of \sect{ImprovedIntegrandSection}, every term in the numerators of level-0 diagrams has 
at least eight external momenta; thus, the leading term corresponds already to logarithmic divergences in $D=24/5$.  No further
expansions of propagators is necessary, implying that the integration of level-0 diagrams in the vacuum expansion requires vacuum 
integrals with at most four dots and an IBP system relating integrals with up to five dots. This is an enormous simplification over the 
earlier integrand.  

Although simpler, the contact diagrams of the new representation of the four-point five-loop $\NeqEight$ integrand contain 
nonvanishing ${\cal N}^{(4)}$ numerator components and thus up to quartic power divergences. Extraction of their logarithmic
divergences requires therefore an expansion to fourth order. 
One might therefore expect vacuum graphs with up to eight dots, which would ruin the simplification of the naive double-copy terms. 
It turns out however that ${\cal N}^{(m)}$ with $m\le 7$  are nonzero only in contact terms in which at least $(8-m)$ external 
lines are attached with four- or higher-point vertex. In the absence of any expansion, the vacuum limit of these graphs has only 
at most $(m-4)$ dots; expanding to $(8-m)$-th order \eqref{expansion_general} to extract the logarithmic divergence yields therefore 
at most four dots. 
%%%
To illustrate this phenomenon, consider the toy example 
\begin{equation}
\frac{2 \ell_5 \cdot k_1}{\ell_5^2 (\ell_5 + k_1)^2} 
= \frac{1}{\ell_5^2} - \frac{1}{(\ell_5 + k_1)^2} \,,
\label{ldotk}
\end{equation}
which we embed in a term that is logarithmically divergent, {\it i.e.} the numerator on the left-hand side is part of the numerator
component ${\cal N}^{(8)}$ of some graph. As discussed before, such terms require no expansion and yield vacuum graphs with 
four dots. 
The terms on the right-hand side mimic the way contact terms are constructed by canceling propagators. Because each numerator 
on the right-hand side is missing a power of external momentum compared to the left-hand side, it is now of ${\cal N}^{(7)}$ type and we need to series expand the denominator to first order in external momenta (which may be either $k_1$ or the other external momenta of the graph).   This series expansion produces exactly one doubled propagator.
This however it does not increase the number of repeated propagators compared to the left-hand side because in going from the 
left- to right-hand side we lost a repeated propagator when setting the external momentum $k_1$ to zero.
The net effect is that the total number of dots in any vacuum graphs arising from the expansion of the contact diagrams does not increase beyond the four that arise from naive double-copy diagrams.

Closing the IBP system by including the diagrams with an additional repeated propagator, we obtain 845,323 independent integrals. 
We will discuss the construction of this system and its solution in section~\ref{AllIntegrationSection}.

%%%%%%%%% FIGURE %%%%%%%%%%%%%%% 
 \begin{figure}[tb]
\begin{center}
{ \includegraphics[scale=0.48]{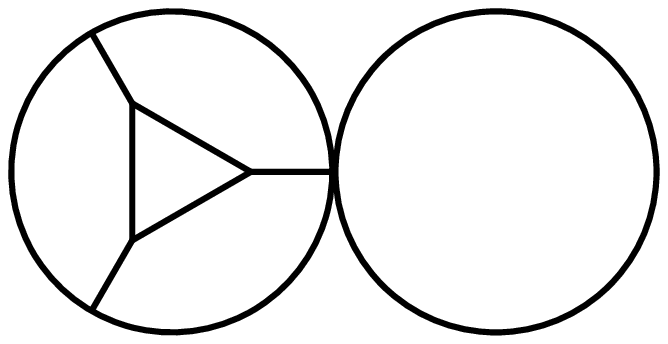}} \hskip .5 cm 
{ \includegraphics[scale=0.48]{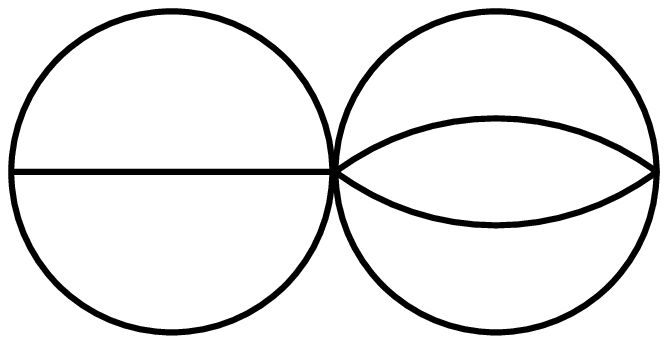} }
\end{center}
\vskip -.6 cm 
\caption{\small Sample factorized vacuum integrals that do not contribute because
of the absence of subdivergences.}
\label{FactorizedVacuumFigure}
\end{figure}
%%%%%%%%%%%%%%%%%%%%%%%%%%%%%%%%

A further important simplification is that since we are working near a fractional dimension, $D = 24 / 5 - 2\epsilon$, 
which in any case is below the critical dimensions at lower-loop orders, no subdivergences are possible. 
Only genuine five-loop vacuum integrals, which do not factorize into lower-loop integrals, can contribute to the 
logarithmic ultraviolet divergence.  Factorized integrals, such as those shown
in \fig{FactorizedVacuumFigure}, are finite in this dimension and can be ignored.

The result of the expansion in external momenta is a collection of vacuum tensor integrals, in which the numerator factors are 
polynomials in Mandelstam invariants of external momenta, inverse propagators and scalar products of loop and external momenta.
For each integral the numerator is separately homogeneous in the loop and external momentum dependence.
These integrals can be further reduced by making use of Lorentz invariance---specifically, that any vacuum tensor integral is a linear combination of products of metric tensors---to separate the dependence on external momenta from that on loop momenta.
More precisely, under integration we can replace a two-tensor
which is dotted into external momentum by
\begin{equation}
\label{TwoTensor}
\ell_i^{\mu} \ell_j^{\nu} \, \rightarrow \, \frac{1}{D} \, \eta^{\mu\nu} \, \ell_i\cdot \ell_j\,,
\end{equation}
and a four-tensor by
\begin{equation}
\ell_i^{\mu} \ell_j^{\nu}\ell_k^{\rho} \ell_l^{\sigma} \, \mapsto \,
\frac{1}{D (D-1) (D+2)} \left(
  A \, \eta^{\mu\nu} \, \eta^{\rho\sigma}
+ B \, \eta^{\mu\rho} \, \eta^{\nu\sigma}
+ C \, \eta^{\mu\sigma} \, \eta^{\nu\rho} \right)\,,
\label{FourTensor}
\end{equation}
where
\begin{align}
A &= (D+1)\ell_i\cdot \ell_j\, \ell_k\cdot \ell_l
 - \ell_i\cdot \ell_k \, \ell_j\cdot \ell_l-\ell_i\cdot \ell_l \, \ell_j\cdot \ell_k\,, \nn\\
B &= -\ell_i\cdot \ell_j \, \ell_k\cdot \ell_l
 +(D+1) \ell_i\cdot \ell_k \, \ell_j\cdot \ell_l-\ell_i\cdot \ell_l \, \ell_j\cdot \ell_k\,, \nn\\
C &= -\ell_i\cdot \ell_j \, \ell_k\cdot \ell_l
 - \ell_i\cdot \ell_k \, \ell_j\cdot \ell_l+(D+1)\ell_i\cdot \ell_l \, \ell_j\cdot \ell_k\,. 
\label{Fourtensorcoeffs}
\end{align}
Since in both cases the highest divergence is quartic, the expansion in small external momenta is to
at most fourth order. Thus, there can be at most four scalar products of loop and external momenta 
and consequently reduction formulas of tensor integrals of rank six or higher are not necessary.

%%%%%%%%%%%%%% FIGURE %%%%%%%%%%%
\begin{figure}
\includegraphics[scale=0.65]{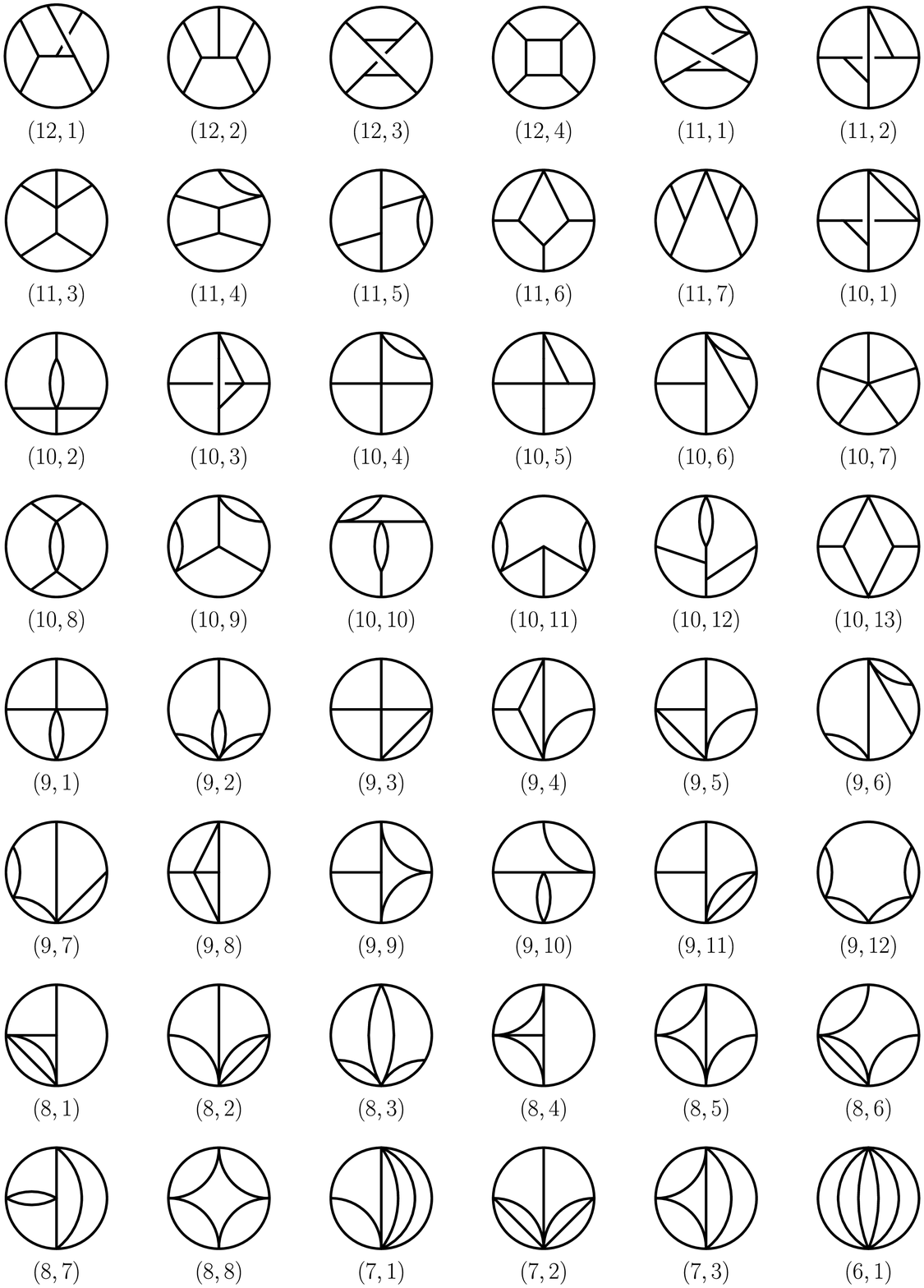} 
%\vspace{20pt}
\caption{All 48 independent vacuum propagator structures, that do not factorize
  into products of lower-loop diagrams. The first number in the diagram label is 
  the number of propagators and the second is the diagram number at that level.}
\label{VacuumIntegralsFigure}
\end{figure}
%%%%%%%%%%%%%%%%%%%%%%%%%%%%%%%

%%%%%%%%%%%%%% FIGURE %%%%%%%%%%%
\begin{figure}[tb]
\includegraphics[scale=.7]{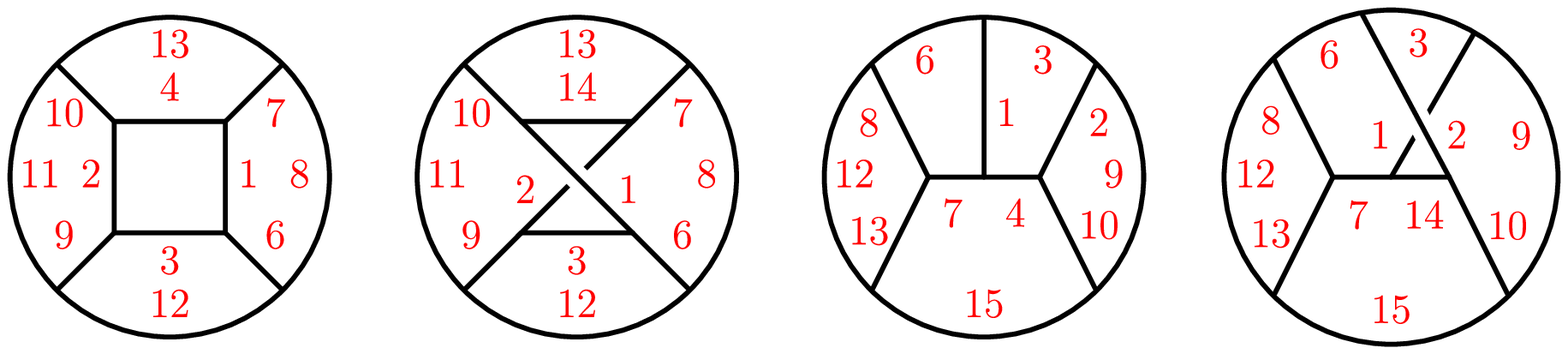} 
%\vspace{20pt}
\caption{The parent vacuum integrals---vacuum integrals with only cubic vertices---with 12 distinct propagators and their
  labels.
  }
\label{VacuumBasis5Loop12PropFigure}
\end{figure}
%%%%%%%%%%%%%%%%%%%%%%%%%%%%%%%

\subsection{Labeling the vacuum diagrams}

After applying Lorentz invariance to reduce the expanded integrals to a
collection of scalar vacuum integrals, with possible numerators and repeated
propagators, we need to organize them into a standard form and
eliminate further redundancies.  The relevant graph topologies are shown in 
\fig{VacuumIntegralsFigure}.
A particularly good labeling scheme has
been devised by Luthe~\cite{LutheThesis}.  Straightforward counting
shows that every vacuum integrand in \fig{VacuumIntegralsFigure} has
15 independent Lorentz dot products between loop momenta.  Depending
on the integral, these dot products are either inverse propagators or
irreducible numerators i.e.~quadratic combinations of loop momenta
that are linearly independent of the propagators.  Remarkably, a
global labeling scheme for momenta can be found for vacuum integrals
at five loops.  We define, following Ref.~\cite{LutheThesis},
\begin{align}
&q_1 = \ell_1, \quad q_2 = \ell_2, \quad q_3 = \ell_3, \quad q_4 = \ell_4, \quad q_5=\ell_5, \quad q_6 = \ell_1 - \ell_3, \quad q_7 = \ell_1 - \ell_4, \nonumber \\
&q_8 = \ell_1 - \ell_5, \quad q_9 = \ell_2 - \ell_3, \quad q_{10} = \ell_2 - \ell_4, \quad q_{11} = \ell_2 - \ell_5, \quad q_{12} = \ell_3 - \ell_5, \nonumber \\
& q_{13} = \ell_4 - \ell_5, \quad q_{14} = \ell_1 + \ell_2 - \ell_4, \quad q_{15} = \ell_3 - \ell_4 \, .
\label{UniformMomenta}
\end{align}
For example, the labeling of the four parent vacuum integrals---vacuum
integrals with only cubic vertices---in this scheme is shown in
\fig{VacuumBasis5Loop12PropFigure}, where the propagator labeled with
$i$ corresponds to $q_i^2$.
The irreducible numerators are $q_i^2$ for the three $i$ labels missing from that diagram.
For daughter diagrams, \ie the 44 diagrams in \fig{VacuumIntegralsFigure} with fewer
than 12 distinct propagators, the number of irreducible numerators is larger, so that the 
total number of independent Lorentz dot products between loop momenta remains the same.
For each daughter diagram there are several possible labelings, inherited from its parents. 
We pick a standard one and map to it all other occurrences of the diagram.

After applying momentum conservation we can rewrite any term in the
integrand of a vacuum integral using the 15 invariants.  With
this labeling scheme we can specify each integral by a list of
the indices representing the exponent of each of the 15 $q_i^2$,
\begin{equation}
\frac{1}{(q_1^2)^{a_1} (q_2^2)^{a_2} (q_3^2)^{a_3} \cdots 
   (q_{14}^2)^{a_{14}} (q_{15}^2)^{a_{15}} }
\Leftrightarrow  F(a_1, a_2, a_3, \ldots,  a_{14}, a_{15}) \, ,
\end{equation}
where a negative power indicates an irreducible numerator rather than
a propagator denominator.  This description is agnostic to whether
the integral is planar or nonplanar, or which diagram the integral is
a daughter of.  Along with the symmetry relations presented next, it
elegantly control the large redundancies introduced by the vacuum
expansion.

In terms of these $F$s, the four diagrams in
\fig{VacuumBasis5Loop12PropFigure} with no irreducible numerators and no
repeated propagators are
\begin{align}
&F(1,1,1,1,0,  1,1,1,1,1,  1,1,1,0,0)\,,  \hskip 1 cm   F(1,1,1,0,0,  1,1,1,1,1,  1,1,1,1,0)\,, \nonumber \\
&F(1,1,1,1,0,  1,1,1,1,1,  0,1,1,0,1)\,,  \hskip 1 cm   F(1,1,1,0,0,  1,1,1,1,1,  0,1,1,1,1) \,.
\end{align}

%%%%%%%%%%%%%%%
\subsection{Symmetry relations among vacuum integrals}
\label{SymmetrySubsection}

%%%%%%%%%%%%%% FIGURE %%%%%%%%%%% 
\begin{figure}
\includegraphics[scale=.6, trim=0 37 0 0]{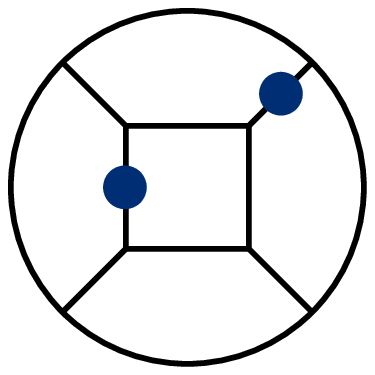} \; \raisebox{.3 cm}{\Large =} \;
\includegraphics[scale=.6, trim=0 37 0 0]{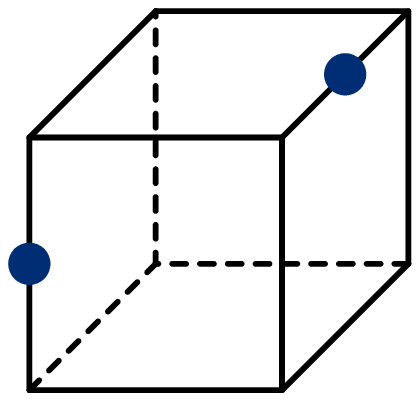}\;
\raisebox{.3 cm}{\Large $\rightarrow$} \;\;
\includegraphics[scale=.6, trim=0 41 0 0]{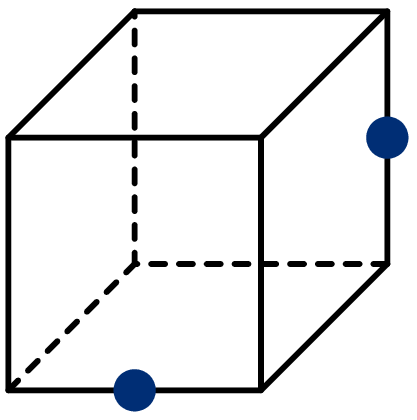}\; \raisebox{.3 cm}{\Large =} \;
\includegraphics[scale=.6, trim=0 37 0 0]{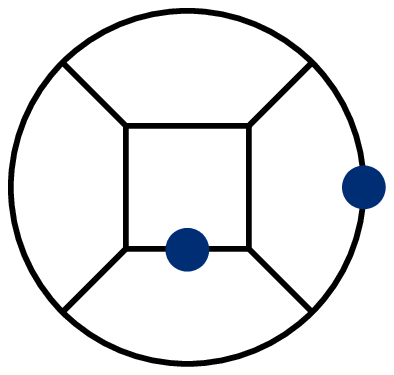}
\vskip .6 cm 
\caption{Moving dots via symmetry in diagram $(12, 4)$ corresponding to the cube.
}
\label{MoveDotsFigure}
\end{figure}
%%%%%%%%%%%%%%%%%%%%%%%%%%%%%%%%%    

In order to efficiently express all integrals in terms of a basis it is useful to first eliminate redundant integrals that are
identical under relabelings.  \fig{MoveDotsFigure} shows an example of
using graph symmetries to rearrange into a canonical format dots that might appear in diagram
$(12,4)$, the cube.  In terms of the $F$s, this symmetry maps
\begin{equation}
F(1,2,1,1,0,  1,2,1,1,1,  1,1,1,0,0) \rightarrow
F(1,1,2,1,0,  1,1,2,1,1,  1,1,1,0,0) \,.
\end{equation}
When irreducible numerators are present, the situation is a bit
more complex because we also need to map the numerators according to
the symmetry transformation. This can generate many contributions when we
re-express the numerators back in terms of the basis $q_i^2$ monomials.
A simple example we encounter is
\begin{align}
&F(1, 1, 1, -1, 0, 3, 2, 0, 0, 0, 0, 2, 2, 1,  0) \rightarrow F(3, 1, 1,  0, 0, 0, 0, 2, 1, 1, 0, 2, 0, 1,  0) \nonumber \\
& - F(3, 1, 2, -1, 0, 0, 0, 2, 1, 1, 0, 2, 0, 1,  0)
+ F(3, 1, 2,  0, 0, 0, 0, 2, 1, 1, 0, 2, 0, 1, -1) \,.
\label{NumeratorSymmetryRelation}
\end{align}
The vast majority of these numerator relabeling relations
often involve iterating the process many times, generating relations
between hundreds of different integrals.

%%%%%%%%%%%%% FIGURE %%%%%%%%%%%%
\begin{figure}[th]
  \centering
  \includegraphics[scale=.65]{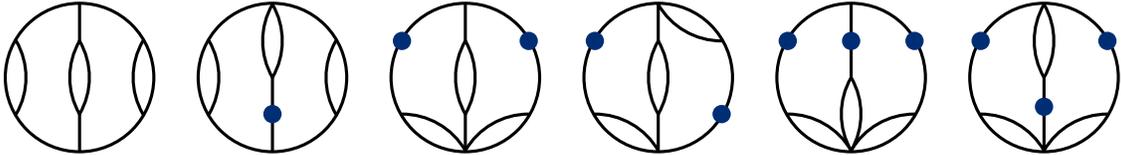}
  \caption{Example of non-isomorphic graphs that all correspond to the same
  Feynman integral.}
  \label{dotsandbubblesFigure}
\end{figure}
%%%%%%%%%%%%%%%%%%%%%%%%%%%%%%%%%

Graph isomorphism is not sufficient to remove all the trivial redundancy,
since certain non-isomorphic graphs can represent the same Feynman integral.
Such relations typically involve ``sliding'' a bubble subdiagram along the 
propagators that connect it to the rest of the graph.  In addition to a different graph 
structure, these transformations can change the number of dots, as illustrated 
in the example in \fig{dotsandbubblesFigure}.
We implement these non-isomorphism graph relations via a graph
transformation that swaps bubble subdiagrams and propagators,
corresponding to the swaps which map the diagrams in
e.g.\ \fig{dotsandbubblesFigure} into each other.  We will refer to this as ``enhanced graph isomorphisms''.
This method efficiently identifies equivalent five-loop vacuum integrals not related by graph isomorphisms.

A less efficient alternative, which we use in parts of the calculation as a consistency check, is to
compute the Symanzik polynomials and bring them to a canonical form
\cite{Pak2011xt,Hoff2016pot}. 
This uses analytic properties of Feynman integrals without resorting to their
graph representation.

Implementing the isomorphism and non-isomorphism relations, we map all integrals to a set of canonical ones.  
There are 3,079,716 scalar vacuum integrals with up to five dots and unit numerator, which map onto 94,670
canonical configurations, as demonstrated in~\fig{MoveDotsFigure}.

%%%%%%%%%%%%%% FIGURE %%%%%%%%%%%%% 
\begin{figure}
$(\ell_1\cdot\ell_2)$\;\includegraphics[scale=.7, trim=0 37 0 0]{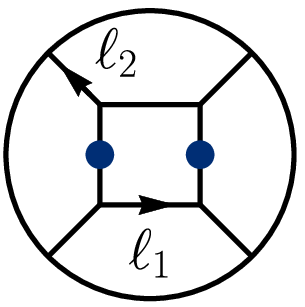} $\; { \hbox{\large $\rightarrow$}} \;-(\ell_1\cdot\ell_3)$\;\includegraphics[scale=.7, trim=0 37 0 0]{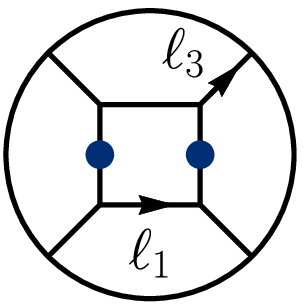} 
\vspace{20pt}
\caption{Numerator relations from residual automorphisms that keep the dot positions invariant. 
}
\label{NumerTransFigure}
\end{figure}
%%%%%%%%%%%%%%%%%%%%%%%%%%%    

In the presence of momentum-dependent numerator factors there also
exist symmetry relations due to automorphisms that preserve both the
graph structure and the position of the dots but change the numerator.
This is distinct from relations of the type in
\eqn{NumeratorSymmetryRelation} which do not relate canonical
integrals, but are used to move dots to canonical positions.  An
example of one particularly simple such relation is given in
\fig{NumerTransFigure}. Transformations of this type generate linear
relations between canonical integrals, which are similar to IBP
relations. Because of this, it is convenient to include and analyze
them together with the IBP relations in \sect{AllIntegrationSection}.

%%%%%%%%%%%%%%%%%%%%%%%%%%%%%
\section{Simplified ultraviolet integration}
\label{TopLevelIntegrationSection}

In this section we discuss the large-loop-momentum integration of the
original form \cite{FiveLoopN8Integrand} of the five-loop four-point $\NeqEight$ supergravity
integrand.  Although, an assumption will be required, this will not only provide a strong cross
check of the complete result obtained in the next section, but will also point
to more powerful ways of extracting the ultraviolet properties of
supergravity theories, especially when combined with the observations
of \sect{PatternsSection}.  As explained in the previous section,
after series expanding and simplifying the original form of the integrand we encounter
vacuum integrals with up to six dots, or repeated propagators, and irreducible numerators. 
Together with the additional dot needed to close the system, this causes a rather unwieldy IBP system.
We will see here that 
the problem can be enormously simplified by targeting parent vacuum integrals---vacuum integrals with only cubic vertices 
or, equivalently, vacuum integrals that have maximal cuts, or also as vacuum integrals with the maximum number of distinct propagators.
The relevant parent vacuum integrals are shown in \fig{VacuumBasis5Loop12PropFigure}.  
We solve the integration-by-parts system on the maximal cuts of the vacuum
integrals, using modern algebraic geometry methods that combine
unitarity cuts with IBP reduction for Feynman
integrals~\cite{Gluza2010ws, CutIntegrals, IBPAdvances, BaikovCuts,NumericalUnitarity}.  

Besides enormously simplifying reduction to a set of master integrals
by focusing on the vacuum integrals with maximal cuts, targeting parent vacuum integrals also has
the added benefit of allowing us to immediately drop large classes of
contact terms from the integrand, including all contact terms obtained
from the \N5 and \N6 levels, even before expanding into vacuum
diagrams.  Any term where a propagator is completely canceled in the vacuum graph can be
dropped. 

In manipulating the vacuum integrals, there are two important issues that must be addressed. 
The first one is the separation of the infrared and ultraviolet divergences. 
This is an important ingredient in various studies of ultraviolet properties, such as the analysis of $\NeqFour$,  $\NeqFive$ 
and  $\NeqEight$ supergravity at three and four loops~\cite{N4GravThreeLoops, N4GravFourLoop, N5GravFourLoops, ThreeFourloopN8}, and the   computation the five-loop beta
function in QCD~\cite{FiveloopQCDBeta}. 
Although there are no physical infrared singularities in $D>4$, our procedure of series expanding around small external 
momenta introduces spurious ones.
We will show in detail in the next section that in an infrared-regularized setup for integrals with no ultraviolet subdivergences, 
terms in the IBP system that are proportional to the infrared regulator involve only ultraviolet-finite integrals. Thus, since we are 
interested only in the ultraviolet poles, we can effectively reduce the vacuum integrals without explicitly introducing an infrared 
regulator.
For the rest of this section, when we discuss linear relations between
integrals, it should be understood that we actually mean linear
relations between the ultraviolet poles of the integrals.

A second issue is that the vacuum expansion of our integrand
contains propagators with raised powers, which is in contradiction with the naive
unitarity cut procedure of replacing propagators by on-shell delta
functions. 
Fortunately, two solutions to this problem are available in the literature. One option~\cite{Sogaard2014ila} is to 
define the cut as the contour integral around propagator poles; this effectively identifies the cut as the residue of 
the propagator pole even for higher-order poles.
Another, proposed in  Ref.~\cite{Zhang2016kfo}, is to use dimension shifting~\cite{Tarasov1996br} such that all propagators
appear only once at the cost of shifting the integration dimension and raising the power of numerators, before
imposing the maximal-cut conditions to discard integrals with canceled propagators.
Here we will use the second strategy.

Starting with the integrand of Ref.~\cite{FiveLoopN8Integrand},
the end result of dimension shifting procedure is a set of vacuum integrals in
$D=-36/ 5 -2\epsilon$ with a total $30$ powers of the irreducible numerators. For
example, for the crossed-cube vacuum diagram shown in the second diagram of \fig{VacuumBasis5Loop12PropFigure},
we have integrals of the form
\begin{equation}
\label{eq:xcubeMaxCutIntegrals}
\int \prod_{k= 1}^5 \frac{d^{D} \ell_k}{(2 \pi)^D} \,
 \frac{(q_4^2)^{A_4} \, (q_{5}^2)^{A_{5}} \, (q_{15}^2)^{A_{15}}}{q_1^2 \, q_2^2 \, q_3^2 \, \widehat q_4^2 \,
\widehat q_5^2\, q_6^2\, q_7^8 \, q_8^2 \,q_9^2 \, q_{10}^2  \, q_{11}^2 \, 
q_{12}^2 \, q_{13}^2\, q_{14}^2 \, \widehat q_{15}^2 } \,,
\end{equation}
where $D = -36/5-2\epsilon$ and the ``hats'' in the denominator mean to
skip those propagators.  The $q_i$ are the uniform momenta defined in
\eqn{UniformMomenta}.  Here the three irreducible numerators are
$q_4^2$, $q_{5}^2$ and $q_{15}^2$; these cannot be written as the
linear combinations of the 12 propagator denominators, as explained in the previous section. To 
obtain a logarithmic divergence in the shifted dimension $-36 / 5$,  we need 30 
powers of numerator factors
\begin{equation}
\label{eq:numRankLogDiv}
A_4 + A_5 + A_{15} = 30\,, \hskip 1 cm \hbox{with } A_4 \ge 0\,, A_5 \ge 0 \,, A_{15} \ge 0 \,.
\end{equation}
In total there are $496$ different combinations of $A_j$ that
satisfy Eq.~\eqref{eq:numRankLogDiv}.  With the new integrand of
\sect{ImprovedIntegrandSection} the power counting is greatly improved
so we need only shift to $D= -16/5 -2\epsilon$ with 20 powers of numerators.
This gives 231 integrals to evaluate.

Consider the cross-cube diagram shown in the second diagram in
\fig{VacuumBasis5Loop12PropFigure}.  
The IBP identities relating the 496 integrals are of the form
\begin{equation}
\label{eq:defIBP}
\int \prod_k \frac{d^D \ell_k}{(2\pi)^D} \frac{\partial}{\partial \ell_i^\mu} \frac{v_i^\mu}{\prod_j d_j} = 0 \,,
\end{equation}
where $v_i^\mu$ has polynomial dependence on external and internal momenta and the $d_j$ are 
the various propagators. We refer to
\begin{equation}
v_i^\mu \frac{\partial}{\partial \ell_i^\mu} \,,
\end{equation}
as the IBP-generating vector, while the rest of Eq.\ \eqref{eq:defIBP},
\begin{equation}
\int \prod_k \frac{d^D \ell_k}{(2 \pi)^D} \frac{1}{\prod_j d_j} \,,
\end{equation}
is referred to as the seed integral.  Integration by parts as above
re-introduces auxiliary integrals with propagators raised to higher powers, since
the derivatives can act on the propagator denominators. 
Lowering again the propagator powers through dimension shifting leads still to new integrals because, 
while of the same topology at the starting ones, they are now in a different dimension.

To eliminate these auxiliary integrals Gluza, Kadja and Kosower~\cite{Gluza2010ws} formulated
IBP relations without doubled propagators, using
special IBP-generating vectors that satisfy
\begin{equation}
\label{eq:GKKCondition}
v_i^\mu \frac{\partial}{\partial \ell_i^\mu} d_j = f_j \, d_j  \,,
\end{equation}
for \emph{all} values of $j$ with $f_j$ restricted to be polynomials
(in external and loop momenta). This cancels any squared propagator
generated by derivatives, and does not introduce spurious new
denominators since $f_j$ are polynomials.  Since the original
publication, strategies for solving \eqn{eq:GKKCondition} have been
explored in Refs.~\cite{Gluza2010ws, IBPAdvances,NumericalUnitarity}. 
We use the strategy in Ref.~\cite{NumericalUnitarity} to obtain a complete set of vectors
 $v_i^\mu$ using computational algebraic geometry algorithms implemented in {\sc SINGULAR}~\cite{Singular}. 
They in turn give the complete set of IBP relations among the 496 cross cube integrals discussed above \eqref{eq:xcubeMaxCutIntegrals}, \eqref{eq:numRankLogDiv} and implies that all of them are expressed in terms 
of a single integral---the second diagram in \fig{VacuumBasis5Loop12PropFigure}.
A similar analysis solves the analogous problem for the 496 integrals of cube topology and expresses them in 
terms of the integral corresponding to the first graph in \fig{VacuumBasis5Loop12PropFigure}.
The IBP systems restricted to integrals with maximal cuts for the parent topologies with internal triangles, 
corresponding to the third and fourth graph in  \fig{VacuumBasis5Loop12PropFigure}, sets all integrals to zero, implying that 
they are all reducible to integrals that do not have maximal cuts.

As a cross-check for the crossed-cube topology, we have also analytically solved for the integrals in closed form by
contour integration~\cite{BaikovCuts} using the Baikov representations~\cite{BaikovRep}, without making use of integral relations 
of the type \eqref{eq:defIBP}. We refer the reader to Ref.~\cite{FiveLoopN8Integrand} for the details of the analogous
computation in $D= 22/5$.  In that case, all parent vacuum diagrams cancel, as expected.

By inverting the dimension shifting relations we can re-express the
final result in terms of parent master integral in the original
dimension $D = 24 /5 -2\epsilon$.  The final result for the leading 
ultraviolet behavior is remarkably
simple:
\begin{align}
{\cal M}_4^{(5)} \Bigr|^{\rm parent\hbox{-}level}_{\rm leading} & =
- \frac{629}{25} \left(\frac{\kappa}{2}\right)^{12}
(s^2 + t^2 + u^2)^2  stu M_4^\tree\,
\left({\ts \frac{1}{3}} \inlinefig{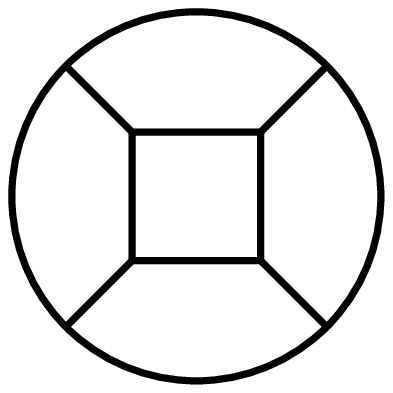} 
          + \inlinefig{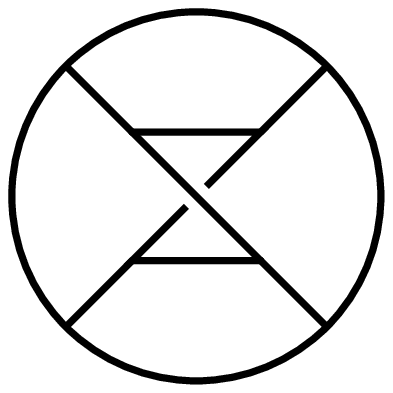} \right) .
\label{FiveLoopN8UVTopLevel}
\end{align}
We obtain identical result, whether we start from the integrand of
Ref.~\cite{FiveLoopN8Integrand} or the improved one in
\sect{ImprovedIntegrandSection}.  This provides a highly nontrivial
check on the cut construction and the integral reduction procedure.
Most importantly, as we show in the next section, the result in
\eqn{FiveLoopN8UVTopLevel} is complete, even though we kept only the parent
master integrals, which have no canceled propagators. 
As we shall see in \sect{PatternsSection}, this seems unlikely to be accidental.

%%%%%%%%%%%%%%%%%%%%%%%%%%%%%
\section{Full ultraviolet integration}
\label{AllIntegrationSection}

In this section, we extract the ultraviolet divergence of the
five-loop four-point $\NeqEight$ supergravity amplitude without making
any assumptions on the class of vacuum integrals that contribute.  To
keep the IBP system under control, we use the improved representation
of the integrand found in \sect{ImprovedIntegrandSection}, expanded at
large loop momentum, as described in \sect{VacuumExpansionSection}.
We organize the IBP relations using and SL$(L)$ reparametrization
symmetry of $L$ loop momenta~\cite{IntegralRelations}. We also
incorporate the integral relations resulting from graph automorphisms
that change kinematic numerator factors, a simple example of which is
shown in \fig{NumerTransFigure}.

%%%%%%%%%%%%%% FIGURE %%%%%%%%%%%%% 
\begin{figure}
\includegraphics[scale=.5]{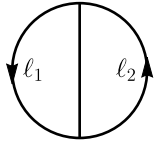}
\vskip -.3 cm 
\caption{Two-loop example for illustrating SL$(L)$ symmetry. }
\label{TwoloopIBPFigure}
\end{figure}
%%%%%%%%%%%%%%%%%%%%%%%%%%%    

\subsection{IBP for ultraviolet poles modulo finite integrals}

Since standard IBP reduction is usually performed for full integrals
in dimensional regularization, there is a large amount of unnecessary
computation for our purpose of extracting only the ultraviolet
poles.\footnote{We have already performed expansion in the
  ultraviolet region to produce vacuum integrals, but even the (infrared-regulated)
  vacuum integrals contain finite parts that are not of interest to
  us here.} We now review setting up a simplified IBP system that only
gives linear relations between the leading ultraviolet poles of
different vacuum integrals~\cite{IntegralRelations}.

As a warm up, consider the toy example of two-loop vacuum
integrals in $D=5-2\epsilon$ shown in \fig{TwoloopIBPFigure}. 
This example will mimic the supergravity situation because
there are no (one-loop) subdivergences due to the properties of
dimensional regularization. We define such two-loop integrals as
\begin{equation}
V_{A,B,C} = \int \frac{d^{D} \ell_1}{(2 \pi)^D} \, \frac{d^{D} \ell_2}{(2\pi)^D} \, 
\frac{1}{[(\ell_1)^2-m^2]^A \,  [(\ell_2)^2-m^2]^B \, [(\ell_1 - \ell_2)^2-m^2]^C}\,,
\end{equation}
where we require $A+B+C=5$ since we are interested in logarithmically
divergent integrals. In this case, there are no irreducible
numerators.

Consider GL(2) transformations of the loop momenta $\Delta \ell_i \equiv \Omega_{ij} \ell_j $,
which generate IBP relations of the form,
\begin{equation}
0 = \int \frac{d^{D} \ell_1}{(2 \pi)^D} \, \frac{d^{D} \ell_2}{(2\pi)^D} \, \frac{\partial}{\partial \ell_i^\mu} 
 \frac{\Omega_{ij} \ell_j^\mu}{[(\ell_1)^2-m^2]^A \,  [(\ell_2)^2-m^2]^B \, [(\ell_1 - \ell_2)^2-m^2]^C} \,,
\end{equation}
where $D = 5 - 2\eps$.
We first look at the SL(2) subalgebra which excludes the trace part of the GL(2) generators. For example, the SL(2) generator
\begin{equation}
  \renewcommand*{\arraystretch}{0.9}
  \Omega_{ij} = \begin{pmatrix}
    \;1 & 0 \; \\
 \;0 & -1 \;
\end{pmatrix} \,,
\end{equation}
produces the IBP relation
\begin{align}
0 &= \int \frac{d^{D} \ell_1}{(2 \pi)^D} \frac{d^{D} \ell_2}{(2\pi)^D} \left( \ell_1^\mu \frac{\partial}{\partial \ell_1^\mu} 
- \ell_2^\mu \frac{\partial}{\partial \ell_2^\mu} \right) 
\frac{1}{(\ell_1^2-m^2)^A (\ell_2^2-m^2)^B \left[ (\ell_1-\ell_2)^2-m^2 \right]^C} \nn \\
&=  (-2A+2B) V_{A,B,C} - 2C\, V_{A-1,B,C+1} + 2 C V_{A,B-1,C+1} \nn \\
&\hskip 4 cm \null  + m^2 \left( -2A \, V_{A+1,B,C} + 2B \, V_{A,B+1,C} \right) \,, 
\label{eq:twoLoopIBPABC}
\end{align}
where we used $A+B+C=5$. The second-to-last line of the
above equation contains integrals that are logarithmically divergent in the ultraviolet, while the
last line contains integrals that are ultraviolet finite by power
counting---as indicated by simple considerations of
dimensional analysis, since the last line is proportional to $m^2$.
Absence of subdivergences implies that \emph{overall} power counting is
sufficient for showing whether an integral is ultraviolet finite. Therefore, for the
purpose of extracting ultraviolet divergences, we can disregard the last line
of the above equations, and instead work with an IBP system
\emph{modulo finite integrals}. Since the generators of the SL$(2)$ subalgebra are
traceless, the IBP relations we generate have no explicit dependence
on the dimension $D$.

Inspecting \eqn{eq:twoLoopIBPABC} we see that, setting $m=0$ from the beginning removes the last line of 
that equation while preserving the relation between integrals exhibiting ultraviolet poles. 
Thus, even though setting $m=0$ turns these vacuum integrals into scaleless integrals that vanish in dimensional
regularization, the SL$(2)$ subalgebra nonetheless generates the correct IBP relations between between ultraviolet poles.
In contrast, including the trace generator,
\begin{equation}
  \renewcommand*{\arraystretch}{0.9}
\Omega_{ij} = \begin{pmatrix}
 \;1\; & \; \;0 \;\;\\
 \;0\; & \; \;1 \;\;
\end{pmatrix} ,
\end{equation}
which extends SL$(2)$ to GL$(2)$, requires nonvanishing $m$. Indeed, this generator produces the IBP relations
\begin{align}
0 &= \int \frac{d^{D} \ell_1}{(2 \pi)^D}\,  \frac{d^{D} \ell_2}{(2\pi)^D} \frac{\partial}{\partial \ell_i^\mu} 
\frac{\ell_i^\mu}{[(\ell_1)^2-m^2]^A \,  [(\ell_2)^2-m^2]^B \, [(\ell_1 - \ell_2)^2-m^2]^C} \nonumber \\
&= -4\epsilon V_{A,B,C} -10 m^2 ( V_{A+1,B,C} + V_{A,B+1,C} + V_{A,B,C+1} ) \, .
\end{align}
If we set $m=0$, the above relations imply that $V_{A,B,C}=0$.
The factor $(-4\epsilon)$ is expected because the diagonal transformation probes the
scaling weight of the integral, which would be exactly zero in
$D=5$. As long as the IBP relations corresponding to the trace part of GL(2) are
omitted, the IBP system no longer sets to zero massless vacuum integrals and
correctly reflects the ultraviolet poles of these integrals without
contamination from IR poles.

The above argument straightforwardly carries over to the five-loop vacuum
integrals in $D=24/5 - 2\epsilon$, since no subdivergences exist
in this dimension. The resulting IBP system only involves logarithmically divergent vacuum integrals, 
and does not include any finite integrals or power-divergent integrals (which do not produce poles in
dimensional regularization). 
This enormously reduces the size of the linear system to be solved.

A useful property of the SL$(L)$-generated IBP system is that, even though
each vacuum integral depends on the dimension $D$ implicitly, the relations between them
do not contain any explicit dependence on $D$ \cite{IntegralRelations}.
This fact appears to help explain the observations in \sect{PatternsSection}.

\subsection{The IBP system at five loops}

The complete set of integral topologies---suppressing dots or
numerators---that we need to consider for the reduction of the vacuum
integrals of the five-loop four-point $\NeqEight$ supergravity
amplitude is shown in \fig{VacuumIntegralsFigure}. This list does not
include any diagram that factorizes, such as those illustrated in
\fig{FactorizedVacuumFigure}.
It also removes integrals related to kept ones by identities between
integrals not isomorphic to each other, such as those illustrated in
\fig{dotsandbubblesFigure}.

By acting with the SL$(5)$ generators on all logarithmically divergent
canonical integrals with up to four dots, we find IBP relations between
vacuum integrals with up to five dots, the additional dot following
from acting with derivatives on propagators. While such integrals do
not appear in the expansion of the integrand in $D=24/5$, they are
necessary for finding the relations between integrals with four dots.
We also include relations between integrals generated by graph automorphisms which transform nontrivially the numerator 
factors, as illustrated in~\fig{NumerTransFigure}. 
In these relations, all the integrals are mapped to canonical
integrals using enhanced graph isomorphisms as described in \sect{SymmetrySubsection}.
Because of their similarity with the IBP relations it is convenient to solve them simultaneously.
The solution to this system of equations expresses all needed vacuum integrals in terms of master integrals.

%%%%%%%%%%%%%% FIGURE %%%%%%%%                                                                                  
\begin{figure}
\includegraphics[scale=.55]{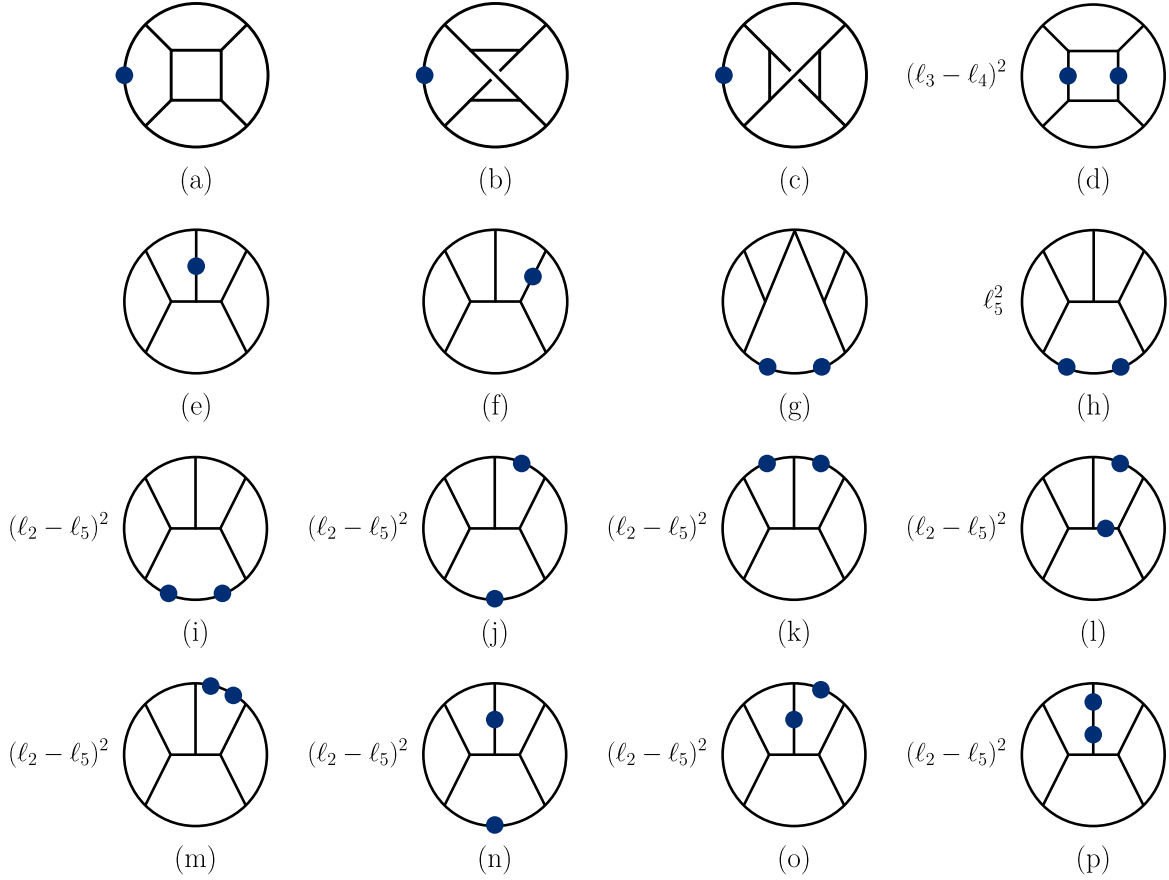}
\vskip -.3 cm
\caption{The sixteen  master integrals to which any five-loop vacuum
  integrals in $\NeqFour$ super-Yang--Mills with up to two dots can be
  reduced.  The dots represent repeated propagators.  The labels of the diagrams match
  those of \fig{VacuumBasis5Loop12PropFigure}.
    }
\label{16masterYMbasisFigure}
\end{figure}
%%%%%%%%%%%%%%%%%%%%%%%%%%%%%% 

As a warm up to setting up and solving the IBP system for the supergravity problem in $D=24/5$, we solved the much simpler
cases of $\NeqEight$ supergravity in $D=22/5$ and $\NeqFour$ super-Yang--Mills theory in $D = 26/5$.  The integrals which 
appear in both these simpler  cases have at most two dots and thus, the IBP system contains integrals with up to three dots.
In the case of $\NeqEight$ supergravity in $D=22/5$, the three-dot system has 44,428 different integrals, and about 
$1.7\times10^5$ linear relations generated.  The simpler numerator factors of  $\NeqFour$ super-Yang--Mills make this case much simpler, 
containing only 5,975 distinct integrals and about 9,900 linear relations between them.  
The solution of the latter system expresses all the two-dot vacuum integrals, divergent in $D=26/5$,
in terms of the 16 master vacuum integrals displayed in \fig{16masterYMbasisFigure}.

%%%%%%%%%%%%%% FIGURE %%%%%%%%%%%%% 
\begin{figure}
\includegraphics[scale=.6]{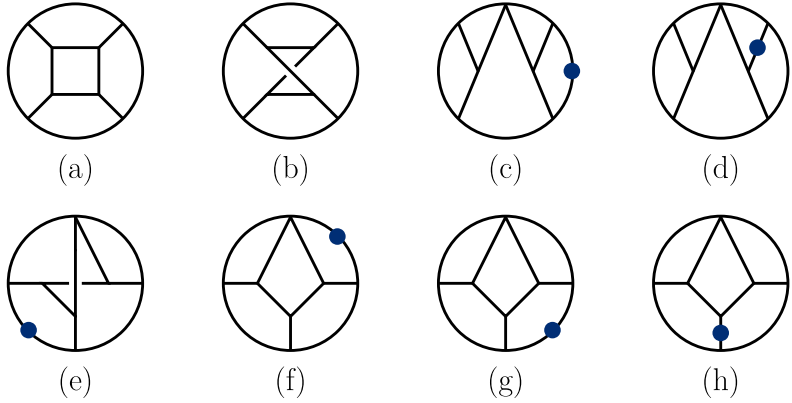}
\vskip -.3 cm 
\caption{The eight master integrals to which any five-loop vacuum
  integrals in $\NeqEight$ supergravity with up to four dots can be
  reduced.  The dots represent repeated propagators.
}
\label{8masterbasisFigure}
\end{figure}
%%%%%%%%%%%%%%%%%%%%%%%%%%%    

For the main problem of $\NeqEight$ supergravity in $D=24/5$ with the improved integrand obtained in 
\sect{ImprovedIntegrandSection}, we have to reduce integrals with up to four dots. There are 141,592 distinct integrals of this type.
The relevant five-dot system has 3,687,534 integrals of which 845,323 are distinct. The SL$(5)$ transformations generate 
about $2.8\times10^6$  IBP relations, while numerator-changing isomorphisms generate about $9\times10^5$ further relations. 
This system is straightforward to solve using sparse Gaussian elimination and finite-field methods~\cite{FiniteFields}; we used the linear system 
solver {\sc LinBox}~\cite{LinBox}, and confirmed the solution with {\sc FinRed}\footnote{We thank Andreas von Manteuffel 
and Robert  Schabinger for providing us with this program.}~\cite{ManteuffelSchabingerFiniteFields}. 
The result is that all vacuum integrals for the expansion of $\NeqEight$ supergravity amplitude in $D=24/5$ are expressed 
as linear combinations of the eight master integrals shown in \fig{8masterbasisFigure}.

\subsection{Result for ultraviolet divergences}

As a first test for the full calculation, we used the reduction of the vacuum integrals to verify that our integrand
exhibits the known ultraviolet properties in $D=22/5$. We find that, as expected, all vacuum 
integrals cancel after IBP reduction, the five-loop four-point $\NeqEight$ amplitude is
ultraviolet finite,
\begin{equation}
{\cal M}_4^{(5)} \Bigr|^{D = 22/5}_{\rm leading } = 0\,.
\end{equation}
With our new integrand there are few potential contributions because
the naive double-copy terms are manifestly ultraviolet finite in
$D=22/5$ and only the contact terms give potential contributions.  A
similar check is performed for the earlier form of the integrand in
Ref.~\cite{FiveLoopN8Integrand}, but that case only 
confirms the cancellation of the vacuum diagrams with the maximum
cuts imposed.

As another test of our approach, we also recovered the leading
divergence of $\NeqFour$ super-Yang--Mills theory in its five-loop
critical dimension, $D=26/5$, originally found in~\cite{FiveLoopN4}. 
Starting from our
improved $\NeqFour$ super-Yang--Mills integrand of
\sect{ImprovedIntegrandSection}, extracting the leading divergence in
terms vacuum integrals and then substituting their expressions in
terms of master integrals, we obtain
\begin{align}
{\cal A}_4^{(5)}\Bigr|_{\rm leading} & \hskip -.2 cm =
  \frac{144}{5}  g^{12} s t A^\tree
N_c^3 \,  \left(\hskip -.1 cm  N_c^2 \inlinefig{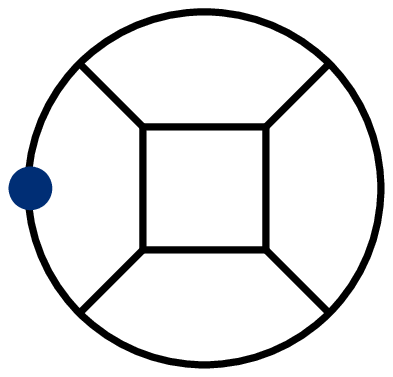}
+ 48 \left(\!\!{\ts\frac{1}{4}} \inlinefig{Vacuum5loopsYMV1.eps}
         \! + {\ts\frac{1}{2}} \inlinefig{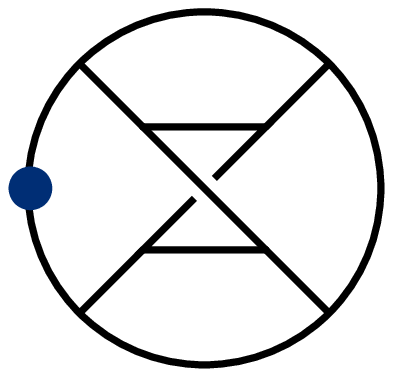}
        \! + {\ts\frac{1}{4}} \inlinefig{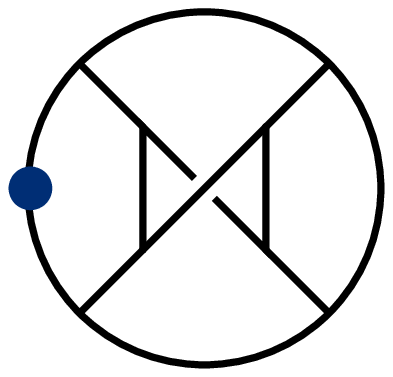}
\right) \! \right)\nonumber \\
& \hskip 4 cm \null \times 
\Bigl( t\, \f^{a_1 a_2 b} \f^{b a_3 a_4}
          + s \, \f^{a_2 a_3 b} \f^{b a_4 a_1} \Bigr) \,.
\label{FiveLoopN8Result}
\end{align}
The $\f^{a b c}$ are the group structure constants, as normalized
below \eqn{CubicRepresentation}, and the $s$ and $t$ are the usual
Mandelstam invariants. Here $A^\tree\equiv A^\tree(1,2,3,4)$ is the
color-ordered tree amplitude with the indicated ordering of external
legs. This reproduces the result of Ref.~\cite{FiveLoopN4}, providing
a nontrivial check of both our gauge-theory integrand construction and IBP
reductions methods.

Interestingly, the thirteen  master integrals in \fig{16masterYMbasisFigure} that have vanishing coefficients 
in \eqn{FiveLoopN8Result} violate a ``no-one-loop-triangle'' rule.\footnote{When counting the number of 
propagators around a loop, each dot should be counted as well.}
Indeed,  diagrams (e)-(p) contain one-loop triangle subdiagrams while diagram (d) contains a loop 
momentum-dependent numerator in one-loop box subdiagrams, which upon expanding and reducing 
of that one-loop subintegral also leads to triangle subintegrals.
Another interesting feature of these results is that the relative
factors of the subleading-color term are given by the symmetry factors
of the corresponding integrals.
In the next section, we will show that these observations are part of a more general pattern.

Extracting the leading ultraviolet terms for $\NeqEight$ supergravity in $D=24/5$
follows the same strategy. 
After reducing the vacuum integrals obtained from our improved integrand to the basis of 
master integrals we find  
\begin{align}
{\cal M}_4^{(5)} \Bigr|_{\rm leading } & \hskip -.2 cm =
- \frac{16\times 629}{25} \left(\frac{\kappa}{2}\right)^{12}
(s^2 + t^2 + u^2)^2  stu M_4^\tree\,
\left({\ts \frac{1}{48}} \inlinefig{Vacuum5loopsV1.eps}
      + {\ts \frac{1}{16}}  \inlinefig{Vacuum5loopsV2.eps}  \right) .
\label{FiveLoopN8UV}
\end{align}
This is the same result as obtained in the previous section by assuming that only vacuum diagrams
with maximal-cuts contribute, and proves that \eqn{FiveLoopN8UVTopLevel} is
complete. 
As in the case of the reduction of the expansion of the four-point five-loop $\NeqFour$ 
super-Yang--Mills amplitude, all master integrals containing triangle subdiagrams,
or with numerators which upon further one-loop reduction lead to triangle subdiagrams,
enter with vanishing coefficients.
Moreover, similarly to the  subleading color in the gauge-theory case, the relative coefficients 
between the integrals are the symmetry factors of the vacuum diagrams.  As we discuss in
the next section, these observations do not appear to be accidental.

The two Wick-rotated vacuum integrals in \eqn{FiveLoopN8UV} are both positive definite, proving that 
no further hidden cancellations are present. We evaluated numerically, using  {\sc FIESTA}~\cite{FIESTA}, 
the two master integrals entering Eq.~\eqref{FiveLoopN8UV}, given by diagrams (a) and (b) in \fig{8masterbasisFigure}, and 
find
\begin{equation}
V_5^{\rm (a)} = \frac{1}{(4\pi)^{12}} \, \frac{0.563}{\eps} \,, \hskip 1 cm 
V_5^{(\rm b)}  =\frac{1}{(4\pi)^{12}} \, \frac{0.523}{\eps} \, .
\end{equation}
The dimensional-regularization parameter is $\eps = 
(24/5-D)/2$.
Using \eqn{FiveLoopN8UV}, the numerical value of the divergence is 
\begin{align}
{\cal M}_4^{(5)} \Bigr|_{\rm leading} & \hskip -.2 cm =
-  17.9  \left(\frac{\kappa}{2}\right)^{12}
\frac{1}{(4\pi)^{12}} 
(s^2 + t^2 + u^2)^2  stu M_4^\tree\,
 \frac{1}{\eps}  \,.
\end{align}
We leave as a problem for the future the question of obtaining an
exact analytic expression instead of the numerical one found here.

%%%%%%%%%%%%%%%%%%%%%%%%%%%%%
\section{Observations on ultraviolet consistency}
\label{PatternsSection} 

Given the wealth of results from previous 
papers~\cite{GSB, BDDPR, ThreeFourloopN8, CompleteFourLoopSYM, ck4l, SixLoopPlanarYM,  FiveLoopN4}, 
as well as those from \sect{AllIntegrationSection}, we are in the position to search for useful structures that can 
lead to a more economic identification of the leading ultraviolet behavior of $\NeqFour$ super-Yang-Mills theory 
and $\NeqEight$ supergravity.  
In this section we analyze the available results in both these theories, observing remarkable consistency and recursive 
properties, whereby leading $L$-loop ultraviolet divergences in the $L$-loop critical dimension appear to be tightly constrained 
by the lower-loop vacuum diagrams describing leading behavior in the lower-loop critical dimension.

First we collect the known results for the leading ultraviolet behavior of both $\NeqFour$ super-Yang--Mills theory and $\NeqEight$ 
supergravity.  We then demonstrate that appropriately-defined subdiagrams of the vacuum diagrams are simply related to
the vacuum diagrams describing lower-loop leading ultraviolet behavior.  

Within the generalized-unitarity method, higher-loop scattering amplitudes  are constructed in terms of lower-loop ones.
The one-particle cut, setting on shell a single propagator, provides a direct link between $L$-loop $n$-point amplitudes
and $(L-1)$-loop $(n+2)$-point amplitudes. 
One may therefore suspect that there may exist a relation between the leading ultraviolet properties of these amplitudes in their respective critical dimensions, which echoes the relation between the complete amplitudes. 
We will find, however, more surprising consistency relations between the leading ultraviolet behavior of $L$- and 
$(L-1)$-loop amplitudes with the same number of external legs for $L\le 6$ for $\NeqFour$ super-Yang-Mills theory and for 
$L\le 5$ for $\NeqEight$ supergravity. 
The nontrivial manipulations necessary for extracting the leading ultraviolet divergence adds to the surprising features of these
relations. Indeed, without appropriate choices of integral bases, they would be obfuscated.  
They point to the possibility of a principle governing perturbative consistency in the ultraviolet.  We close by noting the possibility 
that one may exploit these patterns to directly make detailed predictions of ultraviolet properties at higher loop orders.

\subsection{Review of results}

After IBP reduction, we obtain a simple description of the leading
ultraviolet behavior in terms of a set of master vacuum integrals 
defined as
\begin{equation}
V = -i^{L + \sum_j A_j} \int \prod_{i = 1}^L \frac{d^D \ell_i}{(2 \pi)^D}  \prod_{j} \frac{1}{(p_j^2 -m^2)^{A_j}} \,,
\end{equation}
where the $p_i$ are linear combinations of the independent loop momenta and the $A_i$ are the propagators' exponents. 
The number of dots on propagator $j$ is $A_j -1$ for $A_j \ge 2 $
The indices can  be negative, in which case they represent irreducible numerators, as discussed in \sect{AllIntegrationSection}.  
While there is no need to explicitly introduce a mass regulator for carrying out the IBP reductions, 
we do so here to make the integrals well defined in the infrared.

%%%%%%%%%%%%%%%%% TABLE %%%%%%%%%%%%%%%%%%%%%%%%%%%%%%%%%
\begin{table}[tb]
\begin{center}
\begin{tabular}{|c|c|c|}
\hline
 Loops & $D_c$ for $\NeqFour$ sYM & $D_c$ for $\NeqEight$ sugra \\
\hline
1 &  8  & 8 \\
2 &  7  & 7 \\
3 &  6  & 6 \\
4 &  11/2  & 11/2 \\
5 &  26/5  & 24/5 \\
6 &  5  & --- \\
\hline
\end{tabular}
\caption{The critical dimensions where ultraviolet divergences first occur in 
$\NeqFour$ super Yang--Mills theory and $\NeqEight$ supergravity, as determined by explicit calculations.}
 \label{CriticalDimTable}
\end{center}
\end{table}
%%%%%%%%%%%%%%%%% TABLE %%%%%%%%%%%%%%%%%%%%%%%%%%%%%%%%%

Collecting the results from Refs.~\cite{GSB, BDDPR, ThreeFourloopN8,
  ck4l} and from \eqn{FiveLoopN8UV}, the leading ultraviolet behavior
of $\NeqEight$ supergravity at each loop order through five loops is
described by vacuum diagrams as
\begin{align}
{\cal M}_4^{(1)}\Bigl|_{\rm leading} & = -3  \, \fancyM  \left(\frac{\kappa}{2} \right)^4
 \,\inlinefig{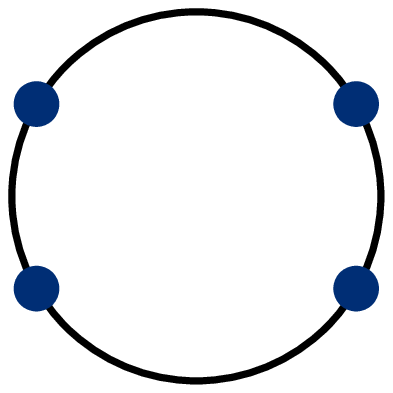} \,,\nn \\
{\cal M}_4^{(2)}\Bigl|_{\rm leading} & = - 8  \, \fancyM  \left(\frac{\kappa}{2} \right)^6
(s^2+t^2+ u^2)
\, \left({\ts \frac{1}{4}}  \inlinefig{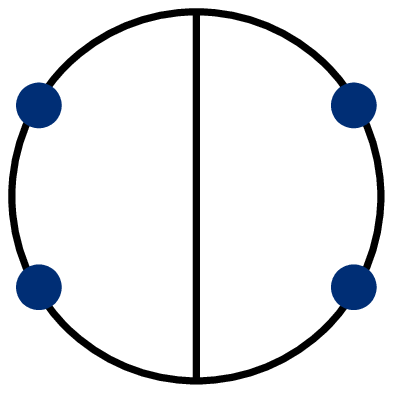} 
   + {\ts \frac{1}{4}} \inlinefig{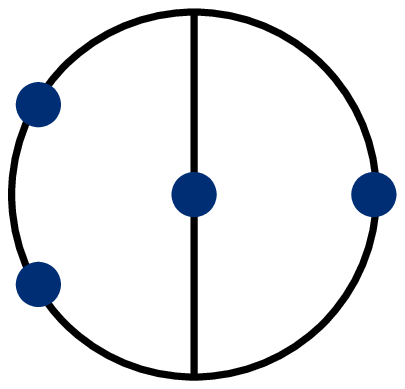} \right) , \nn \\
{\cal M}_4^{(3)}\Bigr|_{\rm leading} & =
- 60\,  \fancyM  \left(\frac{\kappa}{2} \right)^8\, s t u \,
\left( {\ts \frac{1}{6}}  \inlinefig{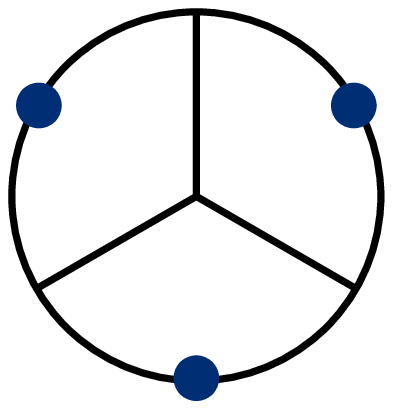} +
 \, {\ts \frac{1}{2}}  \inlinefig{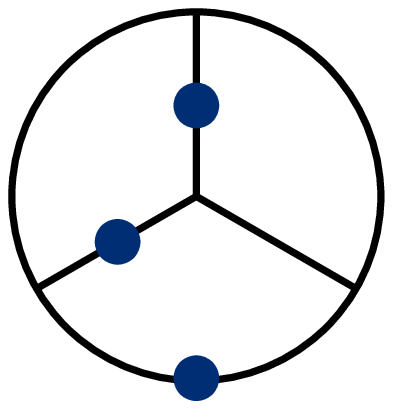} \right) , \nn \\
{\cal M}_4^{(4)}\Bigr|_{\rm leading} & = -\frac{23}{2}  \, \fancyM  \,
\Bigl(\frac{\kappa}{2}\Bigr)^{10}
\, ( s^2 + t^2 + u^2)^2 \, 
\,\left({\ts \frac{1}{4}} \inlinefig{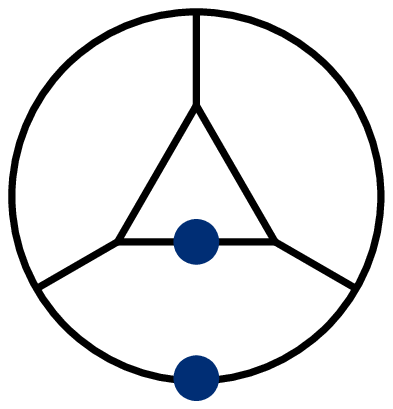} 
  + {\ts \frac{1}{2}} \inlinefig{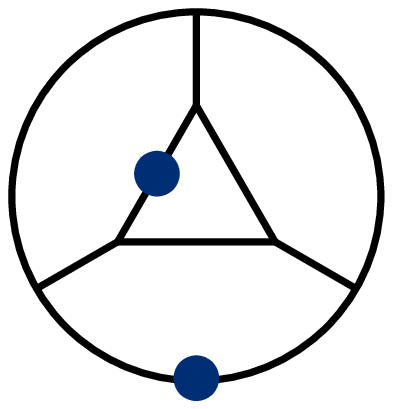}  
  + {\ts \frac{1}{4}} \inlinefig{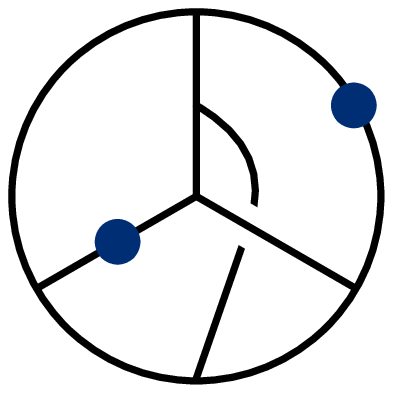}  \right), \nn\\
{\cal M}_4^{(5)} \Bigr|_{\rm leading} & =
- \frac{16\times 629}{25} \,  \fancyM  \left(\frac{\kappa}{2}\right)^{12}
(s^2 + t^2 + u^2)^2  \,
\left({\ts \frac{1}{48}} \inlinefig{Vacuum5loopsV1.eps} 
    + {\ts \frac{1}{16}} \inlinefig{Vacuum5loopsV2.eps}  \right) ,
\label{SGVacuum}
\end{align}
where the universal factor is $\fancyM \equiv s t u \, M_4^\tree(1,2,3,4)$.
For each loop order, the critical dimension is different and is summarized in \tab{CriticalDimTable}.

We also collect all known vacuum graph expressions of the leading ultraviolet behavior
  in the maximally supersymmetric $SU(N_c)$ Yang-Mills theory~\cite{GSB,
    BDDPR, CompleteFourLoopSYM, ck4l, SixLoopPlanarYM, FiveLoopN4},
\begin{align}
{\cal A}_4^{(1)} \Bigr|_{\rm leading}
& \hskip -.2 cm =  g^4 \fancyA \left(N_c (\f^{a_1 a_2 b} \f^{b a_3 a_4} 
  + \f^{a_2 a_3 b} \f^{b a_4 a_1}) - 3 B^{a_1 a_2 a_3 a_4} \right) 
  \inlinefig{Vacuum1loopsV1.eps} \,, \nn \\
{\cal A}_4^{(2)} \Bigr|_{\rm leading}
& \hskip -.2 cm = - \, g^6 \, \fancyA
\left[  F^{a_1 a_2 a_3 a_4} \left( N_c^2 \, \inlinefig{Vacuum2loopsV1.eps}
    + 48 \left( {\ts \frac{1}{4}} \inlinefig{Vacuum2loopsV1.eps} 
              + {\ts \frac{1}{4}} \inlinefig{Vacuum2loopsV2.eps}
             \right)\! \right)  \right.
\nn \\ & \hskip 3.2 cm \null
 \left. \null + 48 \, N_c \, G^{a_1 a_2 a_3 a_4} \left({\ts \frac{1}{4}} 
         \inlinefig{Vacuum2loopsV1.eps} 
       + {\ts \frac{1}{4}} \inlinefig{Vacuum2loopsV2.eps}
             \right) \right] , \nonumber \\
{\cal A}_4^{(3)} \Bigr|_{\rm leading} 
& \hskip -.2 cm =  2 \, g^8 \,\fancyA \, N_c F^{a_1 a_2 a_3 a_4} 
  \left( N_c^2 \, \inlinefig{Vacuum3loopsV1.eps}
    + 72 \, \left({\ts \frac{1}{6}} 
    \inlinefig{Vacuum3loopsV1.eps}
    + {\ts \frac{1}{2}} \,
       \inlinefig{Vacuum3loopsV2.eps}
     \right) \!\right) , 
\label{YMVacuum} \\
{\cal A}_4^{(4)} \Bigr|_{\rm leading} 
&\hskip -.2 cm = -  6 \, g^{10} \, \fancyA \, N_c^2  F^{a_1 a_2 a_3 a_4} \left( N_c^2 \, 
         \inlinefig{Vacuum4loopsV1.eps}
   + 48 \left({\ts \frac{1}{4}} 
            \inlinefig{Vacuum4loopsV1.eps} 
       + {\ts \frac{1}{2}} \inlinefig{Vacuum4loopsV2.eps} 
       + {\ts \frac{1}{4}} \inlinefig{Vacuum4loopsV3.eps}
          \right)\! \right) , \nonumber \\
{\cal A}_4^{(5)}\Bigr|_{\rm leading} & \hskip -.2 cm =
 \!  \frac{144}{5}  g^{12} \fancyA
N_c^3 \, F^{a_1 a_2 a_3 a_4}  \left(\hskip -.1 cm  N_c^2 \inlinefig{Vacuum5loopsYMV1.eps}
+ 48 \left(\!\!{\ts\frac{1}{4}} \inlinefig{Vacuum5loopsYMV1.eps} 
         \! + {\ts\frac{1}{2}} \inlinefig{Vacuum5loopsYMV2.eps} 
        \! + {\ts\frac{1}{4}} \inlinefig{Vacuum5loopsYMV3.eps} 
\right) \! \right)  , \nonumber \\
A^{(6)}_4\Big|_{{\, \rm leading}} & \hskip -.2 cm =
- 120 g^{14} \fancyA  F^{a_1 a_2 a_3 a_4} N_c^6
       \left( {\ts \frac{1}{2}}  
          \inlinefig{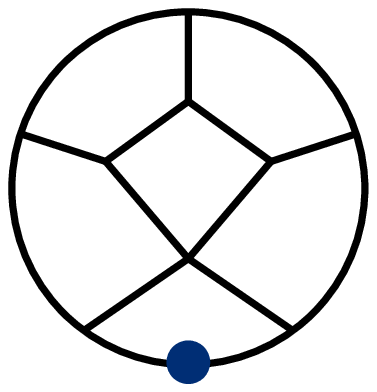} 
      + {\ts \frac{1}{4}} (\ell_1 + \ell_2)^2 
          \inlinefig{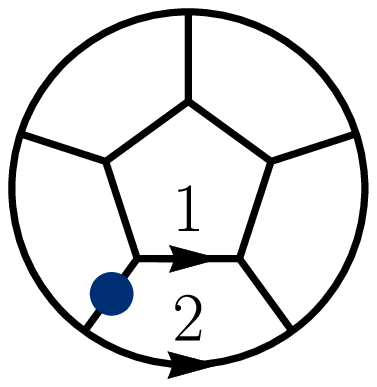}
       - {\ts \frac{1}{20}}
          \inlinefig{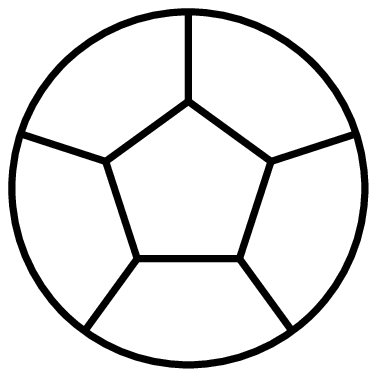} \right) \nn \\ & \hskip 5cm    
          + {\cal O}(N_c^4)  \, ,
\nonumber
\end{align}
where the universal factor is $ \fancyA \equiv s t  \, A_4^\tree(1,2,3,4)$,  and 
\begin{align}
F^{a_1 a_2 a_3 a_4} & \equiv  t\, \f^{a_1 a_2 b} \f^{b a_3 a_4} 
          + s \, \f^{a_2 a_3 b} \f^{b a_4 a_1} \,,\nn \\
G^{a_1 a_2 a_3 a_4} & \equiv 
         s \, \delta^{a_1 a_2} \delta^{a_3 a_4}
       + t \, \delta^{a_4 a_1} \delta^{a_2 a_3}
       + u \, \delta^{a_1 a_3} \delta^{a_2 a_4} \,,\nn \\
B^{a_1 a_2 a_3 a_4} &\equiv \f^{a_1 b_1 b_2} \f^{a_2 b_2 b_3} \f^{a_3 b_3 b_4} \f^{a_4 b_4 b_1} \, .
\end{align}
As before, $\f^{a b c}$ are the group structure constants, with normalization given below \eqn{CubicRepresentation}.  
As in the gravity case, the critical dimension at each loop order is different, and is included in \tab{CriticalDimTable}.

Inspecting Eqs.~\eqref{SGVacuum}~and~\eqref{YMVacuum} we already note a remarkable 
property in both  the supergravity and subleading color gauge-theory expressions: the relative coefficients between vacuum
integrals in these representations, ignoring signs, are given by the symmetry factors of the corresponding vacuum graphs. 
For example, at five loops in \eqn{SGVacuum}, the first vacuum graph has 48 automorphisms
and the second has 16 automorphisms, matching the relative factors. 
While the amplitude has such coefficients for each integral (see e.g.\ Eq.~\eqref{GRContactFiveLop}), their appearance in 
the leading ultraviolet divergence is unexpected due to both the nontrivial manipulations and the choices of master 
integrals that are required to arrive at the final result. 

Further inspection of Eqs.~\eqref{SGVacuum}~and~\eqref{YMVacuum}  reveals further interesting structures, 
showing that the relative coefficients of vacuum integrals are consistently related between the different loop orders.

\subsection{Observed ultraviolet consistency}

An $L$-loop (vacuum) integral has many $L'<L$ subintegrals. A way to isolate one and expose its associated ultraviolet properties 
is to take its loop momenta to be much larger than the other $(L-L')$ ones. 
We define an $L'$-loop subdiagram of an $L$-loop diagram as the sum over all of its $L'$-loop subintegrals. 
Since each subintegral may have a different critical dimension, the critical dimension of an $L'$-loop subdiagram is the minimum 
of the critical dimensions of all the $L'$-loop subintegrals. 

With this definition, to compare the higher- and lower-loop leading ultraviolet properties of four-point amplitudes we carry out the
following steps:
 \begin{enumerate} 
\item For each $L$-loop vacuum diagram construct its $L'$-loop subdiagram.

\item Keep only those contributions with leading ultraviolet behavior, \ie those that are divergent in the lowest critical dimension

\item Apply IBP identities, as needed, to map the lower-loop vacuum integrals into the same vacuum integral basis as the one used 
        in the  ultraviolet expansion  of the lower-loop amplitude.
\end{enumerate}
As we now show by example, every result in \eqns{SGVacuum}{YMVacuum} supports the observation that the leading ultraviolet behavior at $L$ and $L'$ loops
in their respective critical dimensions are consistent.

To see the power of this observation, consider the all-order constraints from one-loop subdiagrams.  
From \eqns{SGVacuum}{YMVacuum}, we see that the one-loop leading ultraviolet divergence is given by a vacuum integral 
with four propagators.  For the higher-loop vacuums this amounts to the statement that there exists an integral basis such that 
all one-loop subloops of any higher  loop vacuum must contain at least four propagators.\footnote{In an arbitrary integral basis 
this property is not manifest and emerges only after the summation over all one-loop subintegrals of all diagrams and reduction to 
a one-loop integral basis. }
This is equivalent to the no-triangle property of one-loop amplitudes in
both $\NeqEight$ and $\NeqFour$ super-Yang--Mills amplitudes~\cite{NoTriangle}, except that here it applies to the 
reduction to an integral basis of the vacuum integrals describing the leading ultraviolet behavior.  
One-loop subgraphs with more than four propagators give a subleading behavior which we discard according to our
procedure which focus on the leading ultraviolet properties.  Because there is only a single
type of leading one-loop subdiagram, this property of one-loop
sub-graphs places no constraint on the relative coefficients of the
higher loop vacuums.  Nevertheless, the constraint that each one-loop subgraph has at least four propagators 
is extremely powerful.  In particular, as discussed in
\sect{AllIntegrationSection}, the only integrals in our basis of
five-loop vacuum integrals without triangle subdiagrams are the two
five-loop integrals contributing to \eqn{SGVacuum}.  A similar
property holds for $\NeqFour$ super-Yang--Mills theory, where the only
five-loop vacuum integral basis elements without any triangle or bubble subintegrals
are the ones appearing in \eqn{YMVacuum}.  
This is quite a remarkable property because, in an appropriately-chosen integral basis that maximizes the
 number of one-loop triangle and bubble sub-integrals, it severely limits the  vacuum integrals that can appear in 
the final expressions.

While the one-loop properties discussed above should hold for each one-loop subintegral at any loop order, 
understanding the consequences of higher-loop ultraviolet divergences in \eqref{SGVacuum}~and~\eqref{YMVacuum} 
can be best appreciated via a case by case analysis.
We choose three illustrative examples.  We begin by showing the consistency of subleading-color $\NeqFour$
super-Yang--Mills between five and four loops.  We focus on the subleading-color part, because it has a more complex
structure than the leading-color part and it is similar to the supergravity case.  
We then examine the consistency of the four-loop ultraviolet divergences with those at lower loops,
which are the same for the $\NeqFour$ super-Yang--Mills theory at
subleading color and the $\NeqEight$ supergravity. Last, we discuss
the five-to-four loop consistency of our results for the five-loop $\NeqEight$ supergravity.

As mentioned earlier, not all terms in the sum that defines a lower-loop subdiagram have the same critical dimension.
For example, when relating $L$ and $(L-1)$-loop diagrams, excluding a dotted propagator leads to a term with a lower critical 
dimension than one obtained by excluding an undotted one. 
Thus, when focusing on the ultraviolet critical dimension of lower-loop diagrams it suffices to keep only terms obtained by
disconnecting the propagators with the largest number of dots.
Once the subdiagrams are identified, we can compare  them to the lower-loop result by treating the subdiagram 
as a new vacuum diagram where we have kept the leading order in small-momentum expansion for the excluded leg.
This results in lower-loop vacuum diagrams with dots on the propagators where the excluded leg is connected to the
subgraph.

For the $\NeqFour$ super-Yang--Mills five-loop vacuum diagrams, the leading four-loop subdiagrams are all those that
exclude the leg that carries the dot.  Diagrammatically, we write
\begin{equation}
\inlineblob{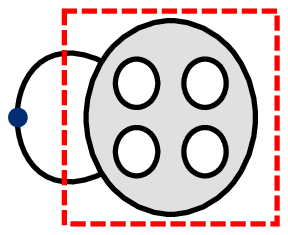} = \frac{1}{4} \inlineblob{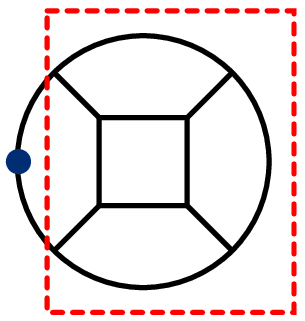} + \frac{1}{2} \inlineblob{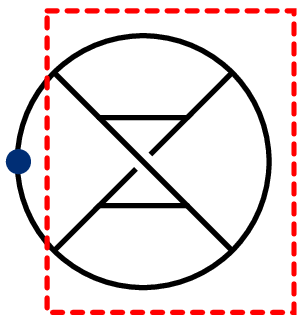} + \frac{1}{4}\inlineblob{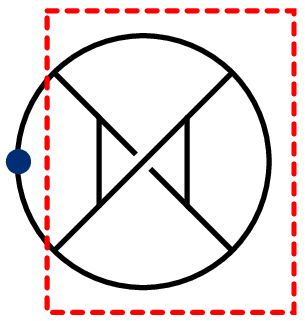} \, .
\label{ym5To4Sub}
\end{equation}
Excluding the propagator outside the dashed box and taking its momentum small compared to the remaining ones 
leads to
\begin{equation}
\inlineblob{5to4loopsubdivdot} \rightarrow \frac{1}{4} \inlineg{4}{1} + \frac{1}{2} \inlineg{4}{2} + \frac{1}{4} \inlineg{4}{3}  \, .
\end{equation}
This exactly matches the subleading-color four-loop vacuum diagrams describing their
relative coefficients in \eqn{YMVacuum}.

Showing the consistency of the four loop expression with lower loops follows similar steps. Now there are 
two dotted legs that can be excluded.  Summing over the two expansions of each subdiagram, we find
\begin{align}
\inlineg{4}{1} &\rightarrow 2 \inlineg{3}{1} \,, \hskip 1 cm 
  \inlineg{4}{2}  \rightarrow 2 \inlineg{3}{2}\,, \hskip 1 cm
  \inlineg{4}{3}  \rightarrow 2 \inlineg{3}{2} \,.
\end{align}
Using this we see that the subdiagrams match the relative factors and three-loop vacuum diagrams in \eqn{YMVacuum},
\begin{align}
\inlineblob{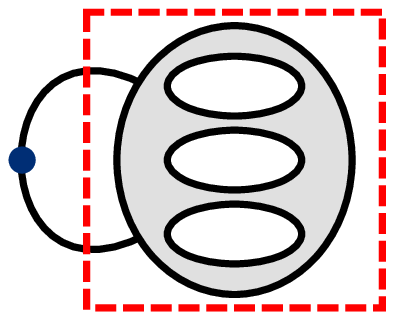} \rightarrow  3 \left( \frac{1}{6} \inlineg{3}{1} + \frac{1}{2} \inlineg{3}{2} \right) \,.
\label{FourToThree}
\end{align}  

Additionally, we can extract the two-loop subdiagrams in the four-loop
divergence by expanding around both dotted propagators.  This gives,
\begin{align}
  \inlineg{4}{1} &\rightarrow \inlineg{2}{1}\,,  \hskip 1 cm 
  \inlineg{4}{2}  \rightarrow \inlineg{2}{2}\,, \hskip 1 cm 
  \inlineg{4}{3}  \rightarrow \inlineg{2}{1}  \,.
\end{align}
Using this we find that with the relative coefficients
from the four-loop expression, these subdiagrams are also consistent
with the leading lower-loop behavior
\begin{equation}
\inlineblob{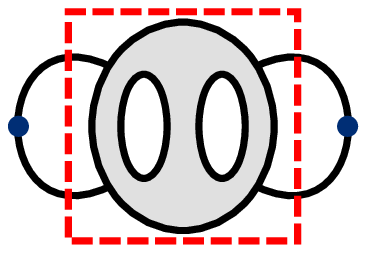} \rightarrow \frac{1}{4} \inlineg{2}{1} +  \frac{1}{4} \inlineg{2}{2} \,.
\end{equation}
It is straightforward to confirm that the same relative coefficients arise by starting
from the three-loop expression in \eqn{FourToThree} and extracting the leading two-loop subdiagrams.

Since master integrals giving the ultraviolet divergence of the
five-loop supergravity amplitude in $D=24/5$ do not have doubled
propagators, all ways of excluding one propagator lead to
integrals of the same critical dimension and must therefore be kept.
The planar diagram is a cube, so all of its edges are equivalent. Summing over all the four-loop subintegrals leads to
\begin{equation}
  \inlineg{5}{1} \rightarrow 12 \inlineg{4}{1} \,.
\end{equation}
The nonplanar diagram has two inequivalent types of legs to exclude.
There are eight legs that, when expanded around, lead to a planar
four-loop subdiagram.  The other four legs lead to a nonplanar
subdiagram.  Thus, after isomorphisms,  the subintegrals of the nonplanar five-loop diagram contribute
\begin{equation}
  \inlineg{5}{2} \rightarrow 8 \inlineg{4}{2} + 4 \inlineg{4}{3} \,.
\end{equation}
After accounting for the relative symmetry factors of $1/48$ and $1/16$ between the two
five-loop diagrams in \eqn{SGVacuum}, we get
\begin{equation}
\inlineblob{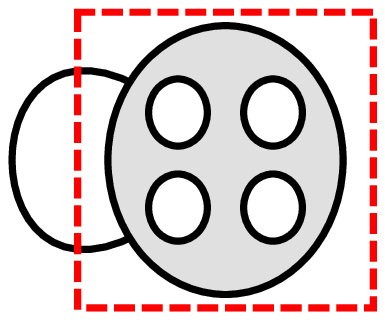} \rightarrow   \frac{1}{4} \inlineg{4}{1} + \frac{1}{2} \inlineg{4}{2} + \frac{1}{4} \inlineg{4}{3}  \,,
  \end{equation}
matching the relative factors between the four-loop vacuum diagrams also given in \eqn{SGVacuum}.

Through four loops super-Yang--Mills subleading-color and supergravity divergences follow the same pattern, being related 
between different loop orders by removing  a dotted propagator. While in both theories the consistency relations hold at five loops 
as well,  they now involve removing a dotted and an undotted propagator, respectively.  The additional propagator in 
the gauge-theory expression raises its critical dimension to $D=26/5$. 
It is remarkable that, even though the various integrals and symmetry factors at five loops differ in the two theories, 
consistency requires that the relative coefficients for four-loop subdiagrams are the same.

Let us elaborate briefly on the structure of the planar $\NeqFour$ super-Yang--Mills vacuum integrals at six loops. Unlike the 
previous examples, the lower-loop integrals given by our construction are not among the five-loop master integrals
in \fig{16masterYMbasisFigure} and a comparison with the five-loop expression \eqref{YMVacuum} requires use of IBP identities.
As in the five-to-four loop relation, the integrals with lowest critical dimension  arise from subdiagrams that exclude the doubled 
propagator in the six-loop vacuum diagrams.  Thus, the leading five-loop subdiagram result is
\begin{equation}
  \inlineblob{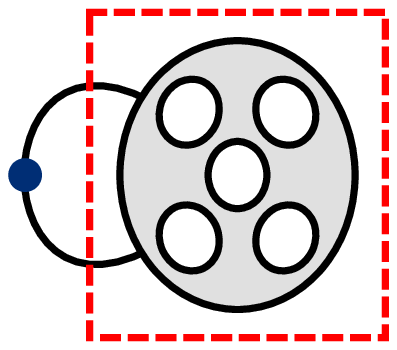} \rightarrow  \frac{1}{2} \inlineblob{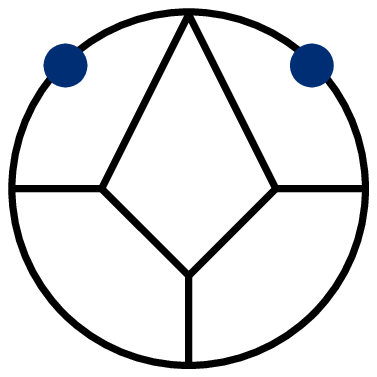}
  + \frac{1}{4} \, (\ell_1 + \ell_2)^2   \inlineblob{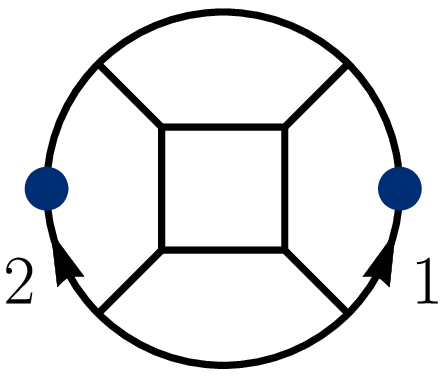}  + \text{subleading color}\,.
\label{FiveLoopSubdiagram}
\end{equation}
Using an integration-by-parts relation (see Eq.~(4) of Ref.~\cite{FiveLoopN4})
\begin{equation}
\frac{1}{2} \inlineblob{Vacuum5loopYMSubDiagrams1}
+ \frac{1}{4} (\ell_1 + \ell_2)^2 \inlineblob{Vacuum5loopYMSubDiagrams2.eps}
= \frac{6}{5} \inlineblob{Vacuum5loopsYMV1.eps} \,,
\end{equation}
to map \eqref{FiveLoopSubdiagram} to the five-loop integral basis,  we find that it is proportional to the five-loop leading 
color term in \eqn{YMVacuum}.  It is gratifying that the subdiagram consistency holds even if not initially obvious.

\subsection{Applications}

The consistency observations discussed above give us additional confidence that we have correctly 
computed the leading ultraviolet behavior of $\NeqEight$ supergravity at five loops by showing that 
in the sense discussed above, it fits the pattern of ultraviolet properties at all lower loops. 
The simple structures at the vacuum diagram level  uncovered here  also offers the exciting possibility of probing 
seemingly out of reach ultraviolet properties at even higher loops. Apart from the possibility of imposing them
on an ansatz for the leading ultraviolet terms of gauge and gravity amplitudes, we can use them to 
simplify the IBP system by focusing only on the vacuum integrals that are expected to appear. 
For example, in \sect{TopLevelIntegrationSection} we vastly simplified the five-loop $\NeqEight$ IBP system 
by assuming that only the vacuum integrals with maximal cuts survive in the final result. 
As emphasized above, this condition follows from demanding consistency of the five-loop vacuum master diagrams
with one-loop subdiagrams, which rules out one-loop triangle subgraphs and all but two five-loop master vacuum 
diagrams in the basis of \fig{8masterbasisFigure}. 
More importantly this condition eliminates nearly all integrals from the IBP system as well as a substantial part of 
the expansion of the integrand.  The same strategy should  continue to be fruitful at even higher loop orders.  
Alternatively, it may be possible to completely bypass the construction of the integrand,
its ultraviolet expansion and integration, and instead extrapolate the final result in terms of vacuum diagrams 
to higher loop orders.  We leave this task for future study.

We emphasize that the observed ultraviolet consistency is a property of the leading behavior after
simplifying the integrals via Lorentz invariance and integration-by-parts relations. 
It relies on nontrivial simplifications that occur in the integral
reduction and is manifest because we judiciously chose the vacuum integral bases.  
A key property of our IBP systems is that the space-time dimension enters only implicitly
through the critical dimension where the integrals are logarithmically divergent. 
Had there been explicit dependence on the dimension, one would naturally expect a nontrivial dependence on 
dimension in the relative coefficients of master integrals and thus, given the differing critical dimensions at different 
loop orders, it would disrupt any systematic cross-loop-order relations.
Simplifications based on Lorentz invariance in \eqns{TwoTensor}{FourTensor} were used, and introduce explicit dependence on 
dimension.  It is rather striking that this dependence drops out once the IBP
relations are used and consequently it does not complicate relations between vacuum diagrams and their subdiagrams.  
These properties are worth investigating.

%%%%%%%%%%%%%%%%%%%%%%%%%%%%%%%
\section{Conclusions and outlook}
\label{ConclusionSection}

In this paper we determined the ultraviolet behavior of the five-loop
four-point amplitude of $\NeqEight$ supergravity, finding the
critical dimension where it first diverges to be $D_c=24/5$.  In analyzing 
the results we made the rather striking observation that the
vacuum diagrams that describe the leading ultraviolet behavior
satisfy certain nontrivial relations to the analogous lower-loop
vacuum diagrams.  

Previous work found examples of enhanced ultraviolet cancellations
that render ultraviolet finite~\cite{N4GravThreeLoops, N5GravFourLoops}
certain amplitudes in $\NeqFour$ and $\NeqFive$ supergravity in
$D=4$, despite the possibility of counterterms allowed by all known
symmetry considerations~\cite{VanishingVolume,NoExplanation}.
Related arguments suggest that $\NeqEight$ supergravity should diverge
at five loops in $D=24/5$~\cite{BjornssonAndGreen}.
While one might have suspected that there could be corresponding
enhanced cancellations in $\NeqEight$ supergravity at five loops, our
results conclusively demonstrate that, at this loop order, there are
no further cancellations of ultraviolet divergences beyond those
identified by symmetry
arguments.  

The divergence we find in $D=24/5$ at five loops corresponds to a $D^8 R^4$ counterterm.
This counterterm is especially interesting because it corresponds to a potential
$D=4$ divergence believed to be consistent with
the $E_{7(7)}$ duality symmetry of maximal supergravity.
It is, however, not clear that our result in $D=24/5$ points towards a
seven-loop divergence in $D=4$, because the existence of counterterms does
not transfer trivially between dimensions and loop orders.
For example, one might be tempted to argue for a three-loop divergence
in $\NeqFour$ or $\NeqFive$ supergravity in $D=4$ based on the
existence~\cite{HalfMaxD5} of a nonvanishing one-loop $R^4$
counterterm in $D=8$ in both theories; we know however that both
theories are finite at three
loops~\cite{N4GravThreeLoops,N5GravFourLoops}.
Another result that indicates that further investigation of the
ultraviolet structure of supergravities in four dimensions is
warranted is the suspected link between anomalies and divergences in
supergravity theories on the one hand, and the anticipated lack of
anomalies in theories with ${\cal N} \ge 5$ supersymmetry on the
other~\cite{Anomaly, N4GravFourLoop}.  Of course, not every divergence necessarily has
an anomaly behind it. Nevertheless, it is surprising that ${\cal N}=5$
supergravity at four loops in $D=4$ appear to have additional
cancellations beyond those predicted by symmetry
considerations~\cite{N5GravFourLoops}, while $\NeqEight$ supergravity
at five loops in $D = 24/5$ does not.

The ultraviolet properties of the amplitude were extracted, following
standard methods~\cite{Vladimirov}, by expanding the integrand at
large loop momenta or equivalently small external momenta, to identify the
logarithmic divergences in various dimensions. The result was
then reduced to a combination of master integrals; to this end we made use of
modern ideas of organizing the system of IBP identities in terms of an
SL($L$) symmetry~\cite{IntegralRelations} (where $L$ is the number of
loops) and restricting to integrals with leading ultraviolet behavior.
In addition to integrating the complete expansion of a new integrand in
both $D=22/5$ and $D=24/5$, we also integrated the expansion of the
previously-obtained integrand \cite{FiveLoopN8Integrand} in these
dimensions, under the assumption that the only master integrals that appear
in the final result
have maximal cuts. These results, obtained by using unitarity-compatible
integration-by-parts techniques~\cite{Gluza2010ws, IBPAdvances}, agree
with those of the full integration of the simpler integrand, thus providing
a highly nontrivial check of our calculations.

The agreement of the two approaches highlights an important trend: the
only integrals that contribute to the divergence of the four-point
$1\le L\le 5$ amplitudes in their critical dimensions are those with
maximal cuts at the vacuum level.  At higher loops we
expect a systematic application of similar considerations to lead to a
drastic reduction in the computational complexity.  An approach based
on exploiting these observations may make it possible to directly
determine the critical dimension of the six- and seven-loop
$\NeqEight$ supergravity amplitudes.

An even greater efficiency gain may lie in the observed ultraviolet
consistency relations described in \sect{PatternsSection}.  That is,
$L'$-loop subdiagrams of the leading ultraviolet divergence in the
$L$-loop critical dimension reproduce, upon reduction to master
integrals, the combination of vacuum diagrams describing the leading
ultraviolet behavior in the $L'$-loop critical dimension.  Moreover,
in an appropriate basis, the relative coefficients of the vacuum
master integrals are given by the order of the
automorphism groups of the diagrams.
We also observed similar patterns in the vacuum diagrams of $\NeqFour$
super-Yang--Mills theory through six loops, suggesting that they will
continue to hold to higher loop orders in both theories.
While these observations are likely connected to standard consistency
relations between multi-loop amplitudes and their subamplitudes, in our case they
remain a conjecture due to the nontrivial steps needed to relate an
amplitude to a basis of master vacuum graphs in the critical dimension.
These vacuum diagram patterns should be very helpful to identify those terms in 
higher-loop amplitudes that are important for determining the leading ultraviolet
behavior, and for enormously simplifying the integration-by-parts system.
By enforcing the patterns described here, it may even be possible 
to obtain detailed
higher-loop information including a determination of the critical 
dimensions, bypassing the construction of complete loop integrands.

In summary, the success of the newly-developed generalized double-copy
construction~\cite{GeneralizedDoubleCopy, FiveLoopN8Integrand}, and
integration tools~\cite{Gluza2010ws, CutIntegrals, IBPAdvances,
  FiniteFields, ManteuffelSchabingerFiniteFields,
  LinBox,IntegralRelations} used in our five-loop calculations, as
well as our observed vacuum subdiagram consistency constraints,
indicates that problems as challenging as seven-loop $\NeqEight$
supergravity in four dimensions may now be within reach of direct
investigations.

%%%%%%%%%%%%%%%%%%%%%%%%%

\begin{acknowledgments}
We thank Lance Dixon, Michael Enciso, Enrico Herrmann, Harald Ita, David Kosower,
Chia-Hsien Shen, Jaroslav Trnka, Arkady Tseytlin and Yang Zhang
for many useful and interesting
discussions.  This work is supported by the Department of Energy under
Award Numbers DE-SC0009937 and DE-SC0013699.  We acknowledge the
hospitality of KITP at UC Santa Barbara in the program ``Scattering
Amplitudes and Beyond'', during early stages of this work.  While at
KITP this work was also supported by US NSF under Grant
No.~PHY11-25915.  J.\,J.\,M.\,C. is supported by the European Research
Council under ERC-STG-639729, {\it preQFT: Strategic Predictions for
  Quantum Field Theories}.  The research of H.\,J. is supported in
part by the Swedish Research Council under grant 621-2014-5722, the
Knut and Alice Wallenberg Foundation under grant KAW 2013.0235, and
the Ragnar S\"{o}derberg Foundation under grant S1/16. Z.\,B., J.\,J.\,M.\,C.,
A.\,E. and J.\,P.-M. thank the Institute for Gravitation and the Cosmos
for hospitality while this work was being finished. A.\,E. and J.\,P.-M. also 
thank the Mani L. Bhaumik Institute for generous support.
\end{acknowledgments}

%%%%%%%%%%%%%%%%%%%%%%%%%
%\begingroup
%\let\clearpage\relax

%\endgroup


\begin{thebibliography}{99}

%+% 1 ref
\bibitem{Supergravity}
 D.~Z.~Freedman, P.~van Nieuwenhuizen and S.~Ferrara,
  %``Progress Toward a Theory of Supergravity,''
  Phys.\ Rev.\ D {\bf 13}, 3214 (1976);\\
%doi:10.1103/PhysRevD.13.3214
  %%CITATION = doi:10.1103/PhysRevD.13.3214;%%
S.~Deser and B.~Zumino,
  %``Consistent Supergravity,''
  Phys.\ Lett.\ B {\bf 62}, 335 (1976)
  [Phys.\ Lett.\  {\bf 62B}, 335 (1976)].
%  doi:10.1016/0370-2693(76)90089-7
  %%CITATION = doi:10.1016/0370-2693(76)90089-7;%%

%+% 1 ref
\bibitem{SurprisingCancellations}
Z.~Bern, L.~J.~Dixon and R.~Roiban,
%``Is N = 8 supergravity ultraviolet finite?,''
Phys.\ Lett.\ B {\bf 644}, 265 (2007)
%doi:10.1016/j.physletb.2006.11.030
[hep-th/0611086];\\
%%CITATION = doi:10.1016/j.physletb.2006.11.030;%%
%
Z.~Bern, J.~J.~Carrasco, D.~Forde, H.~Ita and H.~Johansson,
%``Unexpected Cancellations in Gravity Theories,''
Phys.\ Rev.\ D {\bf 77}, 025010 (2008)
%doi:10.1103/PhysRevD.77.025010
[arXiv:0707.1035 [hep-th]];\\
%%CITATION = doi:10.1103/PhysRevD.77.025010;%%
%
E.~Herrmann and J.~Trnka,
%``Gravity On-shell Diagrams,''
JHEP {\bf 1611}, 136 (2016)
%doi:10.1007/JHEP11(2016)136
[arXiv:1604.03479 [hep-th]].
%%CITATION = doi:10.1007/JHEP11(2016)136;%%

\bibitem{NeqFourSugra}
E.~Cremmer, J.~Scherk and S.~Ferrara,
%``SU(4) Invariant Supergravity Theory,''
Phys.\ Lett.\  {\bf 74B}, 61 (1978);\\
%  doi:10.1016/0370-2693(78)90060-6
%%CITATION = doi:10.1016/0370-2693(78)90060-6;%%
 %
A.~K.~Das,
%``SO(4) Invariant Extended Supergravity,''
Phys.\ Rev.\ D {\bf 15}, 2805 (1977);\\
%  doi:10.1103/PhysRevD.15.2805
%%CITATION = doi:10.1103/PhysRevD.15.2805;%%
%
E.~Cremmer and J.~Scherk,
%``Algebraic Simplifications in Supergravity Theories,''
Nucl.\ Phys.\ B {\bf 127}, 259 (1977).
%  doi:10.1016/0550-3213(77)90214-0
%%CITATION = doi:10.1016/0550-3213(77)90214-0;%%

\bibitem{NeqFiveSugra}
B.~de Wit and H.~Nicolai,
%``Extended Supergravity With Local SO(5) Invariance,''
Nucl.\ Phys.\ B {\bf 188} (1981) 98.
%  doi:10.1016/0550-3213(81)90107-3
%%CITATION = doi:10.1016/0550-3213(81)90107-3;%%


%+% 9 refs
\bibitem{N4GravThreeLoops}
 Z.~Bern, S.~Davies, T.~Dennen and Y.-t.~Huang,
%``Absence of Three-Loop Four-Point Divergences in N=4 Supergravity,''
Phys.\ Rev.\ Lett.\  {\bf 108}, 201301 (2012)
%doi:10.1103/PhysRevLett.108.201301
[arXiv:1202.3423 [hep-th]].
 %%CITATION = doi:10.1103/PhysRevLett.108.201301;%%

%+% 10 refs
\bibitem{N5GravFourLoops}
 Z.~Bern, S.~Davies and T.~Dennen,
%``Enhanced ultraviolet cancellations in $\mathcal N=5$ supergravity at four loops,''
Phys.\ Rev.\ D {\bf 90}, no. 10, 105011 (2014)
%doi:10.1103/PhysRevD.90.105011
[arXiv:1409.3089 [hep-th]].
%%CITATION = doi:10.1103/PhysRevD.90.105011;%%

%+% 2 refs
\bibitem{HalfMaxD5}
 Z.~Bern, S.~Davies, T.~Dennen and Y.-t.~Huang,
%``Ultraviolet Cancellations in Half-Maximal Supergravity 
% as a Consequence of the Double-Copy Structure,''
Phys.\ Rev.\ D {\bf 86}, 105014 (2012)
%doi:10.1103/PhysRevD.86.105014  
[arXiv:1209.2472 [hep-th]].
%%CITATION = doi:10.1103/PhysRevD.86.105014;%%

%+% 8 refs
\bibitem{IntegralRelations}
 Z.~Bern, M.~Enciso, J.~Parra-Martinez and M.~Zeng,
%``Manifesting enhanced cancellations in supergravity: 
% integrands versus integrals,''
JHEP {\bf 1705}, 137 (2017)
%doi:10.1007/JHEP05(2017)137
[arXiv:1703.08927 [hep-th]].
%%CITATION = doi:10.1007/JHEP05(2017)137;%%

%+% 2 refs
\bibitem{NoExplanation}
 G.~Bossard, P.~S.~Howe and K.~S.~Stelle,
%``Anomalies and divergences in N=4 supergravity,''                             
Phys.\ Lett.\ B {\bf 719}, 424 (2013)
[arXiv:1212.0841 [hep-th]];\\
%%CITATION = ARXIV:1212.0841;%%
%
G.~Bossard, P.~S.~Howe and K.~S.~Stelle,
%``Invariants and divergences in half-maximal supergravity theories,''          
JHEP {\bf 1307}, 117 (2013)
[arXiv:1304.7753 [hep-th]];\\
%%CITATION = ARXIV:1304.7753;%%
%
Z.~Bern, S.~Davies and T.~Dennen,
%``The Ultraviolet Structure of Half-Maximal Supergravity with Matter Multiplets at Two and Three Loops,''
Phys.\ Rev.\ D {\bf 88}, 065007 (2013)
%doi:10.1103/PhysRevD.88.065007
[arXiv:1305.4876 [hep-th]].
%%CITATION = doi:10.1103/PhysRevD.88.065007;%%

%+% 2 refs
\bibitem{Anomaly}
 N.~Marcus,
  %``Composite Anomalies in Supergravity,''
  Phys.\ Lett.\  {\bf 157B}, 383 (1985);\\
%doi:10.1016/0370-2693(85)90385-5
  %%CITATION = doi:10.1016/0370-2693(85)90385-5;%%
%
J.~J.~M.~Carrasco, R.~Kallosh, R.~Roiban and A.~A.~Tseytlin,
%``On the U(1) duality anomaly and the S-matrix of N=4 supergravity,'' 
  JHEP {\bf 1307}, 029 (2013)
  %doi:10.1007/JHEP07(2013)029   
  [arXiv:1303.6219 [hep-th]];\\
  %%CITATION = doi:10.1007/JHEP07(2013)029;%%
%
 R.~Kallosh,
  %``Cancellation of Conformal and Chiral Anomalies in $\mathcal{N} \geq 5$ supergravities,''
  Phys.\ Rev.\ D {\bf 95}, no. 4, 041701 (2017)
%doi:10.1103/PhysRevD.95.041701
  [arXiv:1612.08978 [hep-th]];\\
  %%CITATION = doi:10.1103/PhysRevD.95.041701;%%
%
 D.~Z.~Freedman, R.~Kallosh, D.~Murli, A.~Van Proeyen and Y.~Yamada,
  %``Absence of U(1) Anomalous Superamplitudes in $\mathcal{N}\geq 5$ Supergravities,''
  JHEP {\bf 1705}, 067 (2017)
%doi:10.1007/JHEP05(2017)067
  [arXiv:1703.03879 [hep-th]];\\
  %%CITATION = doi:10.1007/JHEP05(2017)067;%%
%
  Z.~Bern, A.~Edison, D.~Kosower and J.~Parra-Martinez,
  %``Curvature-squared multiplets, evanescent effects, and the U(1) anomaly in $N=4$ supergravity,''
  Phys.\ Rev.\ D {\bf 96}, no. 6, 066004 (2017)
  %doi:10.1103/PhysRevD.96.066004
  [arXiv:1706.01486 [hep-th]];\\
  %%CITATION = doi:10.1103/PhysRevD.96.066004;%%
%
Z.~Bern, J.~Parra-Martinez and R.~Roiban,
%``Cancelling the U(1) Anomaly in the S-matrix of N=4 Supergravity,''
arXiv:1712.03928 [hep-th].
%%CITATION = ARXIV:1712.03928;%%

%+% 6 refs
\bibitem{N4GravFourLoop}
 Z.~Bern, S.~Davies, T.~Dennen, A.~V.~Smirnov and V.~A.~Smirnov,
%``Ultraviolet Properties of N=4 Supergravity at Four Loops,''
Phys.\ Rev.\ Lett.\  {\bf 111}, no. 23, 231302 (2013)
%doi:10.1103/PhysRevLett.111.231302
[arXiv:1309.2498 [hep-th]].
%%CITATION = doi:10.1103/PhysRevLett.111.231302;%%

%+% 1 ref
\bibitem{NeqEightSugra}
E.~Cremmer, B.~Julia and J.~Scherk,
  %``Supergravity Theory in Eleven-Dimensions,''
  Phys.\ Lett.\ B {\bf 76}, 409 (1978);\\
%doi:10.1016/0370-2693(78)90894-8
  %%CITATION = doi:10.1016/0370-2693(78)90894-8;%%
%
 E.~Cremmer and B.~Julia,
%``The N=8 Supergravity Theory. 1. The Lagrangian,''
Phys.\ Lett.\  {\bf 80B}, 48 (1978);
%doi:10.1016/0370-2693(78)90303-9;
 %%CITATION = doi:10.1016/0370-2693(78)90303-9;%%
%
% E.~Cremmer and B.~Julia,
  %``The SO(8) Supergravity,''
Nucl.\ Phys.\ B {\bf 159}, 141 (1979).
%doi:10.1016/0550-3213(79)90331-6
%%CITATION = doi:10.1016/0550-3213(79)90331-6;%%

%+% 6 refs
\bibitem{GSB}
 M.~B.~Green, J.~H.~Schwarz and L.~Brink,
%``N=4 Yang--Mills and N=8 Supergravity as Limits of String Theories,''
Nucl.\ Phys.\ B {\bf 198}, 474 (1982).
%doi:10.1016/0550-3213(82)90336-4
%%CITATION = doi:10.1016/0550-3213(82)90336-4;%%

%+% 1 ref
\bibitem{N8Predictions}
P.~S.~Howe and U.~Lindstrom,
%``Higher Order Invariants in Extended Supergravity,''
Nucl.\ Phys.\ B {\bf 181}, 487 (1981);\\
%  doi:10.1016/0550-3213(81)90537-X
%%CITATION = doi:10.1016/0550-3213(81)90537-X;%%
%
R.~E.~Kallosh,
%``Counterterms in extended supergravities,''
Phys.\ Lett.\  {\bf 99B}, 122 (1981);\\
%  doi:10.1016/0370-2693(81)90964-3
 %%CITATION = doi:10.1016/0370-2693(81)90964-3;%%
%  
 M.~T.~Grisaru and W.~Siegel,
%``Supergraphity. 2. Manifestly Covariant Rules and Higher Loop Finiteness,''
Nucl.\ Phys.\ B {\bf 201}, 292 (1982)
Erratum: [Nucl.\ Phys.\ B {\bf 206}, 496 (1982)];\\
%doi:10.1016/0550-3213(82)90433-3, 10.1016/0550-3213(82)90282-6
%%CITATION = doi:10.1016/0550-3213(82)90433-3, 10.1016/0550-3213(82)90282-6;%%
%
N.~Marcus and A.~Sagnotti,
%``The Ultraviolet Behavior of $N=4$ {Yang--Mills} and the Power Counting of Extended Superspace,''
Nucl.\ Phys.\ B {\bf 256}, 77 (1985);\\
%doi:10.1016/0550-3213(85)90386-4
%%CITATION = doi:10.1016/0550-3213(85)90386-4;%%
%
G.~Chalmers,
%``On the finiteness of N = 8 quantum supergravity,''                                                                         
hep-th/0008162; \\
%%CITATION = HEP-TH 0008162;%%
%
N.~Berkovits,
%``New higher-derivative R**4 theorems,''                                                                                     
Phys.\ Rev.\ Lett.\  {\bf 98}, 211601 (2007)
[hep-th/0609006];\\
%%CITATION = PRLTA,98,211601;%%
%
M.~B. Green, J.~G. Russo and P. Vanhove,
%``Non-renormalisation conditions in type II string theory and maximal                                                        
%supergravity,''                                                                                                              
JHEP {\bf 0702}, 099 (2007)
[hep-th/0610299];\\
%%CITATION = JHEPA,0702,099;%%
%
M.~B.~Green, J.~G.~Russo and P.~Vanhove,
%``Ultraviolet properties of maximal supergravity,''                                                                          
Phys.\ Rev.\ Lett.\  {\bf 98}, 131602 (2007)
[hep-th/0611273];\\
%%CITATION = PRLTA,98,131602;%%
%
G.~Bossard, P.~S.~Howe and K.~S.~Stelle,
%``The ultra-violet question in maximally supersymmetric field theories,''                                                    
Gen.\ Rel.\ Grav.\  {\bf 41}, 919 (2009)
[0901.4661 [hep-th]];\\
%%CITATION = GRGVA,41,919;%%
%
R.~Kallosh,
%``N=8 Supergravity on the Light Cone,''                                                                                      
0903.4630 [hep-th];\\
%%CITATION = ARXIV:0903.4630;%%
%
N.~Berkovits, M.~B.~Green, J.~G.~Russo and P.~Vanhove,
%``Non-renormalization conditions for four-gluon scattering in supersymmetric string and field theory,''
JHEP {\bf 0911}, 063 (2009)
%doi:10.1088/1126-6708/2009/11/063
[arXiv:0908.1923 [hep-th]].
%%CITATION = doi:10.1088/1126-6708/2009/11/063;%%




%+% 3 refs
\bibitem{SevenLoopGravity}
 M.~B.~Green, J.~G.~Russo and P.~Vanhove,
%``String-theory dualities and supergravity divergences,''                      
JHEP {\bf 1006}, 075 (2010) [arXiv:1002.3805 [hep-th]];\\
%
G.~Bossard, P.~S.~Howe and K.~S.~Stelle,
%``On duality symmetries of supergravity invariants,''                          
JHEP {\bf 1101}, 020 (2011) [arXiv:1009.0743 [hep-th]];\\
%%CITATION = ARXIV:1009.0743;%%
%
N.~Beisert, H.~Elvang, D.~Z.~Freedman, M.~Kiermaier, A.~Morales
and S.~Stieberger,
%``E7(7) constraints on counterterms in N=8 supergravity,''                     
Phys.\ Lett.\ B {\bf 694}, 265 (2010)
[arXiv:1009.1643 [hep-th]];\\
%%CITATION = ARXIV:1009.1643;%%
%
P.~Vanhove,
%``The Critical ultraviolet behaviour of N=8 supergravity amplitudes,''
arXiv:1004.1392 [hep-th].
%%CITATION = ARXIV:1004.1392;%%

%+% 4 refs
\bibitem{BjornssonAndGreen}
J.~Bj\"{o}rnsson and M.~B.~Green,
%``5 loops in 24/5 dimensions,''                                                
JHEP {\bf 1008}, 132 (2010) [arXiv:1004.2692 [hep-th]];\\
%%CITATION = JHEPA,1008,132;%%
%
J.~Bj\"ornsson,
%``Multi-loop amplitudes in maximally supersymmetric pure spinor                
% field theory,''                                                               
JHEP {\bf 1101}, 002 (2011)
[arXiv:1009.5906 [hep-th]].
%%CITATION = ARXIV:1009.5906;%%

%+% 4 refs
\bibitem{VanishingVolume}
G.~Bossard, P.~S.~Howe, K.~S.~Stelle and P.~Vanhove,
%``The vanishing volume of D=4 superspace,''                                    
Class.\ Quant.\ Grav.\  {\bf 28}, 215005 (2011)
[arXiv:1105.6087 [hep-th]].
%%CITATION = ARXIV:1105.6087;%%

%+% 6 refs
\bibitem{BDDPR}
 Z.~Bern, L.~J.~Dixon, D.~C.~Dunbar, M.~Perelstein and J.~S.~Rozowsky,
  %``On the relationship between Yang--Mills theory and gravity and its implication for ultraviolet divergences,''
  Nucl.\ Phys.\ B {\bf 530}, 401 (1998)
%doi:10.1016/S0550-3213(98)00420-9
  [hep-th/9802162].
  %%CITATION = doi:10.1016/S0550-3213(98)00420-9;%%

%+% 7 refs
\bibitem{ThreeFourloopN8}
 Z.~Bern, J.~J.~Carrasco, L.~J.~Dixon, H.~Johansson, D.~A.~Kosower and R.~Roiban,
%``Three-Loop Superfiniteness of N=8 Supergravity,''
Phys.\ Rev.\ Lett.\  {\bf 98}, 161303 (2007)
%doi:10.1103/PhysRevLett.98.161303
[hep-th/0702112];\\
%%CITATION = doi:10.1103/PhysRevLett.98.161303;%%
%
Z.~Bern, J.~J.~M.~Carrasco, L.~J.~Dixon, H.~Johansson and R.~Roiban,
%``Manifest Ultraviolet Behavior for the Three-Loop Four-Point Amplitude of N=8 Supergravity,''
Phys.\ Rev.\ D {\bf 78}, 105019 (2008)
%doi:10.1103/PhysRevD.78.105019
[arXiv:0808.4112 [hep-th]];\\
%%CITATION = doi:10.1103/PhysRevD.78.105019;%%
%
Z.~Bern, J.~J.~Carrasco, L.~J.~Dixon, H.~Johansson and R.~Roiban,
%``The Ultraviolet Behavior of N=8 Supergravity at Four Loops,''
Phys.\ Rev.\ Lett.\  {\bf 103}, 081301 (2009)
%doi:10.1103/PhysRevLett.103.081301
[arXiv:0905.2326 [hep-th]].
%%CITATION = doi:10.1103/PhysRevLett.103.081301;%%

%+% 11 refs
\bibitem{ck4l}
 Z.~Bern, J.~J.~M.~Carrasco, L.~J.~Dixon, H.~Johansson and R.~Roiban,
%``Simplifying Multiloop Integrands and Ultraviolet Divergences of Gauge Theory and Gravity Amplitudes,''
Phys.\ Rev.\ D {\bf 85}, 105014 (2012)
%doi:10.1103/PhysRevD.85.105014
[arXiv:1201.5366 [hep-th]].
%%CITATION = doi:10.1103/PhysRevD.85.105014;%%

%+% 4 refs
\bibitem{CompleteFourLoopSYM}
 Z.~Bern, J.~J.~M.~Carrasco, L.~J.~Dixon, H.~Johansson and R.~Roiban,
  %``The Complete Four-Loop Four-Point Amplitude in N=4 Super-Yang--Mills Theory,''
  Phys.\ Rev.\ D {\bf 82}, 125040 (2010)
  %doi:10.1103/PhysRevD.82.125040
  [arXiv:1008.3327 [hep-th]].
  %%CITATION = doi:10.1103/PhysRevD.82.125040;%%


\bibitem{N4YM}
F.~Gliozzi, J.~Scherk and D.~I.~Olive, %``Supersymmetry, Supergravity Theories and the Dual Spinor Model,''
  Nucl.\ Phys.\ B {\bf 122}, 253 (1977);\\
%  doi:10.1016/0550-3213(77)90206-1
  %%CITATION = doi:10.1016/0550-3213(77)90206-1;%%
%
L.~Brink, J.~H.~Schwarz and J.~Scherk, 
%``Supersymmetric Yang-Mills Theories,''
 Nucl.\ Phys.\ B {\bf 121}, 77 (1977).
%  doi:10.1016/0550-3213(77)90328-5
 %%CITATION = doi:10.1016/0550-3213(77)90328-5;%%


%+% 1 ref
\bibitem{FinitenessN4YM}
 S.~Mandelstam,
  %``Light-cone Superspace and the Finiteness of the $N=$ 4 Model,''
  J.\ Phys.\ Colloq.\  {\bf 43}, no. C3, 331 (1982);\\
%doi:10.1051/jphyscol:1982367
  %%CITATION = doi:10.1051/jphyscol:1982367;%%
  %4 citations counted in INSPIRE as of 30 Mar 2018
%
 S.~Mandelstam,
  %``Light Cone Superspace and the Ultraviolet Finiteness of the N=4 Model,''
  Nucl.\ Phys.\ B {\bf 213}, 149 (1983);\\
%doi:10.1016/0550-3213(83)90179-7
  %%CITATION = doi:10.1016/0550-3213(83)90179-7;%%
%
L.~Brink, O.~Lindgren and B.~E.~W.~Nilsson,
%``The Ultraviolet Finiteness of the N=4 Yang--Mills Theory,''
Phys.\ Lett.\  {\bf 123B}, 323 (1983);\\
%doi:10.1016/0370-2693(83)91210-8
  %%CITATION = doi:10.1016/0370-2693(83)91210-8;%%
%
 P.~S.~Howe, K.~S.~Stelle and P.~K.~Townsend,
  %``The Relaxed Hypermultiplet: An Unconstrained N=2 Superfield Theory,''
  Nucl.\ Phys.\ B {\bf 214}, 519 (1983).
%doi:10.1016/0550-3213(83)90249-3
  %%CITATION = doi:10.1016/0550-3213(83)90249-3;%%

%+% 4 refs
\bibitem{N4SugraMatter}
 Z.~Bern, S.~Davies and T.~Dennen,
  %``The Ultraviolet Structure of Half-Maximal Supergravity with Matter Multiplets at Two and Three Loops,''
  Phys.\ Rev.\ D {\bf 88}, 065007 (2013)
%doi:10.1103/PhysRevD.88.065007
  [arXiv:1305.4876 [hep-th]].
  %%CITATION = doi:10.1103/PhysRevD.88.065007;%%

%+% 3 refs
\bibitem{GeneralizedUnitarity}
 Z.~Bern, L.~J.~Dixon, D.~C.~Dunbar and D.~A.~Kosower,
%``One loop n point gauge theory amplitudes, unitarity and collinear limits,''
Nucl.\ Phys.\ B {\bf 425}, 217 (1994)
%doi:10.1016/0550-3213(94)90179-1
[hep-ph/9403226];\\
%%CITATION = doi:10.1016/0550-3213(94)90179-1;%%
%
Z.~Bern, L.~J.~Dixon, D.~C.~Dunbar and D.~A.~Kosower,
%``Fusing gauge theory tree amplitudes into loop amplitudes,''
Nucl.\ Phys.\ B {\bf 435}, 59 (1995)
%doi:10.1016/0550-3213(94)00488-Z
[hep-ph/9409265];\\
%%CITATION = doi:10.1016/0550-3213(94)00488-Z;%%
%
Z.~Bern, L.~J.~Dixon and D.~A.~Kosower,
  %``One loop amplitudes for e+ e- to four partons,''
  Nucl.\ Phys.\ B {\bf 513}, 3 (1998)
%doi:10.1016/S0550-3213(97)00703-7
  [hep-ph/9708239];\\
  %%CITATION = doi:10.1016/S0550-3213(97)00703-7;%%
%
 R.~Britto, F.~Cachazo and B.~Feng,
  %``Generalized unitarity and one-loop amplitudes in N=4 super-Yang-Mills,''
  Nucl.\ Phys.\ B {\bf 725}, 275 (2005)
%doi:10.1016/j.nuclphysb.2005.07.014
  [hep-th/0412103].
  %%CITATION = doi:10.1016/j.nuclphysb.2005.07.014;%%

%+% 4 refs
\bibitem{MaximalCutMethod}
 Z.~Bern, J.~J.~M.~Carrasco, H.~Johansson and D.~A.~Kosower,
%``Maximally supersymmetric planar Yang--Mills amplitudes at five loops,''
Phys.\ Rev.\ D {\bf 76}, 125020 (2007)
%doi:10.1103/PhysRevD.76.125020
[arXiv:0705.1864 [hep-th]].
%%CITATION = doi:10.1103/PhysRevD.76.125020;%%

%+% 9 refs
\bibitem{GeneralizedDoubleCopy}
 Z.~Bern, J.~J.~Carrasco, W.~M.~Chen, H.~Johansson and R.~Roiban,
%``Gravity Amplitudes as Generalized Double Copies of Gauge-Theory Amplitudes,''
Phys.\ Rev.\ Lett.\  {\bf 118}, no. 18, 181602 (2017)
%doi:10.1103/PhysRevLett.118.181602
[arXiv:1701.02519 [hep-th]].
%%CITATION = doi:10.1103/PhysRevLett.118.181602;%%

%+% 45 refs
\bibitem{FiveLoopN8Integrand}
 Z.~Bern, J.~J.~M.~Carrasco, W.~M.~Chen, H.~Johansson, R.~Roiban and M.~Zeng,
%``Five-loop four-point integrand of $N=8$ supergravity as a generalized double copy,''
Phys.\ Rev.\ D {\bf 96}, no. 12, 126012 (2017)
%doi:10.1103/PhysRevD.96.126012
[arXiv:1708.06807 [hep-th]].
%%CITATION = doi:10.1103/PhysRevD.96.126012;%%

%+% 3 refs
\bibitem{KLT}
 H.~Kawai, D.~C.~Lewellen and S.~H.~H.~Tye,
%``A Relation Between Tree Amplitudes of Closed and Open Strings,''             
Nucl.\ Phys.\ B {\bf 269}, 1 (1986).
%doi:10.1016/0550-3213(86)90362-7;                                            
%%CITATION = doi:10.1016/0550-3213(86)90362-7;%%

%+% 5 refs
\bibitem{BCJ}
 Z.~Bern, J.~J.~M.~Carrasco and H.~Johansson,
%``New Relations for Gauge-Theory Amplitudes,''
Phys.\ Rev.\ D {\bf 78}, 085011 (2008)
%doi:10.1103/PhysRevD.78.085011
[arXiv:0805.3993 [hep-ph]].
%%CITATION = doi:10.1103/PhysRevD.78.085011;%%

%+% 10 refs
\bibitem{BCJLoop}
 Z.~Bern, J.~J.~M.~Carrasco and H.~Johansson,
%``Perturbative Quantum Gravity as a Double Copy of Gauge Theory,''
Phys.\ Rev.\ Lett.\  {\bf 105}, 061602 (2010)
%doi:10.1103/PhysRevLett.105.061602
[arXiv:1004.0476 [hep-th]].
%%CITATION = doi:10.1103/PhysRevLett.105.061602;%%

%+% 15 refs
\bibitem{FiveLoopN4}
 Z.~Bern, J.~J.~M.~Carrasco, H.~Johansson and R.~Roiban,
%``The lFive-Loop Four-Point Amplitude of N=4 super-Yang--Mills Theory,''
Phys.\ Rev.\ Lett.\  {\bf 109}, 241602 (2012)
%doi:10.1103/PhysRevLett.109.241602
[arXiv:1207.6666 [hep-th]].
%%CITATION = doi:10.1103/PhysRevLett.109.241602;%%

%+% 5 refs
\bibitem{AttachedFile}
See the ancillary files of this manuscript.  The file {\tt N4YM\_5loop.m} contains the improved super-Yang--Mills integrand.
The files {\tt Level0Diagrams.m}$,\, \ldots,\, ${\tt Level6Diagrams.m} contain the 
improved supergravity integrands.

%+% 2 refs
\bibitem{FiveloopQCDBeta}
 P.~A.~Baikov, K.~G.~Chetyrkin and J.~H.~K\"uhn,
 %``Five-Loop Running of the QCD coupling constant,''
Phys.\ Rev.\ Lett.\  {\bf 118}, no. 8, 082002 (2017)
%doi:10.1103/PhysRevLett.118.082002
[arXiv:1606.08659 [hep-ph]];\\
%%CITATION = doi:10.1103/PhysRevLett.118.082002;%%
%
F.~Herzog, B.~Ruijl, T.~Ueda, J.~A.~M.~Vermaseren and A.~Vogt,
%``The five-loop beta function of Yang--Mills theory with fermions,''
JHEP {\bf 1702}, 090 (2017)
%doi:10.1007/JHEP02(2017)090
[arXiv:1701.01404 [hep-ph]];\\
%%CITATION = doi:10.1007/JHEP02(2017)090;%%
%
T.~Luthe, A.~Maier, P.~Marquard and Y.~Schroder,
%``Complete renormalization of QCD at five loops,''
JHEP {\bf 1703}, 020 (2017)
%doi:10.1007/JHEP03(2017)020
[arXiv:1701.07068 [hep-ph]].
%%CITATION = doi:10.1007/JHEP03(2017)020;%%

%+% 1 ref
\bibitem{IBPRefs}
 K.G.~Chetyrkin and F.V.~Tkachov,
%``Integration by parts: the algorithm to calculate beta functions 
%in 4 loops,''                                                                   
Nucl.\ Phys.\ B {\bf 192}, 159 (1981);\\
%%CITATION = NUPHA,B192,159;%%
%
K.~G.~Chetyrkin and F.~V.~Tkachov,
%``Integration by Parts: The Algorithm to Calculate beta Functions in 4 Loops,''
Nucl.\ Phys.\ B {\bf 192}, 159 (1981);\\
%doi:10.1016/0550-3213(81)90199-1;
%%CITATION = doi:10.1016/0550-3213(81)90199-1;%%
%
S.~Laporta,
%``High precision calculation of multiloop Feynman integrals by difference equations,''
Int.\ J.\ Mod.\ Phys.\ A {\bf 15}, 5087 (2000)
%doi:10.1016/S0217-751X(00)00215-7, 10.1142/S0217751X00002157
[hep-ph/0102033];\\
%%CITATION = doi:10.1016/S0217-751X(00)00215-7, 10.1142/S0217751X00002157;%%
%
S.~Laporta and E.~Remiddi,
%``The Analytical value of the electron (g-2) at order alpha**3 in QED,''
Phys.\ Lett.\ B {\bf 379}, 283 (1996)
%doi:10.1016/0370-2693(96)00439-X
[hep-ph/9602417];\\
%%CITATION = doi:10.1016/0370-2693(96)00439-X;%%
%
C.~Anastasiou and A.~Lazopoulos,
%``Automatic integral reduction for higher order perturbative calculations,''
JHEP {\bf 0407}, 046 (2004)
%doi:10.1088/1126-6708/2004/07/046
[hep-ph/0404258];\\
%%CITATION = doi:10.1088/1126-6708/2004/07/046;%%
%
A.~V.~Smirnov,
%``FIRE5: a C++ implementation of Feynman Integral REduction,''
Comput.\ Phys.\ Commun.\  {\bf 189}, 182 (2015)
%doi:10.1016/j.cpc.2014.11.024
[arXiv:1408.2372 [hep-ph]];\\
%%CITATION = doi:10.1016/j.cpc.2014.11.024;%%
%
A.~von Manteuffel and C.~Studerus,
%``Reduze 2 - Distributed Feynman Integral Reduction,''
arXiv:1201.4330 [hep-ph];\\
%%CITATION = ARXIV:1201.4330;%%
R.~N.~Lee,
%``Presenting LiteRed: a tool for the Loop InTEgrals REDuction,''
arXiv:1212.2685 [hep-ph];\\
%%CITATION = ARXIV:1212.2685;%%
%
B.~Ruijl, T.~Ueda and J.~A.~M.~Vermaseren,
%``Forcer, a FORM program for the parametric reduction of four-loop massless propagator diagrams,'' 
arXiv:1704.06650 [hep-ph];\\
%%CITATION = ARXIV:1704.06650;%%
%
P.~Maierhoefer, J.~Usovitsch and P.~Uwer,
%``Kira - A Feynman Integral Reduction Program,''
arXiv:1705.05610 [hep-ph].
%%CITATION = ARXIV:1705.05610;%%

%+% 1 ref
\bibitem{SmirnovBook}
 V.~A.~Smirnov,
  {\it Analytic tools for Feynman integrals,}
  Springer Tracts Mod.\ Phys.\  {\bf 250}, 1 (2012).
%doi:10.1007/978-3-642-34886-0
  %%CITATION = doi:10.1007/978-3-642-34886-0;%%

%+% 6 refs
\bibitem{Gluza2010ws}
  J.~Gluza, K.~Kajda and D.~A.~Kosower,
  %``Towards a Basis for Planar Two-Loop Integrals,''
  Phys.\ Rev.\ D {\bf 83}, 045012 (2011)
  %doi:10.1103/PhysRevD.83.045012
  [arXiv:1009.0472 [hep-th]].
  %%CITATION = doi:10.1103/PhysRevD.83.045012;%%

%+% 3 refs
\bibitem{CutIntegrals}
D.~A.~Kosower and K.~J.~Larsen,
%``Maximal Unitarity at Two Loops,''                                          
Phys.\ Rev.\ D {\bf 85}, 045017 (2012)
%doi:10.1103/PhysRevD.85.045017
[arXiv:1108.1180 [hep-th]];\\
%%CITATION = doi:10.1103/PhysRevD.85.045017;%%
%
%CaronHuot:2012ab
S.~Caron-Huot and K.~J.~Larsen,
%``Uniqueness of two-loop master contours,''                                  
JHEP {\bf 1210}, 026 (2012)
%doi:10.1007/JHEP10(2012)026
[arXiv:1205.0801 [hep-ph]];\\
%%CITATION = doi:10.1007/JHEP10(2012)026;%%
%
M.~S{\o}gaard,
  %``Global Residues and Two-Loop Hepta-Cuts,''                                 
  JHEP {\bf 1309}, 116 (2013)
%doi:10.1007/JHEP09(2013)116
  [arXiv:1306.1496 [hep-th]];\\
%%CITATION = doi:10.1007/JHEP09(2013)116;%%
%
H.~Johansson, D.~A.~Kosower and K.~J.~Larsen,
%``Maximal Unitarity for the Four-Mass Double Box,''                          
Phys.\ Rev.\ D {\bf 89}, no. 12, 125010 (2014)
%doi:10.1103/PhysRevD.89.125010
[arXiv:1308.4632 [hep-th]];\\
%%CITATION = doi:10.1103/PhysRevD.89.125010;%%
%
M.~S{\o}gaard and Y.~Zhang,
%``Multivariate Residues and Maximal Unitarity,''                             
JHEP {\bf 1312}, 008 (2013)
%doi:10.1007/JHEP12(2013)008
[arXiv:1310.6006 [hep-th]];\\
%%CITATION = doi:10.1007/JHEP12(2013)008;%%
%
M.~S{\o}gaard and Y.~Zhang,
%``Unitarity Cuts of Integrals with Doubled Propagators,''                    
JHEP {\bf 1407}, 112 (2014)
%doi:10.1007/JHEP07(2014)112
[arXiv:1403.2463 [hep-th]];\\
%%CITATION = doi:10.1007/JHEP07(2014)112;%%
%
S.~Abreu, R.~Britto, C.~Duhr and E.~Gardi,
%``Cuts from residues: the one-loop case,''                                   
JHEP {\bf 1706}, 114 (2017)
%doi:10.1007/JHEP06(2017)114
[arXiv:1702.03163 [hep-th]].
%%CITATION = doi:10.1007/JHEP06(2017)114;%%

%+% 5 refs
\bibitem{IBPAdvances}
R.~M.~Schabinger,
%``A New Algorithm For The Generation Of Unitarity-Compatible Integration By Parts Relations,''
JHEP {\bf 1201}, 077 (2012)
%doi:10.1007/JHEP01(2012)077
[arXiv:1111.4220 [hep-ph]];\\
%%CITATION = doi:10.1007/JHEP01(2012)077;%%
%
H.~Ita,
%``Two-loop Integrand Decomposition into Master Integrals and Surface Terms,''
Phys.\ Rev.\ D {\bf 94}, no. 11, 116015 (2016),
%doi:10.1103/PhysRevD.94.116015
[arXiv:1510.05626 [hep-th]];\\
%%CITATION = doi:10.1103/PhysRevD.94.116015;%%
%
K.~J.~Larsen and Y.~Zhang,
%``Integration-by-parts reductions from unitarity cuts and algebraic geometry,''
Phys.\ Rev.\ D {\bf 93}, no. 4, 041701 (2016),
%doi:10.1103/PhysRevD.93.041701
[arXiv:1511.01071 [hep-th]];\\
%%CITATION = doi:10.1103/PhysRevD.93.041701;%%
%
A.~Georgoudis, K.~J.~Larsen and Y.~Zhang,
%``Azurite: An algebraic geometry based package for finding bases of loop integrals,''
Comput.\ Phys.\ Commun.\  {\bf 221}, 203 (2017)
%doi:10.1016/j.cpc.2017.08.013
[arXiv:1612.04252 [hep-th]];\\
%%CITATION = doi:10.1016/j.cpc.2017.08.013;%%
%
Z.~Bern, M.~Enciso, H.~Ita and M.~Zeng,
%``Dual Conformal Symmetry, Integration-by-Parts Reduction, Differential Equations and the Nonplanar Sector,''
Phys.\ Rev.\ D {\bf 96}, no. 9, 096017 (2017)
%  doi:10.1103/PhysRevD.96.096017
[arXiv:1709.06055 [hep-th]];\\
  %%CITATION = doi:10.1103/PhysRevD.96.096017;%%
%
D.~A.~Kosower,
%``Direct Solution of Integration-by-Parts Systems,''
arXiv:1804.00131 [hep-ph].
%%CITATION = ARXIV:1804.00131;%%

%+% 2 refs
\bibitem{Zhang2016kfo}
 Y.~Zhang,
%``Lecture Notes on Multi-loop Integral Reduction and Applied Algebraic Geometry,''
arXiv:1612.02249 [hep-th].
%%CITATION = ARXIV:1612.02249;%%

%+% 1 ref
\bibitem{OConnelHighPowerocunt}
 G.~Mogull and D.~O'Connell,
%``Overcoming Obstacles to Colour-Kinematics Duality at Two Loops,''
JHEP {\bf 1512}, 135 (2015)
%doi:10.1007/JHEP12(2015)135
[arXiv:1511.06652 [hep-th]].
%%CITATION = doi:10.1007/JHEP12(2015)135;%%

%+% 4 refs
\bibitem{FivePointN4}
 J.~J.~Carrasco and H.~Johansson,
%``Five-Point Amplitudes in N=4 Super-Yang--Mills Theory and N=8 Supergravity,''
Phys.\ Rev.\ D {\bf 85}, 025006 (2012)
%doi:10.1103/PhysRevD.85.025006
[arXiv:1106.4711 [hep-th]].
%%CITATION = doi:10.1103/PhysRevD.85.025006;%%

%+% 1 ref
\bibitem{BCJonCuts}
 Z.~Bern, S.~Davies and J.~Nohle,
  %``Double-Copy Constructions and Unitarity Cuts,''
  Phys.\ Rev.\ D {\bf 93}, no. 10, 105015 (2016)
%doi:10.1103/PhysRevD.93.105015
  [arXiv:1510.03448 [hep-th]].
  %%CITATION = doi:10.1103/PhysRevD.93.105015;%%

%+% 3 refs
\bibitem{Square}
 Z.~Bern, T.~Dennen, Y.-t.~Huang and M.~Kiermaier,
%``Gravity as the Square of Gauge Theory,''
Phys.\ Rev.\ D {\bf 82}, 065003 (2010)
%doi:10.1103/PhysRevD.82.065003
[arXiv:1004.0693 [hep-th]].
%%CITATION = doi:10.1103/PhysRevD.82.065003;%%

%+% 1 ref
\bibitem{WeinzierlBCJ}
 M.~Tolotti and S.~Weinzierl,
%``Construction of an effective Yang--Mills Lagrangian with manifest BCJ duality,''
JHEP {\bf 1307}, 111 (2013)
%doi:10.1007/JHEP07(2013)111
[arXiv:1306.2975 [hep-th]].
%%CITATION = doi:10.1007/JHEP07(2013)111;%%

%+% 2 refs
\bibitem{BjerrumMomKernel}
M.~Kiermaier,
Amplitudes 2010, Queen Mary, University of London, \url{http://www.strings.ph.qmul.ac.uk/~theory/Amplitudes2010/Talks/MK2010.pdf};\\
%
N.~E.~J.~Bjerrum-Bohr, P.~H.~Damgaard, T.~Sondergaard and P.~Vanhove,
%``The Momentum Kernel of Gauge and Gravity Theories,''
JHEP {\bf 1101}, 001 (2011)
%doi:10.1007/JHEP01(2011)001 
[arXiv:1010.3933 [hep-th]];\\
%%CITATION = doi:10.1007/JHEP01(2011)001;%%
%
C.~R.~Mafra, O.~Schlotterer and S.~Stieberger,
%``Explicit BCJ Numerators from Pure Spinors,''
JHEP {\bf 1107}, 092 (2011)
%doi:10.1007/JHEP07(2011)092 
[arXiv:1104.5224 [hep-th]];\\
%%CITATION = doi:10.1007/JHEP07(2011)092;%%
%
Y.~J.~Du and C.~H.~Fu,
%``Explicit BCJ numerators of nonlinear sigma model,''
JHEP {\bf 1609}, 174 (2016)
%doi:10.1007/JHEP09(2016)174
[arXiv:1606.05846 [hep-th]];\\
%%CITATION = doi:10.1007/JHEP09(2016)174;%%
%
N.~E.~J.~Bjerrum-Bohr, J.~L.~Bourjaily, P.~H.~Damgaard and B.~Feng,
%``Manifesting Color-Kinematics Duality in the Scattering Equation Formalism,''
JHEP {\bf 1609}, 094 (2016)
%doi:10.1007/JHEP09(2016)094
[arXiv:1608.00006 [hep-th]];\\
%%CITATION = doi:10.1007/JHEP09(2016)094;%%
%
Y.~J.~Du and F.~Teng,
%``BCJ numerators from reduced Pfaffian,''
JHEP {\bf 1704}, 033 (2017)
%doi:10.1007/JHEP04(2017)033 
[arXiv:1703.05717 [hep-th]];\\
%%CITATION = doi:10.1007/JHEP04(2017)033;%%
%
Y.~J.~Du, B.~Feng and F.~Teng,
%``Expansion of All Multitrace Tree Level EYM Amplitudes,''
arXiv:1708.04514 [hep-th].
%%CITATION = ARXIV:1708.04514;%%
+% 1 ref                                                                                                                                                       

%+% 2 refs
\bibitem{Review}
 J.~J.~M.~Carrasco and H.~Johansson,
%``Generic multiloop methods and application to N=4 super-Yang--Mills,''
J.\ Phys.\ A {\bf 44}, 454004 (2011)
%doi:10.1088/1751-8113/44/45/454004
[arXiv:1103.3298 [hep-th]];\\
%%CITATION = doi:10.1088/1751-8113/44/45/454004;%%
%
J.~J.~M.~Carrasco,
%``Gauge and Gravity Amplitude Relations,''
%doi:10.1142/9789814678766\_0011
arXiv:1506.00974 [hep-th];\\
%%CITATION = doi:10.1142/9789814678766_0011;%%
%
M.~Chiodaroli,
%``Simplifying amplitudes in Maxwell-Einstein and Yang--Mills-Einstein supergravities,''
arXiv:1607.04129 [hep-th];\\
%%CITATION = ARXIV:1607.04129;%%
%
C.~Cheung,
  %``TASI Lectures on Scattering Amplitudes,''
  arXiv:1708.03872 [hep-ph].
  %%CITATION = ARXIV:1708.03872;%%

%+% 1 ref
\bibitem{OneloopN8}
 Z.~Bern, L.~J.~Dixon, M.~Perelstein and J.~S.~Rozowsky,
%``Multileg one loop gravity amplitudes from gauge theory,''
Nucl.\ Phys.\ B {\bf 546}, 423 (1999)
%doi:10.1016/S0550-3213(99)00029-2
[hep-th/9811140].
%%CITATION = doi:10.1016/S0550-3213(99)00029-2;%%

%+% 2 refs
\bibitem{abelianZ}
 J.~J.~M.~Carrasco, C.~R.~Mafra and O.~Schlotterer,
%``Abelian Z-theory: NLSM amplitudes and $\alpha$'-corrections from the open string,''
JHEP {\bf 1706}, 093 (2017)
%doi:10.1007/JHEP06(2017)093
[arXiv:1608.02569 [hep-th]].
%%CITATION = doi:10.1007/JHEP06(2017)093;%%

%+% 1 ref
\bibitem{OtherExamples}
 R.~H.~Boels, B.~A.~Kniehl, O.~V.~Tarasov and G.~Yang,
  %``Color-kinematic Duality for Form Factors,''
  JHEP {\bf 1302}, 063 (2013)
  %doi:10.1007/JHEP02(2013)063
  [arXiv:1211.7028 [hep-th]];\\
  %%CITATION = doi:10.1007/JHEP02(2013)063;%%
%
Z.~Bern, S.~Davies, T.~Dennen, Y.-t.~Huang and J.~Nohle,
  %``Color-Kinematics Duality for Pure Yang--Mills and Gravity at One and Two Loops,''
  Phys.\ Rev.\ D {\bf 92}, no. 4, 045041 (2015)
%doi:10.1103/PhysRevD.92.045041
  [arXiv:1303.6605 [hep-th]];\\
  %%CITATION = doi:10.1103/PhysRevD.92.045041;%%
%
C.~R.~Mafra and O.~Schlotterer,
%``Two-loop five-point amplitudes of super Yang-Mills and supergravity in pure spinor superspace,''
JHEP {\bf 1510}, 124 (2015)
%doi:10.1007/JHEP10(2015)124
[arXiv:1505.02746 [hep-th]];\\
%%CITATION = doi:10.1007/JHEP10(2015)124;%%
%
S.~He, R.~Monteiro and O.~Schlotterer,
  %``String-inspired BCJ numerators for one-loop MHV amplitudes,''
  JHEP {\bf 1601}, 171 (2016)
%doi:10.1007/JHEP01(2016)171
  [arXiv:1507.06288 [hep-th]];\\
  %%CITATION = doi:10.1007/JHEP01(2016)171;%%
%
E.~Herrmann and J.~Trnka,
  %``Gravity On-shell Diagrams,''
  JHEP {\bf 1611}, 136 (2016)
%doi:10.1007/JHEP11(2016)136
  [arXiv:1604.03479 [hep-th]];\\
 %%CITATION = doi:10.1007/JHEP11(2016)136;%%
%
G.~Yang,
%``Color-kinematics duality and Sudakov form factor at five loops for N=4 supersymmetric Yang--Mills theory,''
Phys.\ Rev.\ Lett.\  {\bf 117}, no. 27, 271602 (2016)
%doi:10.1103/PhysRevLett.117.271602
[arXiv:1610.02394 [hep-th]];\\
%%CITATION = doi:10.1103/PhysRevLett.117.271602;%%
%
R.~H.~Boels, T.~Huber and G.~Yang,
%``Four-Loop Nonplanar Cusp Anomalous Dimension in N=4 Supersymmetric Yang-Mills Theory,''
Phys.\ Rev.\ Lett.\  {\bf 119}, no. 20, 201601 (2017)
%doi:10.1103/PhysRevLett.119.201601
[arXiv:1705.03444 [hep-th]];\\
 %%CITATION = doi:10.1103/PhysRevLett.119.201601;%%
%
H.~Johansson, G.~K{\"a}lin and G.~Mogull,
  %``Two-loop supersymmetric QCD and half-maximal supergravity amplitudes,''
  JHEP {\bf 1709}, 019 (2017)
  %doi:10.1007/JHEP09(2017)019
  [arXiv:1706.09381 [hep-th]].
  %%CITATION = doi:10.1007/JHEP09(2017)019;%%

%+% 2 refs
\bibitem{BCJDifficulty}
 Z.~Bern, S.~Davies and J.~Nohle,
%``Double-Copy Constructions and Unitarity Cuts,''
Phys.\ Rev.\ D {\bf 93}, no. 10, 105015 (2016)
%doi:10.1103/PhysRevD.93.105015
[arXiv:1510.03448 [hep-th]];\\
%%CITATION = doi:10.1103/PhysRevD.93.105015;%%
%
G.~Mogull and D.~O'Connell,
%``Overcoming Obstacles to Colour-Kinematics Duality at Two Loops,''
JHEP {\bf 1512}, 135 (2015)
%doi:10.1007/JHEP12(2015)135
[arXiv:1511.06652 [hep-th]].
%%CITATION = doi:10.1007/JHEP12(2015)135;%%

%+% 1 ref
\bibitem{ClassicalSolutions}
 R.~Monteiro, D.~O'Connell and C.~D.~White,
%``Black holes and the double copy,''
JHEP {\bf 1412}, 056 (2014)
%doi:10.1007/JHEP12(2014)056
[arXiv:1410.0239 [hep-th]];\\
 %%CITATION = doi:10.1007/JHEP12(2014)056;%%
%
A.~Luna, R.~Monteiro, D.~O'Connell and C.~D.~White,
%``The classical double copy for Taub-NUT spacetime,''
Phys.\ Lett.\ B {\bf 750}, 272 (2015)
%doi:10.1016/j.physletb.2015.09.021
[arXiv:1507.01869 [hep-th]];\\
%%CITATION = doi:10.1016/j.physletb.2015.09.021;%%
%
G.~Cardoso, S.~Nagy and S.~Nampuri,
%``Multi-centered $ \mathcal{N}=2 $ BPS black holes: a double copy description,''
JHEP {\bf 1704}, 037 (2017)
%doi:10.1007/JHEP04(2017)037
[arXiv:1611.04409 [hep-th]];\\
 %%CITATION = doi:10.1007/JHEP04(2017)037;%%
%
T.~Adamo, E.~Casali, L.~Mason and S.~Nekovar,
%``Scattering on plane waves and the double copy,''
arXiv:1706.08925 [hep-th];\\
%%CITATION = ARXIV:1706.08925;%%
%
N.~Bahjat-Abbas, A.~Luna and C.~D.~White,
%``The Kerr-Schild double copy in curved spacetime,''
JHEP {\bf 1712}, 004 (2017)
%doi:10.1007/JHEP12(2017)004
[arXiv:1710.01953 [hep-th]];\\
%%CITATION = doi:10.1007/JHEP12(2017)004;%%
%
M.~Carrillo-Gonz{\'a}lez, R.~Penco and M.~Trodden,
%``The classical double copy in maximally symmetric spacetimes,''
JHEP {\bf 1804}, 028 (2018)
%doi:10.1007/JHEP04(2018)028
[arXiv:1711.01296 [hep-th]].
%%CITATION = doi:10.1007/JHEP04(2018)028;%%

%+% 1 ref
\bibitem{RadiationSolutions}
 A.~Luna, R.~Monteiro, I.~Nicholson, D.~O'Connell and C.~D.~White,
%``The double copy: Bremsstrahlung and accelerating black holes,''
JHEP {\bf 1606}, 023 (2016)
%doi:10.1007/JHEP06(2016)023
[arXiv:1603.05737 [hep-th]];\\
%%CITATION = doi:10.1007/JHEP06(2016)023;%%
%
W.~D.~Goldberger and A.~K.~Ridgway,
%``Radiation and the classical double copy for color charges,''
Phys.\ Rev.\ D {\bf 95}, no. 12, 125010 (2017)
%doi:10.1103/PhysRevD.95.125010
[arXiv:1611.03493 [hep-th]];\\
%%CITATION = doi:10.1103/PhysRevD.95.125010;%%
%
A.~Luna, R.~Monteiro, I.~Nicholson, A.~Ochirov, D.~O'Connell, N.~Westerberg and C.~D.~White,
  %``Perturbative spacetimes from Yang--Mills theory,''
  JHEP {\bf 1704}, 069 (2017)
  %doi:10.1007/JHEP04(2017)069
  [arXiv:1611.07508 [hep-th]];\\
  %%CITATION = doi:10.1007/JHEP04(2017)069;%%
%
  W.~D.~Goldberger, S.~G.~Prabhu and J.~O.~Thompson,
  %``Classical gluon and graviton radiation from the bi-adjoint scalar double copy,''
  Phys.\ Rev.\ D {\bf 96}, no. 6, 065009 (2017)
  %doi:10.1103/PhysRevD.96.065009
  [arXiv:1705.09263 [hep-th]];\\
  %%CITATION = doi:10.1103/PhysRevD.96.065009;%%
%
  W.~D.~Goldberger, J.~Li and S.~G.~Prabhu,
  %``Spinning particles, axion radiation, and the classical double copy,''
  arXiv:1712.09250 [hep-th];\\
  %%CITATION = ARXIV:1712.09250;%%
%
  W.~D.~Goldberger and A.~K.~Ridgway,
  %``Bound states and the classical double copy,''
  arXiv:1711.09493 [hep-th];\\
  %%CITATION = ARXIV:1711.09493;%%
%
J.~Li and S.~G.~Prabhu,
%``Gravitational radiation from the classical spinning double copy,''
arXiv:1803.02405 [hep-th].
%%CITATION = ARXIV:1803.02405;%%

%+% 1 ref
\bibitem{Donoghue}
 N.~E.~J.~Bjerrum-Bohr, J.~F.~Donoghue, B.~R.~Holstein, L.~Plant\`{e} and P.~Vanhove,
  %``Bending of Light in Quantum Gravity,''
  Phys.\ Rev.\ Lett.\  {\bf 114}, no. 6, 061301 (2015)
%doi:10.1103/PhysRevLett.114.061301
  [arXiv:1410.7590 [hep-th]];\\
  %%CITATION = doi:10.1103/PhysRevLett.114.061301;%%
%
  N.~E.~J.~Bjerrum-Bohr, J.~F.~Donoghue, B.~R.~Holstein, L.~Plant\'{e} and P.~Vanhove,
  %``Light-like Scattering in Quantum Gravity,''
  JHEP {\bf 1611}, 117 (2016)
%doi:10.1007/JHEP11(2016)117
  [arXiv:1609.07477 [hep-th]];\\
  %%CITATION = doi:10.1007/JHEP11(2016)117;%%
%
  N.~E.~J.~Bjerrum-Bohr, B.~R.~Holstein, J.~F.~Donoghue, L.~Plant\'{e} and P.~Vanhove,
  %``Illuminating Light Bending,''
  PoS CORFU {\bf 2016}, 077 (2017)
  [arXiv:1704.01624 [gr-qc]].
  %%CITATION = ARXIV:1704.01624;%%

%+% 1 ref
\bibitem{SugraSyms}
 L.~Borsten, M.~J.~Duff, L.~J.~Hughes and S.~Nagy,
%``Magic Square from Yang--Mills Squared,''
Phys.\ Rev.\ Lett.\  {\bf 112}, no. 13, 131601 (2014)
%doi:10.1103/PhysRevLett.112.131601
[arXiv:1301.4176 [hep-th]];\\
%%CITATION = doi:10.1103/PhysRevLett.112.131601;%%
%
A.~Anastasiou, L.~Borsten, M.~J.~Duff, L.~J.~Hughes and S.~Nagy,
%``A magic pyramid of supergravities,''
JHEP {\bf 1404}, 178 (2014)
%doi:10.1007/JHEP04(2014)178
[arXiv:1312.6523 [hep-th]];\\
%%CITATION = doi:10.1007/JHEP04(2014)178;%%
%
A.~Anastasiou, L.~Borsten, M.~J.~Duff, L.~J.~Hughes and S.~Nagy,
%``Yang--Mills origin of gravitational symmetries,''
Phys.\ Rev.\ Lett.\  {\bf 113}, no. 23, 231606 (2014)
%doi:10.1103/PhysRevLett.113.231606
[arXiv:1408.4434 [hep-th]].
%%CITATION = doi:10.1103/PhysRevLett.113.231606;%%

%+% 1 ref
\bibitem{DoubleCopyTheories}
 J.~J.~M.~Carrasco, M.~Chiodaroli, M.~Gunaydin and R.~Roiban,
%``One-loop four-point amplitudes in pure and matter-coupled $N \le 4$ supergravity,''
JHEP {\bf 1303}, 056 (2013)
%doi:10.1007/JHEP03(2013)056
[arXiv:1212.1146 [hep-th]];\\
%%CITATION = doi:10.1007/JHEP03(2013)056;%%
%
M.~Chiodaroli, M.~Gunaydin, H.~Johansson and R.~Roiban,
%``Scattering amplitudes in $ \mathcal{N}=2 $ Maxwell-Einstein and Yang--Mills/Einstein supergravity,''
JHEP {\bf 1501}, 081 (2015)
%doi:10.1007/JHEP01(2015)081
[arXiv:1408.0764 [hep-th]];\\
  %%CITATION = doi:10.1007/JHEP01(2015)081;%%
%
 M.~Chiodaroli, M.~Gunaydin, H.~Johansson and R.~Roiban,
% ``Explicit Formulae for Yang--Mills-Einstein Amplitudes from the Double Copy,''
  JHEP {\bf 1707}, 002 (2017)
%doi:10.1007/JHEP07(2017)002
  [arXiv:1703.00421 [hep-th]].
  %%CITATION = doi:10.1007/JHEP07(2017)002;%%

%+% 1 ref
\bibitem{DoubleCopyTheoriesFund}
M.~Chiodaroli, M.~Gunaydin, H.~Johansson and R.~Roiban,
%``Spontaneously Broken Yang--Mills-Einstein Supergravities as Double Copies,''
JHEP {\bf 1706}, 064 (2017)
%doi:10.1007/JHEP06(2017)064
[arXiv:1511.01740 [hep-th]];\\
%%CITATION = doi:10.1007/JHEP06(2017)064;%%
%
M.~Chiodaroli, M.~Gunaydin H.~Johansson and R.~Roiban,
%``Complete construction of magical, symmetric and homogeneous N=2 supergravities as double copies of gauge theories,''
Phys.\ Rev.\ Lett.\  {\bf 117}, no. 1, 011603 (2016)
%doi:10.1103/PhysRevLett.117.011603
[arXiv:1512.09130 [hep-th]];\\
%%CITATION = doi:10.1103/PhysRevLett.117.011603;%
%
A.~Anastasiou, L.~Borsten, M.~J.~Duff, M.~J.~Hughes, A.~Marrani, S.~Nagy and M.~Zoccali,
  %``Twin supergravities from Yang--Mills theory squared,''
  Phys.\ Rev.\ D {\bf 96}, no. 2, 026013 (2017)
  %doi:10.1103/PhysRevD.96.026013
  [arXiv:1610.07192 [hep-th]];\\
  %%CITATION = doi:10.1103/PhysRevD.96.026013;%%
%
  A.~Anastasiou, L.~Borsten, M.~J.~Duff, A.~Marrani, S.~Nagy and M.~Zoccali,
  %``Are all supergravity theories Yang--Mills squared?,''
  arXiv:1707.03234 [hep-th];\\
  %%CITATION = ARXIV:1707.03234;%%
%
 M.~Chiodaroli, M.~Gunaydin, H.~Johansson and R.~Roiban,
  %``Gauged supergravities and spontaneous SUSY breaking from the double copy,''
  arXiv:1710.08796 [hep-th].
  %%CITATION = ARXIV:1710.08796;%%

%+% 1 ref
\bibitem{BLGBCJ}
 T.~Bargheer, S.~He and T.~McLoughlin,
%``New Relations for Three-Dimensional Supersymmetric Scattering Amplitudes,''  
Phys.\ Rev.\ Lett.\  {\bf 108}, 231601 (2012)
%doi:10.1103/PhysRevLett.108.231601                                             
[arXiv:1203.0562 [hep-th]];\\
%%CITATION = doi:10.1103/PhysRevLett.108.231601;%%
%
Y.-t.~Huang and H.~Johansson,
%``Equivalent D=3 Supergravity Amplitudes from Double Copies of Three-Algebra and Two-Algebra Gauge Theories,''
Phys.\ Rev.\ Lett.\  {\bf 110}, 171601 (2013)
%doi:10.1103/PhysRevLett.110.171601                                             
[arXiv:1210.2255 [hep-th]];\\
%%CITATION = doi:10.1103/PhysRevLett.110.171601;%%
%
Y.-t.~Huang, H.~Johansson and S.~Lee,
%``On Three-Algebra and Bi-Fundamental Matter Amplitudes and Integrability of Supergravity,''
JHEP {\bf 1311}, 050 (2013)
%doi:10.1007/JHEP11(2013)050                                                    
[arXiv:1307.2222 [hep-th]].
%%CITATION = doi:10.1007/JHEP11(2013)050;%%
%

%+% 1 ref
\bibitem{NLSMBCJ}
 G.~Chen and Y.~J.~Du,
%``Amplitude Relations in Non-linear Sigma Model,''
JHEP {\bf 1401}, 061 (2014)
%doi:10.1007/JHEP01(2014)061
[arXiv:1311.1133 [hep-th]];\\
%%CITATION = doi:10.1007/JHEP01(2014)061;%%
%
F.~Cachazo, S.~He and E.~Y.~Yuan,
%``Scattering Equations and Matrices: From Einstein To Yang--Mills, DBI and NLSM,''
JHEP {\bf 1507}, 149 (2015)
%doi:10.1007/JHEP07(2015)149
[arXiv:1412.3479 [hep-th]];\\
%%CITATION = doi:10.1007/JHEP07(2015)149;%%
%
F.~Cachazo, P.~Cha and S.~Mizera,
%``Extensions of Theories from Soft Limits,''
JHEP {\bf 1606}, 170 (2016)
%doi:10.1007/JHEP06(2016)170
[arXiv:1604.03893 [hep-th]];\\
  %%CITATION = doi:10.1007/JHEP06(2016)170;%%
%
C.~R.~Mafra and O.~Schlotterer,
%``Non-abelian $Z$-theory: Berends-Giele recursion for the $\alpha'$-expansion of disk integrals,''
JHEP {\bf 1701}, 031 (2017)
%doi:10.1007/JHEP01(2017)031
[arXiv:1609.07078 [hep-th]];\\
%%CITATION = doi:10.1007/JHEP01(2017)031;%%
%
J.~J.~M.~Carrasco, C.~R.~Mafra and O.~Schlotterer,
%``Semi-abelian Z-theory: NLSM$+?^{3}$ from the open string,''
JHEP {\bf 1708}, 135 (2017)
%doi:10.1007/JHEP08(2017)135
[arXiv:1612.06446 [hep-th]];\\
%%CITATION = doi:10.1007/JHEP08(2017)135;%%
%
C.~Cheung, C.~H.~Shen and C.~Wen,
%``Unifying Relations for Scattering Amplitudes,''
JHEP {\bf 1802}, 095 (2018)
%doi:10.1007/JHEP02(2018)095
[arXiv:1705.03025 [hep-th]].
%%CITATION = doi:10.1007/JHEP02(2018)095;%%

%+% 1 ref
\bibitem{NLSMaction}
 C.~Cheung and C.~H.~Shen,
  %``Symmetry for Flavor-Kinematics Duality from an Action,''
  Phys.\ Rev.\ Lett.\  {\bf 118}, no. 12, 121601 (2017)
  %doi:10.1103/PhysRevLett.118.121601
  [arXiv:1612.00868 [hep-th]].
  %%CITATION = doi:10.1103/PhysRevLett.118.121601;%%

%+% 1 ref
\bibitem{DiskandHeteroticStringBCJ}
 J.~Broedel, O.~Schlotterer and S.~Stieberger,
%``Polylogarithms, Multiple Zeta Values and Superstring Amplitudes,''
Fortsch.\ Phys.\  {\bf 61}, 812 (2013)
%doi:10.1002/prop.201300019
[arXiv:1304.7267 [hep-th]];\\
%%CITATION = doi:10.1002/prop.201300019;%%
%
S.~Stieberger and T.~R.~Taylor,
%``Closed String Amplitudes as Single-Valued Open String Amplitudes,''
Nucl.\ Phys.\ B {\bf 881}, 269 (2014)
%doi:10.1016/j.nuclphysb.2014.02.005
[arXiv:1401.1218 [hep-th]];\\
%%CITATION = doi:10.1016/j.nuclphysb.2014.02.005;%%
%
Y.-t.~Huang, O.~Schlotterer and C.~Wen,
%``Universality in string interactions,''
JHEP {\bf 1609}, 155 (2016)
%doi:10.1007/JHEP09(2016)155
[arXiv:1602.01674 [hep-th]];\\
%%CITATION = doi:10.1007/JHEP09(2016)155;%%
%
C.~R.~Mafra and O.~Schlotterer,
%``The double-copy structure of one-loop open-string amplitudes,''
arXiv:1711.09104 [hep-th];\\
%%CITATION = ARXIV:1711.09104;%%
%
T.~Azevedo, M.~Chiodaroli, H.~Johansson and O.~Schlotterer,
  %``Heterotic and bosonic string amplitudes via field theory,''
  arXiv:1803.05452 [hep-th].
  %%CITATION = ARXIV:1803.05452;%%

%+% 1 ref
\bibitem{ConformalGravity}
  H.~Johansson and J.~Nohle,
  %``Conformal Gravity from Gauge Theory,’'
   arXiv:1707.02965 [hep-th].
  %%CITATION = ARXIV:1707.02965;%%

%+% 1 ref
\bibitem{FundMatter}
 H.~Johansson and A.~Ochirov,
  %``Pure Gravities via Color-Kinematics Duality for Fundamental Matter,''                                                                                      
  JHEP {\bf 1511}, 046 (2015)
%doi:10.1007/JHEP11(2015)046                                                                                                                                  
  [arXiv:1407.4772 [hep-th]];\\
  %%CITATION = doi:10.1007/JHEP11(2015)046;%%
%
H.~Johansson and A.~Ochirov,
%``Color-Kinematics Duality for QCD Amplitudes,''                                                                                                               
JHEP {\bf 1601}, 170 (2016)
%doi:10.1007/JHEP01(2016)170                                                                                                                                    
[arXiv:1507.00332 [hep-ph]].
%%CITATION = doi:10.1007/JHEP01(2016)170;%%

%+% 1 ref
\bibitem{Nair}
 V.~P.~Nair,
%``A Current Algebra for Some Gauge Theory Amplitudes,''
Phys.\ Lett.\ B {\bf 214}, 215 (1988).
%doi:10.1016/0370-2693(88)91471-2
%%CITATION = doi:10.1016/0370-2693(88)91471-2;%%

%+% 2 refs
\bibitem{SuperSums}
 H.~Elvang, D.~Z.~Freedman and M.~Kiermaier,
%``Recursion Relations, Generating Functions, and Unitarity Sums in N=4 SYM Theory,''
JHEP {\bf 0904}, 009 (2009)
%doi:10.1088/1126-6708/2009/04/009
[arXiv:0808.1720 [hep-th]];\\
  %%CITATION = doi:10.1088/1126-6708/2009/04/009;%%
%
Z.~Bern, J.~J.~M.~Carrasco, H.~Ita, H.~Johansson and R.~Roiban,
%``On the Structure of Supersymmetric Sums in Multi-Loop Unitarity Cuts,''
Phys.\ Rev.\ D {\bf 80}, 065029 (2009)
%doi:10.1103/PhysRevD.80.065029
[arXiv:0903.5348 [hep-th]].
%%CITATION = doi:10.1103/PhysRevD.80.065029;%%

%+% 1 ref
\bibitem{SixDimSuperSpace}
 C.~Cheung and D.~O'Connell,
  %``Amplitudes and Spinor-Helicity in Six Dimensions,''
  JHEP {\bf 0907}, 075 (2009)
%doi:10.1088/1126-6708/2009/07/075
  [arXiv:0902.0981 [hep-th]];\\
  %%CITATION = doi:10.1088/1126-6708/2009/07/075;%%
%
 T.~Dennen, Y.-t.~Huang and W.~Siegel,
  %``Supertwistor space for 6D maximal super Yang--Mills,''
  JHEP {\bf 1004}, 127 (2010)
%doi:10.1007/JHEP04(2010)127
  [arXiv:0910.2688 [hep-th]];\\
  %%CITATION = doi:10.1007/JHEP04(2010)127;%%
%
 Z.~Bern, J.~J.~Carrasco, T.~Dennen, Y.-t.~Huang and H.~Ita,
%``Generalized Unitarity and Six-Dimensional Helicity,''
Phys.\ Rev.\ D {\bf 83}, 085022 (2011)
%doi:10.1103/PhysRevD.83.085022
[arXiv:1010.0494 [hep-th]].
%%CITATION = doi:10.1103/PhysRevD.83.085022;%%

%+% 1 ref
\bibitem{JConstraints}
S.~H.~Henry Tye and Y.~Zhang,
%``Dual Identities inside the Gluon and the Graviton Scattering Amplitudes,''
JHEP {\bf 1006}, 071 (2010)
Erratum: [JHEP {\bf 1104}, 114 (2011)]
%doi:10.1007/JHEP06(2010)071, 10.1007/JHEP04(2011)114
[arXiv:1003.1732 [hep-th]];\\
%%CITATION = doi:10.1007/JHEP06(2010)071, 10.1007/JHEP04(2011)114;%%
%
N.~E.~J.~Bjerrum-Bohr, P.~H.~Damgaard, T.~Sondergaard and P.~Vanhove,
%``Monodromy and Jacobi-like Relations for Color-Ordered Amplitudes,''
JHEP {\bf 1006}, 003 (2010)
%doi:10.1007/JHEP06(2010)003
[arXiv:1003.2403 [hep-th]].
%%CITATION = doi:10.1007/JHEP06(2010)003;%%

%+% 1 ref
\bibitem{BRY}
 Z.~Bern, J.~S.~Rozowsky and B.~Yan,
  %``Two loop four gluon amplitudes in N=4 superYang--Mills,''
  Phys.\ Lett.\ B {\bf 401}, 273 (1997)
  %doi:10.1016/S0370-2693(97)00413-9
  [hep-ph/9702424].
  %%CITATION = doi:10.1016/S0370-2693(97)00413-9;%%

%+% 2 refs
\bibitem{Vladimirov}
 A.~A.~Vladimirov,
%``Method For Computing Renormalization Group Functions In Dimensional
%Renormalization Scheme,''
Theor.\ Math.\ Phys.\  {\bf 43}, 417 (1980)
[Teor.\ Mat.\ Fiz.\  {\bf 43}, 210 (1980)];\\
%%CITATION = TMFZA,43,210;%%
%
N.~Marcus and A.~Sagnotti,
%``A Simple Method For Calculating Counterterms,''
Nuovo Cim.\ A {\bf 87}, 1 (1985);\\
%%CITATION = NUCIA,A87,1;%%
%
M.~Beneke and V.~A.~Smirnov,
%``Asymptotic expansion of Feynman integrals near threshold,''
Nucl.\ Phys.\ B {\bf 522}, 321 (1998)
%  doi:10.1016/S0550-3213(98)00138-2
[hep-ph/9711391].
%%CITATION = doi:10.1016/S0550-3213(98)00138-2;%%

%+% 2 refs
\bibitem{LutheThesis}
 T. Luthe, 2015, ``Fully massive vacuum integrals at 5 loops'', PhD thesis, Bielefeld University.

%+% 1 ref
\bibitem{Pak2011xt}
  A.~Pak,
  %``The Toolbox of modern multi-loop calculations: novel analytic and semi-analytic techniques,''
  J.\ Phys.\ Conf.\ Ser.\  {\bf 368}, 012049 (2012)
%doi:10.1088/1742-6596/368/1/012049
  [arXiv:1111.0868 [hep-ph]].
  %%CITATION = doi:10.1088/1742-6596/368/1/012049;%%

%+% 1 ref
\bibitem{Hoff2016pot}
  J.~Hoff,
  %``The Mathematica package TopoID and its application to the Higgs boson production cross section,''
  J.\ Phys.\ Conf.\ Ser.\  {\bf 762}, no. 1, 012061 (2016)
%doi:10.1088/1742-6596/762/1/012061
  [arXiv:1607.04465 [hep-ph]].
  %%CITATION = doi:10.1088/1742-6596/762/1/012061;%%

%+% 2 refs
\bibitem{BaikovCuts}
 H.~Frellesvig and C.~G.~Papadopoulos,
%``Cuts of Feynman Integrals in Baikov representation,''
JHEP {\bf 1704}, 083 (2017)
%doi:10.1007/JHEP04(2017)083
[arXiv:1701.07356 [hep-ph]];\\
%
J.~Bosma, M.~S{\o}gaard and Y.~Zhang,
%``Maximal Cuts in Arbitrary Dimension,''
JHEP {\bf 1708}, 051 (2017)
%doi:10.1007/JHEP08(2017)051
[arXiv:1704.04255 [hep-th]];\\
%%CITATION = doi:10.1007/JHEP08(2017)051;%%
%
M.~Harley, F.~Moriello and R.~M.~Schabinger,
%``Baikov-Lee Representations Of Cut Feynman Integrals,''                     
JHEP {\bf 1706}, 049 (2017)
%doi:10.1007/JHEP06(2017)049
[arXiv:1705.03478 [hep-ph]].
%%CITATION = doi:10.1007/JHEP06(2017)049;%%

%+% 3 refs
\bibitem{NumericalUnitarity}
S.~Abreu, F.~Febres Cordero, H.~Ita, M.~Jaquier, B.~Page and M.~Zeng,
%``Two-Loop Four-Gluon Amplitudes with the Numerical Unitarity Method,''
Phys.\ Rev.\ Lett.\  {\bf 119}, no. 14, 142001 (2017)
%doi:10.1103/PhysRevLett.119.142001
[arXiv:1703.05273 [hep-ph]];\\
%%CITATION = doi:10.1103/PhysRevLett.119.142001;%%
%
S.~Abreu, F.~Febres Cordero, H.~Ita, B.~Page and M.~Zeng,
%``Planar Two-Loop Five-Gluon Amplitudes from Numerical Unitarity,''
arXiv:1712.03946 [hep-ph].
%%CITATION = ARXIV:1712.03946;%%

%+% 1 ref
\bibitem{Sogaard2014ila}
  M.~S{\o}gaard and Y.~Zhang,
  %``Unitarity Cuts of Integrals with Doubled Propagators,''
  JHEP {\bf 1407}, 112 (2014)
%doi:10.1007/JHEP07(2014)112
  [arXiv:1403.2463 [hep-th]];\\
  %%CITATION = doi:10.1007/JHEP07(2014)112;%%
%
H.~Johansson, D.~A.~Kosower, K.~J.~Larsen and M.~S{\o}gaard,
%``Cross-Order Integral Relations from Maximal Cuts,''
Phys.\ Rev.\ D {\bf 92}, no. 2, 025015 (2015)
% doi:10.1103/PhysRevD.92.025015
[arXiv:1503.06711 [hep-th]];\\
%%CITATION = doi:10.1103/PhysRevD.92.025015;%%
%
G.~Chen, J.~Liu, R.~Xie, H.~Zhang and Y.~Zhou,
%``Syzygies Probing Scattering Amplitudes,''
JHEP {\bf 1609}, 075 (2016)
%doi:10.1007/JHEP09(2016)075
[arXiv:1511.01058 [hep-th]].
%%CITATION = doi:10.1007/JHEP09(2016)075;%%

%+% 1 ref
\bibitem{Tarasov1996br}
 Z.~Bern, L.~J.~Dixon and D.~A.~Kosower,
  %``Dimensionally regulated one loop integrals,''
  Phys.\ Lett.\ B {\bf 302}, 299 (1993)
  Erratum: [Phys.\ Lett.\ B {\bf 318}, 649 (1993)]
%doi:10.1016/0370-2693(93)90469-X, 10.1016/0370-2693(93)90400-C
  [hep-ph/9212308];\\
  %%CITATION = doi:10.1016/0370-2693(93)90469-X, 10.1016/0370-2693(93)90400-C;%%
%
  O.~V.~Tarasov,
  %``Connection between Feynman integrals having different values of the space-time dimension,''
  Phys.\ Rev.\ D {\bf 54}, 6479 (1996)
%doi:10.1103/PhysRevD.54.6479
  [hep-th/9606018].
  %%CITATION = doi:10.1103/PhysRevD.54.6479;%%

%+% 1 ref
\bibitem{Singular}
W. Decker, G.-M.~Greuel, G. Pfister and H. Sch{\"o}nemann: 
\newblock {\sc Singular} {4-1-1} --- {A} computer algebra system for polynomial computations.
\newblock {http://www.singular.uni-kl.de} (2018).

%+% 1 ref
\bibitem{BaikovRep}
 P.~A.~Baikov,
%``Explicit solutions of the three loop vacuum integral recurrence relations,''
Phys.\ Lett.\ B {\bf 385}, 404 (1996)
%doi:10.1016/0370-2693(96)00835-0
[hep-ph/9603267];\\
%%CITATION = doi:10.1016/0370-2693(96)00835-0;%%
%
  P.~A.~Baikov,
  %``Explicit solutions of the multiloop integral recurrence relations and its application,''
  Nucl.\ Instrum.\ Meth.\ A {\bf 389}, 347 (1997)
%doi:10.1016/S0168-9002(97)00126-5
  [hep-ph/9611449];\\
  %%CITATION = doi:10.1016/S0168-9002(97)00126-5;%%
%
  R.~E.~Cutkosky,
  %``Singularities and discontinuities of Feynman amplitudes,''                 
  J.\ Math.\ Phys.\  {\bf 1}, 429 (1960);\\
%doi:10.1063/1.1703676
  %%CITATION = doi:10.1063/1.1703676;%%
%
  A.~G.~Grozin,
  %``Integration by parts: An Introduction,''                                   
  Int.\ J.\ Mod.\ Phys.\ A {\bf 26}, 2807 (2011)
%doi:10.1142/S0217751X11053687
  [arXiv:1104.3993 [hep-ph]].
  %%CITATION = doi:10.1142/S0217751X11053687;%%

%+% 2 refs
\bibitem{FiniteFields}
A.~von Manteuffel and R.~M.~Schabinger,
%``A novel approach to integration by parts reduction,''
Phys.\ Lett.\ B {\bf 744}, 101 (2015)
%doi:10.1016/j.physletb.2015.03.029
arXiv:1406.4513 [hep-ph]];\\
%%CITATION = doi:10.1016/j.physletb.2015.03.029;%%
%
T.~Peraro,
%``Scattering amplitudes over finite fields and multivariate functional reconstruction,''
JHEP {\bf 1612}, 030 (2016)
%doi:10.1007/JHEP12(2016)030
[arXiv:1608.01902 [hep-ph]].
%%CITATION = doi:10.1007/JHEP12(2016)030;%%

%+% 2 refs
\bibitem{LinBox}
{\sc LinBox}: A generic library for exact linear algebra. 
J.-G. Dumas, T. Gautier, M. Giesbrecht, P. Giorgi, B. Hovinen, E. Kaltofen, B. D. Saunders, W. J. Turner, and G. Villard. Mathematical Software. July 2002, 40-50. \url{http://www.linalg.org}.

%+% 2 refs
\bibitem{ManteuffelSchabingerFiniteFields}
A.~von Manteuffel and R.~M.~Schabinger,
%``Quark and gluon form factors to four-loop order in QCD: the $N_f^3$ contributions,''
Phys.\ Rev.\ D {\bf 95}, no. 3, 034030 (2017)
%doi:10.1103/PhysRevD.95.034030
[arXiv:1611.00795 [hep-ph]].
%%CITATION = doi:10.1103/PhysRevD.95.034030;%%

%+% 1 ref
\bibitem{FIESTA}
 A.~V.~Smirnov and M.~N.~Tentyukov,
%``Feynman Integral Evaluation by a Sector decomposiTion Approach (FIESTA),''
Comput.\ Phys.\ Commun.\  {\bf 180}, 735 (2009)
%doi:10.1016/j.cpc.2008.11.006
[arXiv:0807.4129 [hep-ph]];\\
%%CITATION = doi:10.1016/j.cpc.2008.11.006;%%
%
A.~V.~Smirnov, V.~A.~Smirnov and M.~Tentyukov,
%``FIESTA 2: Parallelizeable multiloop numerical calculations,''
Comput.\ Phys.\ Commun.\  {\bf 182}, 790 (2011)
%doi:10.1016/j.cpc.2010.11.025
[arXiv:0912.0158 [hep-ph]];\\
%%CITATION = doi:10.1016/j.cpc.2010.11.025;%%
%
A.~V.~Smirnov,
%``FIESTA 3: cluster-parallelizable multiloop numerical calculations in physical regions,''
Comput.\ Phys.\ Commun.\  {\bf 185}, 2090 (2014)
%doi:10.1016/j.cpc.2014.03.015
[arXiv:1312.3186 [hep-ph]].
%%CITATION = doi:10.1016/j.cpc.2014.03.015;%%

%+% 2 refs
\bibitem{SixLoopPlanarYM}
Z.~Bern, J.~J.~Carrasco, L.~J.~Dixon, M.~R.~Douglas, M.~von Hippel and H.~Johansson,
%``D=5 maximally supersymmetric Yang--Mills theory diverges at six loops,''
Phys.\ Rev.\ D {\bf 87}, no. 2, 025018 (2013)
%doi:10.1103/PhysRevD.87.025018
[arXiv:1210.7709 [hep-th]].
%%CITATION = doi:10.1103/PhysRevD.87.025018;%%

%+% 1 ref
\bibitem{NoTriangle}
 Z.~Bern, N.~E.~J.~Bjerrum-Bohr and D.~C.~Dunbar,
  %``Inherited twistor-space structure of gravity loop amplitudes,''
  JHEP {\bf 0505}, 056 (2005)
%doi:10.1088/1126-6708/2005/05/056
  [hep-th/0501137];\\
  %%CITATION = doi:10.1088/1126-6708/2005/05/056;%%
%
 N.~E.~J.~Bjerrum-Bohr and P.~Vanhove,
  %``Explicit Cancellation of Triangles in One-loop Gravity Amplitudes,''
  JHEP {\bf 0804}, 065 (2008)
%doi:10.1088/1126-6708/2008/04/065
  [arXiv:0802.0868 [hep-th]];\\
  %%CITATION = doi:10.1088/1126-6708/2008/04/065;%%
%
N.~E.~J.~Bjerrum-Bohr and P.~Vanhove,
%``Absence of Triangles in Maximal Supergravity Amplitudes,''
JHEP {\bf 0810}, 006 (2008)
%doi:10.1088/1126-6708/2008/10/006
[arXiv:0805.3682 [hep-th]];\\
  %%CITATION = doi:10.1088/1126-6708/2008/10/006;%%
%
N.~Arkani-Hamed, F.~Cachazo and J.~Kaplan,
%``What is the Simplest Quantum Field Theory?,''
JHEP {\bf 1009}, 016 (2010)
%doi:10.1007/JHEP09(2010)016
[arXiv:0808.1446 [hep-th]].
  %%CITATION = doi:10.1007/JHEP09(2010)016;%%

\end{thebibliography}
\end{document}